\documentclass[aps,pra,one column,superscriptaddress,amsmath,showpacs,tightenlines,longbibliography,10pt]{revtex4-2}
\usepackage{graphicx}

\usepackage{amssymb}
\usepackage{comment}
\usepackage{amsmath}
\usepackage[utf8x]{inputenc}
\usepackage[T1]{fontenc}
\usepackage{dsfont}

\usepackage{url}
\usepackage[colorlinks]{hyperref}
\hypersetup{%
    plainpages=true,
    breaklinks=true,
    hypertexnames=false,
    pageanchor=true,
    colorlinks=true,
    linkcolor={blue},
    citecolor={red},
    urlcolor={blue},
    anchorcolor={black}
    }

\usepackage[position=top,caption=false]{subfig}

\newcommand{\ket}[1]{| #1 \rangle}
\newcommand{\bra}[1]{\langle #1 |}
\newcommand{\bracket}[1]{\langle #1 \rangle}

\def\<{\langle}
\def\>{\rangle}

\newcommand{\cor}[1]{{g_{#1}^{(2)}(0)}}
\newcommand{\corTau}[1]{{g_{#1}^{(2)}(\tau)}}

\newcommand{\sgn}{{\rm sgn}}
\def\SMR{_{\rm SMR}}

\begin{document}

\title{Hybrid photon-phonon blockade}

\author{Shilan Abo}
\affiliation{Institute of Spintronics and Quantum Information,
Faculty of Physics, Adam Mickiewicz University, 61-614 Pozna\'{n},
Poland}

\author{Grzegorz Chimczak}
\affiliation{Institute of Spintronics and Quantum Information,
Faculty of Physics, Adam Mickiewicz University, 61-614 Pozna\'{n},
Poland}

\author{Anna Kowalewska-Kud\l{}aszyk}
\affiliation{Institute of Spintronics and Quantum Information,
Faculty of Physics, Adam Mickiewicz University, 61-614 Pozna\'{n},
Poland}

\author{Jan Pe\v{r}ina Jr.}
\affiliation{Joint Laboratory of Optics of Palack\'{y} University
and Institute of Physics of CAS, Faculty of Science, Palack\'{y}
University, 17. listopadu 12, 771 46 Olomouc, Czech Republic}

\author{Ravindra Chhajlany}
\affiliation{Institute of Spintronics and Quantum Information,
Faculty of Physics, Adam Mickiewicz University, 61-614 Pozna\'{n},
Poland}

\author{Adam Miranowicz}
\email[]{miran@amu.edu.pl} \affiliation{Institute of Spintronics
and Quantum Information, Faculty of Physics, Adam Mickiewicz
University, 61-614 Pozna\'{n}, Poland}

\date{\today}

\begin{abstract}
We describe a novel type of blockade in a hybrid mode generated by
linear coupling of photonic and phononic modes.  We refer to this
effect as hybrid photon-phonon blockade and show how it can be
generated and detected in a driven nonlinear optomechanical
superconducting system. Thus, we study boson-number correlations
in the photon, phonon, and hybrid modes in linearly coupled
microwave and mechanical resonators with a superconducting qubit
inserted in one of them. We find such system parameters for which
we observe eight types of different combinations of either
blockade or tunnelling effects (defined via the sub- and
super-Poissonian statistics, respectively) for photons, phonons,
and hybrid bosons. In particular, we find that the hybrid
photon-phonon blockade can be generated by mixing the photonic and
phononic modes which do not exhibit blockade.
\end{abstract}

\maketitle

\section{Introduction}
Photon blockade (PB)~\cite{Imamoglu1997}, also referred to as
optical state truncation (see reviews in Refs.~\cite{Adam2001,
Leonski2001}), or nonlinear quantum scissors (for a review~see
Ref.~\cite{Leonski2011}) is an optical analogue of Coulomb's
blockade. Specifically, it refers to the effect in which a single
photon, generated in a driven nonlinear system, can block the
generation of more photons. The light generated by an ideal (or
`true') PB exhibits both sub-Poissonian photon-number statistics
and photon antibunching. But even if one of these properties is
satisfied, the term PB is often used.

PB has been demonstrated experimentally in various driven
nonlinear systems with single~\cite{Birnbaum2005, Faraon2008,
Lang2011, Hoffman2011, Reinhard2011, Muller2015, Hamsen2017} and
two~\cite{Snijders2018, Vaneph2018} resonators, in a bimodal
cavity~\cite{Majumdar2012pra}, or even in cavity-free
systems~\cite{Peyronel2012}. Experimental platforms where PB was
observed include: cavity quantum electrodynamics (QED) with
Fabry-Perot cavities~\cite{Birnbaum2005}, photonic
crystals~\cite{Faraon2008}, and whispering-gallery-mode
cavities~\cite{Dayan2008}, as well as circuit
QED~\cite{Lang2011,Hoffman2011}. Note that the possibility of
producing a single-photon state in a driven cavity with a
nonlinear Kerr medium was predicted already
in~Refs.~\cite{Tian1992,Leonski1994,Adam1996}, but only the
publication of Ref.~\cite{Imamoglu1997}, where the term `photon
blockade' was coined, has triggered much interest in studying this
effect both theoretically and experimentally. Arguably, many
studies reported already in the 1970s and 1980s on photon
antibunching and sub-Poissonian light (see, e.g., reviews
in~Refs.~\cite{Paul1982, Teich1988, Kozierowski1980} and
references therein) are actually about PB-related effects,
although  such a relation  (to the optical analogue of Coulomb's
blockade) was not mentioned explicitly there.

In addition to the original idea of using PB as a single-photon
turnstile device with single~\cite{Imamoglu1997, Michler2000,
Dayan2008} or multiple~\cite{Wang2016photon} outputs, PB can have
much wider applications in quantum nonlinear optics at the
single-photon level, including single-photon induced nonlinear
effects, quantum noise reduction via antibunching of photons,
simulations of nonreciprocal nonlinear processes, or studying
chirality at exceptional points for quantum metrology, etc.

A number of generalisations of the standard single-PB effect were
proposed, which include: (i) two- and multi-photon versions of PB,
as first predicted in~Refs.~\cite{Shamailov2010,Adam2013} and
demonstrated experimentally in Refs.~\cite{Hamsen2017,
Chakram2020}; (ii) unconventional PB as predicted in
Ref.~\cite{Liew2010} and experimentally demonstrated in
Refs.~\cite{Snijders2018, Vaneph2018}; (iii) conventional and
unconventional nonreciprocal PB effects as predicted in
Refs.~\cite{Huang2018,Li2019} and (at least partially) confirmed
experimentally in Ref.~\cite{Yang2019}; (iv) state-dependent
PB~\cite{Adam2014}, (v) exceptional PB~\cite{Huang2022}, and (vi)
linear quantum scissors based on conditional measurements for:
single-PB~\cite{Pegg1998, Ozdemir2001, Ozdemir2002}, which was
experimentally demonstrated in Ref.~\cite{Babichev2003}, as well
as two-PB~\cite{Koniorczyk2000}, and
multi-PB~\cite{Adam2005,Adam2014cats} using multiport Mach-Zehnder
interferometers~\cite{Reck1994}. This probabilistic approach to PB
enables also nondeterministic quantum teleportation and more
selective optical-state truncations, e.g, hole burning in the
Hilbert space~\cite{Adam2007}. Concerning example (ii), note that
PB in two driven Kerr resonators was first studied in
Refs.~\cite{Leonski2004, Adam2006}, but only for relatively strong
Kerr nonlinearities. Surprisingly, PB remains in such
two-resonator systems even for extremely weak Kerr nonlinearities,
as first predicted in Ref.~\cite{Liew2010} and explained via
destructive quantum interference in Ref.~\cite{Bamba2011}. This
effect is now referred to as unconventional
PB~\cite{Flayac2018review}.

Here we study phonon blockade~\cite{Liu2010}, which is a
mechanical analogue of the mentioned blockade effects, i.e., the
blockade of quantum vibrational excitations of a mechanical
resonator. This effect has not been demonstrated experimentally
yet. However, a number of experimentally feasible methods have
been proposed for measuring it, including a magnetomotive
technique~\cite{Liu2010}, an indirect measurement of phonon
correlations via optical interferometry~\cite{Didier2011}, or by
coupling a mechanical resonator to a qubit, which is used not only
for inducing the resonator nonlinearity, but also to detect the
blockade effect itself, i.e., by measuring qubit's
states~\cite{XinWang2016phonon}. Among possible applications of
phonon blockade, we mention: testing nonclassicality of meso- or
macroscopic mechanical systems~\cite{Liu2010} and studying
single-phonon optomechanics, in addition to offering a source of
single- or multiple phonons~\cite{Adam2016,Shi2018}.

PB can be changed into light transmission~\cite{Liu2014}, e.g., by
photon-induced tunnelling (PIT)~\cite{Faraon2008}. This is another
nonclassical photon-number correlation phenomenon, in which the
probability of observing more photons in a higher manifold of the
system increases with the generation of the first photon near the
resonance frequency of the system. Multi-PIT effects were also
predicted~\cite{Huang2018}, including those generated by
squeezing~\cite{Kowalewska2019}.

For simplicity, we use here the abbreviation PB, when referring to
the blockade of not only photons, but also of phonons or hybrid
photon-phonon bosons. The precise meaning can be found from its
context, e.g., when we refer to a specific mode, including the
optical ($a$), mechanical ($b$), or hybrid ($c$) modes.
Analogously, PIT denotes a given particle-induced tunnelling among
the three types of excitations.

Nanomechanical resonators can coherently interact with
electromagnetic radiation~\cite{Aspelmeyer2014}, and quantum
correlations between single photons and single phonons were
studied for a single entangled photon-phonon pair~\cite{Xu2019} or
via photon and phonon blockade effects in optomechanical
systems~\cite{Xu2018}. A mechanical switch between PB and PIT has
been studied recently~\cite{Zhai2019}. PB and PIT effects in
systems comprising mechanical and optical resonators, which are
characterised by the same or similar bare frequencies, to our
knowledge, have not been studied experimentally yet, although they
seem to be experimentally feasible and, thus, they are at focus of
this paper.

Crucial signatures of PB and PIT can be observed by measuring the
second-order correlation function, $g^{(2)}(0)$. Specifically for
photons, (i) the condition of $g^{(2)}(0)<1$ defines the
sub-Poissonian photon-number statistics (also referred to as
zero-delay-time photon antibunching), which indicates the
possibility of observing PB, while (ii) the condition
$g^{(2)}(0)>1$, defines the super-Poissonian statistics (also
referred to as zero-delay-time photon bunching), which is a
signature of PIT in a given system. To observe the `true' effects
of PB and PIT, also other criteria should be satisfied, such as
nonzero-delay-time photon antibunching and higher-order
sub-Poissonian photon-number statistics. Indeed, an ideal
conventional PB, which can be served as a single-photon source,
usually should also be verified by studying higher-order
correlation functions, $g^{(n)}(0)$ for $n>2$. For example, in
case of single-PB (1PB) conditions $g^{(2)}(0)<1$ and
$g^{(n)}(0)<1$ for $n>2$ should be fulfilled.

PB can be verified also in other ways via demonstrating, e.g., a
staircase-like dependence of the mean photon number (or measured
power transmitted through a nonlinear resonator) on the energy
spectrum of the photons incident on the
resonator~\cite{Hoffman2011,Liu2014}. Such a dependence is the
photon analogue of the Coulomb staircase. All of the above
criteria are just necessary but not sufficient conditions for
demonstrating PB. A sufficient condition could be, e.g., showing a
high fidelity of a given generated light (with a nonzero mean
photon number) to an ideally truncated two-dimensional state,
which is the closest to the generated one. This approach was
applied in, e.g.,~\cite{Ozdemir2001, Ozdemir2002, Adam2013}. The
latter two types of PB tests are, however, are not applied in this
paper.

Conventional single-PB prevents the absorption of a second photon
with a specific frequency due to the nonlinearity of a given
system. Such a nonlinearity can be described by a Kerr-type
interaction and/or can induced by an atom (real or artificial)
coupled to a resonator. An artificial atom can be realised by,
e.g., a quantum dot~\cite{Michler2000,Santori2001,Ding2016} in
cavity QED~\cite{Muller2015} or a superconducting qubit or qudit
in circuit QED~\cite{Liu2014}.

Unconventional PB, which is induced by destructive interference,
operates better for very low (or even extremely low) mean photon
numbers~\cite{Snijders2018,Vaneph2018}. This can be
disadvantageous by considerably decreasing the probability of
generating a single photon. But, at the same time, it can be an
advantage, because a very small mean photon number usually reduces
the chance of generating multi-photon states and inducing
higher-order coherence. This is not always the case, and even if
the probability of observing two photons is suppressed,
higher-order coherence might be enhanced, leading to the
generation of multi-photon states~\cite{Flayac2018review}.

In this paper, we consider an optomechanical system, which
generates photonic and phononic modes. Then we apply a balanced
linear coupling transformation to the these modes to create hybrid
modes (also referred to as supermodes). We study the interplay
between photons and phonons resulting in their nonclassical number
correlation effects. Thus, we find such system parameters to
observe either PB or PIT in the four modes. In particular, we
predict PB in one of the hybrid modes, but not in the individual
(photon and phonon) modes, i.e., this PB is created from the two
modes, which do not exhibit PB. We refer to this effect as hybrid
photon-phonon blockade, which is the main result reported here.

Specifically, we define hybrid photon-phonon blockade as the
blockade of hybrid-mode bosons (polaritons) obtained by coupling
photons of an optical or microwave mode with phonons of a
mechanical mode by a balanced linear coupler. The idea and
criteria for testing this type of blockade are analogous to those
for other known blockade effects (e.g., of photons, phonons, or
magnons), but it is predicted for another type of bosons. We show
that this hybrid blockade can occur by coupling the modes, which
exhibit neither photon blockade nor phonon blockade.

To show this effect we analyse the system of two linearly-coupled
resonators: a superconducting microwave resonator (SMR), which
might be a transmission line resonator, and a micromechanical
resonator, referred to as a quantum drum (QD), which is
capacitively coupled to the SMR. To generate any kind of PB
(including unconventional PB), one needs to incorporate a
nonlinearity into a given system~\cite{Tian1992, Grangier1998,
Kimble1998}. This can be done by coupling  one of the resonators
(e.g., the SMR) to a qubit (e.g., an artificial superconducting
two-level atom). We also assume that the system is driven either
at the QD or the SMR as described in detail in the next section.

The paper is organised as follows: First, the hybrid
optomechanical system and its Hamiltonians are introduced. We also
define the hybrid photon-phonon modes, which can be generated by
the balanced linear coupling of photonic and phononic modes. Then,
we study the correlation effects in the photonic, phononic, and
one of the hybrid modes in the system driven at either the optical
or mechanical resonator, respectively, for experimentally feasible
parameters specified in Methods. We then predict and analytically
explain the generation of unconventional hybrid-mode blockade via
a non-Hermitian Hamiltonian method. We systematically study
different weaker and stronger criteria for observing blockade and
tunnelling effects in our system. We also find all the eight
combinations of the conventional blockade and tunnelling effects
in the three modes. In particular, we find a surprising effect
that the hybrid-mode photon-phonon blockade can be generated by
mixing the photonic and phononic modes exhibiting tunnelling
effects. In addition to this study of the second-order correlation
effects, we discuss also higher-order effects and their
classification in Methods. Moreover, we discuss two types of
schemes for measuring photon-phonon correlations in hybrid modes.
Finally, we summarise our results and indicate their potential
applications.

\begin{figure}[ht]
\centering
\includegraphics[width=0.4\linewidth]{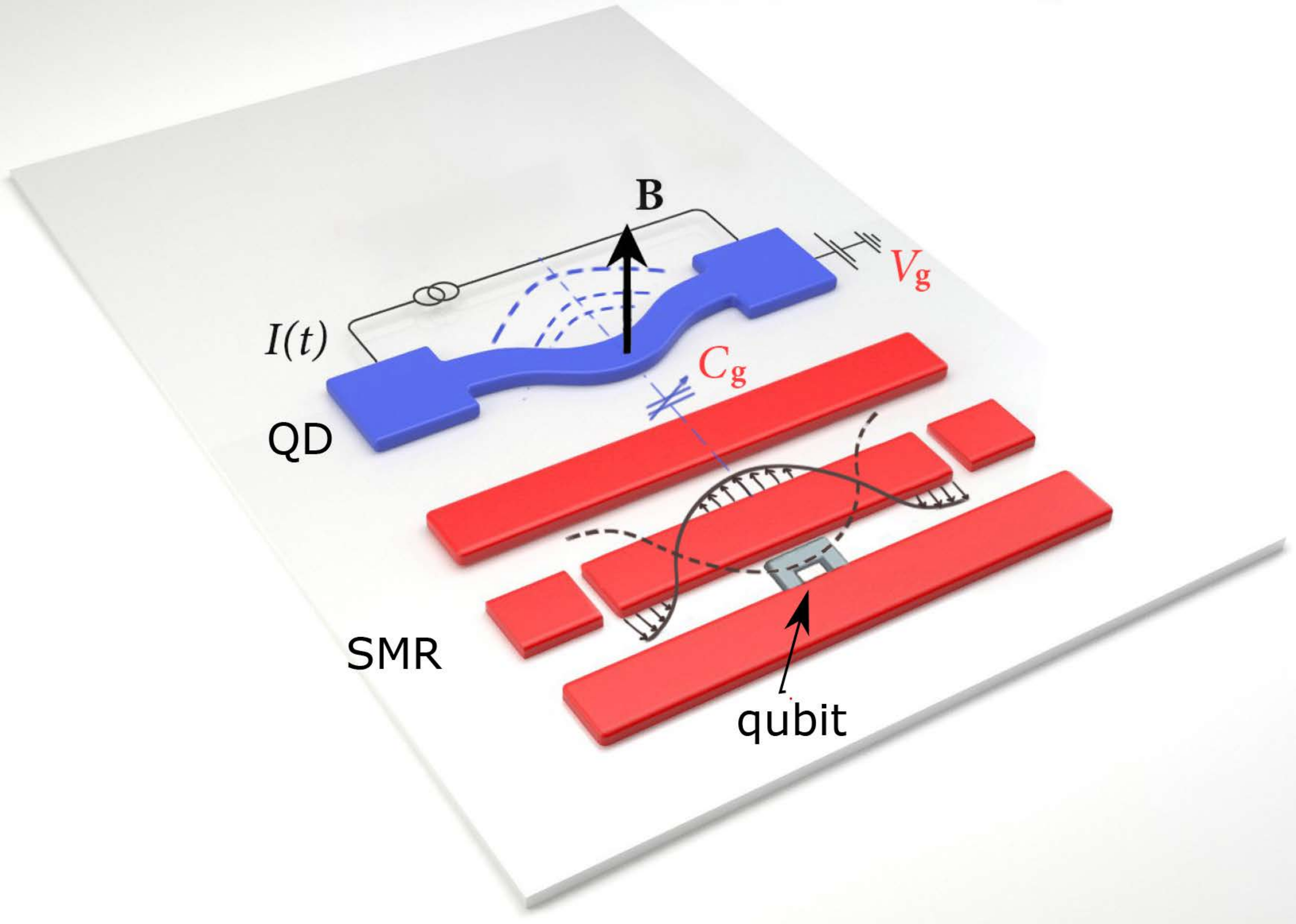}
\caption{Schematics of the discussed circuit-QED-based realisation
of the considered hybrid optomechanical system. It consists of a
superconducting qubit embedded in a superconducting microwave
resonator (SMR), e.g.,  a transmission-line resonator, to induce
its nonlinearity. A quantum micromechanical resonator, which is
referred to as a quantum drum (QD), is coupled to the SMR with a
tunable capacitor $C_g$. We assume that the system is driven
either at the SMR or QD. Dashed semicircular curves visualise that
the QD is oscillating. The driving and motion detection of the QD
can be realised by controlling the static magnetic field $B$,
potential $V_g$, and alternating current $I(t)$, as described in
Ref.~\cite{Liu2010} for detecting phonon blockade.} \label{fig01}
\end{figure}

\section{The system and Hamiltonians}\label{Section:Hamiltonians}

Figure~\ref{fig01} shows the schematics of the studied hybrid
system, which consists of a superconducting two-level artificial
atom (a qubit) embedded in a waveguide and coupled to an SMR,
which might be a transmission-line resonator. This qubit induces
anharmonicity in the SMR, which is crucial for observing PB. Our
setup includes also a microwave-frequency mechanical resonator (a
QD), which is capacitively coupled to the SMR. The nonlinearity of
the QD is induced indirectly by the linear coupling of the QD to
the effectively nonlinear SMR.

The free Hamiltonian of the SMR is $H_a=\hbar \omega_{\SMR}
a^\dagger a$, where $\omega_{\SMR}$ is its resonance frequency
(assumed here of the order of tens of GHz) and $a~(a^{\dagger})$
is the photon annihilation (creation) operator. We can reasonably
assume the SMR quality factor as $Q_{\SMR}\approx 10^4$. The free
Hamiltonian of the QD is $H_b=\hbar \omega_m b^\dagger b$, where
$\omega_m$ is its resonance frequency and $b~(b^{\dagger})~$ is
the phonon annihilation (creation) operator. In our numerical
simulations, we set $\omega_m/2\pi=7.8$~GHz and the QD quality
factor as $Q_m\approx260$. Moreover a two-level quantum system has
the ground state $\ket{g}$ and the excited state $\ket{e}$ with
transition frequency $\omega_q$ (set here of the order of
$\omega_m$ and $\omega_{\SMR}$). The free qubit Hamiltonian is
described as $H_q= \hbar\omega_{q} \sigma_{+} \sigma_{-}$, where
$\sigma_{+}=\ket{e}\bra{g}$ ($\sigma_{-}=\ket{g}\bra{e}$) is the
atomic raising (lowering) operator. Thus, the total free
Hamiltonian of the system is $H_0=H_a+H_b+H_q$. The complete
Hamiltonian (without driving) of our coupled system can be given
by ($\hbar=1$)
\begin{eqnarray}
 H'_{\pm}&=&H_0+g(a^\dagger \sigma_{-} +a\sigma_{+})
 +g_r(b+b^\dagger)a^\dagger a+g_l(a \pm a^{\dagger})(b+b^{\dagger}),
\label{Hprime_pm}
\end{eqnarray}
which includes the three coupling terms: (i) the Jaynes-Cummings
term describing the interaction between the SMR and qubit under
the rotating-wave approximation (RWA); (ii) the radiation-pressure
term with coupling strength $g_r$; and (iii) the Hopfield-type
nonlinear coupling term with strength $g_l$. The $g_l$ coupling
can be realized via a capacitor, as shown in Fig.~\ref{fig01} and
explained in a more detail in Ref.~\cite{Didier2011} for a similar
system. Note that the $g_l$ term describes canonical
position-position (momentum-position) interactions, where $g_l$ is
real (imaginary) for $H'_{+}$ ($H'_{-}$). These interactions can
be interchanged by adding the $\pi/2$ phase to ${a,a^\dagger}$,
i.e., $a\rightarrow i a$ and $a^\dagger\rightarrow -i a^\dagger$.
This extra phase does not change number correlations in the modes
$a$ and $b$. In typical ranges of parameters of analogous
superconducting circuits~\cite{Gu2017}, the $g_l$ term is
dominant, so the radiation-pressure term can be
neglected~\cite{Tian2009}. Moreover, although the counter-rotating
terms $ab\pm b^\dagger a^\dagger$, which appear in the
$g_l$-interaction, play an important role in the ultrastrong and
deep-strong coupling regimes~\cite{Kockum2019}. However, they can
be safely omitted under the RWA, which is valid in the weak and
strong coupling regimes. Indeed, the latter regimes are solely
studied in this paper, as discussed below. Then the Hopfield
nonlinear $g_l$-interaction becomes effectively linearised. Thus,
Hamiltonian~(\ref{Hprime_pm}) reduces to
\begin{eqnarray}
H_{\pm }&=&H_0
 +g(a^\dagger \sigma_{-} +a\sigma_{+})
 +f(a b^{\dagger} \pm a^{\dagger}b),
 \label{H_pm}
\end{eqnarray}
where the linear-coupling strength is denoted by $f$, which
replaces the symbol $g_l$. Analogously to $g_l$, $f$ is real
(imaginary) for $H_{+}$ ($H_{-}$). In the following, for
simplicity, we focus on studying the canonical position-position
interactions between the modes $a$ and $b$, as described by
$H_{+}$. The eigenstates of Hamiltonian $H_{\pm }$ can be referred
to as atomic-optomechanical polaritons or atom-cavity-mechanics
polaritons~\cite{Restrepo2014}. It is clear that Hamiltonians
$H_{\pm }$ conserve the polariton number,
\begin{equation}
  N_{\rm polariton}=a^\dagger a + b^\dagger b + \sigma_+\sigma_-,
  \label{N_polariton}
\end{equation}
which is the total number of excitations. Thus, $H_{\pm }$ can be
diagonalised in each subspace (or manifold) ${\cal H}^{(n)}$ with
exactly $n$ polaritons.

The RWA is fully justified assuming both (i) the weak- or
strong-couplings and (ii) small detunings between the SMR and QD,
and the SMR and qubit (see, e.g., Ref.~\cite{Larson2021}). We
stress that these conditions are fully satisfied for the
parameters applied in all our numerical calculations in this
paper. Thus, the Jaynes-Cummings and frequency-converter (or
linear-coupler) models can be applied. However, the RWA cannot be
applied in the ultrastrong and deep-strong coupling regimes, as
defined by $g> 0.1\,\omega_{i}$ and $g>\omega_{i}$,
respectively~\cite{Kockum2019}, where $i={\rm SMR}, m, q$. In
these regimes, the quantum Rabi and Hopfield models cannot be
reduced to the Jaynes-Cummings and frequency-converter  models,
respectively. However, we study the system for the parameters
specified in Eqs.~(\ref{setA1})--(\ref{setA3}), for which the
ratios of the coupling strengths and frequencies, $f/\omega_i$ and
$g/\omega_i$, are $<0.002$. So, the system is in the weak- or
strong-coupling regimes, and far away from the border line with
the USC regime. Moreover, the chosen detunings are
$|\omega_{\SMR}-\omega_m|/\omega_{\SMR} \le 2.6  \times 10^{-3}$
and $|\omega_{\SMR}-\omega_q|/\omega_{\SMR} < 8 \times 10^{-4}$.
Thus, it is clearly seen that we can safely apply the RWA. Anyway,
as a double test, we have calculated time-dependent second-order
correlation functions for the Hamiltonian $H'_{\pm}$ and $H_{\pm}$
for the parameters set in Eqs.~(\ref{setA1})--(\ref{setA3}) for
various evolution times assuming classical drives (as specified
below) and no dissipation. And we have found that the differences
between the correlation functions calculated for the models with
and without the RWA are negligible on the scale of figures. The
inclusion of dissipation in the system makes such differences even
smaller.

We assume that an optical pump field of frequency $\omega_{p}$ is
applied either to the SMR mode $a$, as described by
\begin{equation}
 H^{(a)}_{\rm drv}(t)=\eta_{a}(e^{i\omega_p t}a+e^{-i\omega_{p}t}a^{\dagger}),
 \label{H_drv_a}
\end{equation}
or to the QD mode $b$, as given by
\begin{equation}
 H^{(b)}_{\rm drv}(t)=\eta_{b}(e^{i\omega_p t}b+e^{-i\omega_{p}t}b^{\dagger}),
 \label{H_drv_b}
\end{equation}
to drive (excite) the system (with coupling strength $\eta_{a}$ or
$\eta_{b}$) from its ground state and to induce the emission of
photons and phonons. Thus, the total Hamiltonian becomes
\begin{eqnarray}
H^{(n)}(t)=H_+ +H^{(n)}_{\rm drv}(t) \hspace{1cm} (n=a,b).
 \label{H_n}
\end{eqnarray}
Direct driving of the QD can be implemented by a weak-oscillating
current, as considered in Refs.~\cite{Liu2010, Didier2011}, where
the drive strength $\eta_b$ is proportional to the current
amplitude $I(t)$ and the magnetic field $B$ shown in
Fig.~\ref{fig01}. The SMR can be driven in circuit-QED systems in
various ways~\cite{Gu2017}.

Note that by driving directly the SMR (or alternatively the QD),
one also indirectly drives the QD (SMR) through the capacitive
coupling $C_g$, as shown in the scheme in Fig.~\ref{fig01}. So, by
referring to the SMR- or QD-driven systems, we indicate only the
resonator, which is directly pumped, although finally both
resonators are driven.

The inclusion of an additional nonlinearity in the QD and/or
applying drives to the qubit(s) and both resonators is not
essential for the prediction of hybrid blockade, but this could
enable achieving stronger photon-phonon antibunching and more
sub-Poissonian statistics.

Considering the case, where the pump field drives only the SMR, to
remove the time dependence of the Hamiltonian $H^{(n)}(t)$ and to
obtain its steady-state solution, we transform the system
Hamiltonian into a reference frame rotating at frequency
$\omega_p$.

We apply the unitary transformation $U_R(t)=\exp\left(-i N_{\rm
polariton}\omega_p t \right)$ to $H^{(n)}$ according to the
general formula
\begin{eqnarray}
  H^{(n)}_{\rm rot}&=&U_R^\dagger H^{(n)} U_R -i U_R^\dagger \frac{\partial}{\partial t}
  U_R. \label{Hrot}
\end{eqnarray}
Thus, $H^{(a)}(t)$ reduces the time-independent SMR-driven
Hamiltonian:
\begin{eqnarray}
H'&\equiv& H^{(a)}_{\rm rot} = \Delta_{\SMR}a^\dagger a+\Delta_{m}
b^\dagger b +\Delta_{q} \sigma_{+} \sigma_{-} \nonumber\\&&
+g(a^\dagger \sigma_{-} +a\sigma_{+})+f(a^{\dagger}b+a
b^{\dagger})+\eta_{a}(a+a^{\dagger}), \label{Hamiltonian1}
\end{eqnarray}
where $\Delta_{i}=\omega_{i}-\omega_{p}$ for $i=a,b,q$. So, in
particular, $\Delta_{b}\equiv \Delta_{m}$~($\Delta_{a}\equiv
\Delta_{\SMR}$) is the mechanical (microwave) resonator frequency
detuning with respect to the pump frequency. Analogously, in the
same rotating frame, $H^{(b)}(t)$ reduces to the QD-driven
Hamiltonian:
\begin{eqnarray}
H''&\equiv& H^{(b)}_{\rm rot} =  \Delta_{\SMR}a^\dagger
a+\Delta_{m} b^\dagger b +\Delta_{q} \sigma_{+} \sigma_{-}
\nonumber\\&& +g(a^\dagger \sigma_{-}
+a\sigma_{+})+f(a^{\dagger}b+a
b^{\dagger})+\eta_{b}(b+b^{\dagger}). \label{Hamiltonian2}
\end{eqnarray}
We recall that Eqs.~(\ref{Hamiltonian1}) and~(\ref{Hamiltonian2})
are directly derived from Eq.~(\ref{H_n}) for $H_+$ given in
Eq.~(\ref{H_pm}). Moreover, $H_+$ is derived from
Eq.~(\ref{Hprime_pm}) assuming the RWA, which is justified for
small detunings in the weak- and strong-coupling regimes, which
are the only numerically studied regimes in this paper, as
emphasised above. Indeed, the studied ranges of parameters
guarantee the system evolution is far from the USC regime. Note
that the Hamiltonian $H'_+$ in~(\ref{Hprime_pm}) for $g_r=0$ with
an additional drive term $H^{(n)}_{\rm drv}(t) $ can be
transformed, according to Eq.~(\ref{Hrot}), to $H^{(n)}_{\rm rot}$
given in Eqs.~(\ref{Hamiltonian1}) and~(\ref{Hamiltonian2}) but
with the additional term $f[ a b \exp(2i\omega_p t)+{\rm h.c.}]$.
In all our numerical calculations we set $\omega_p$ of the order
of GHz. Thus, the effect of this rapidly oscillating term is
negligible compared to all the other terms in the Hamiltonians.
Moreover, we have also assumed that the optomechanical term $g_r$
is negligible. In general, this assumption is not necessary,
because the $g_r$ term can be reduced (in the red-detuned regime)
to an interaction term describing a linear coupler (or a beam
splitter), which can be combined with the $f$ term. Anyway, for
simplicity concerning both theory and potential experiments, we
set $g_r=0$. We have also assumed that the system is driven at
either the mechanical or optical mode to obtain effectively
time-independent Hamiltonians in a rotating frame. This
simplification would not be directly possible by considering the
system driven simultaneously at both modes with different
frequencies.

\begin{figure}[ht]
\centering
\includegraphics[width=0.7\linewidth]{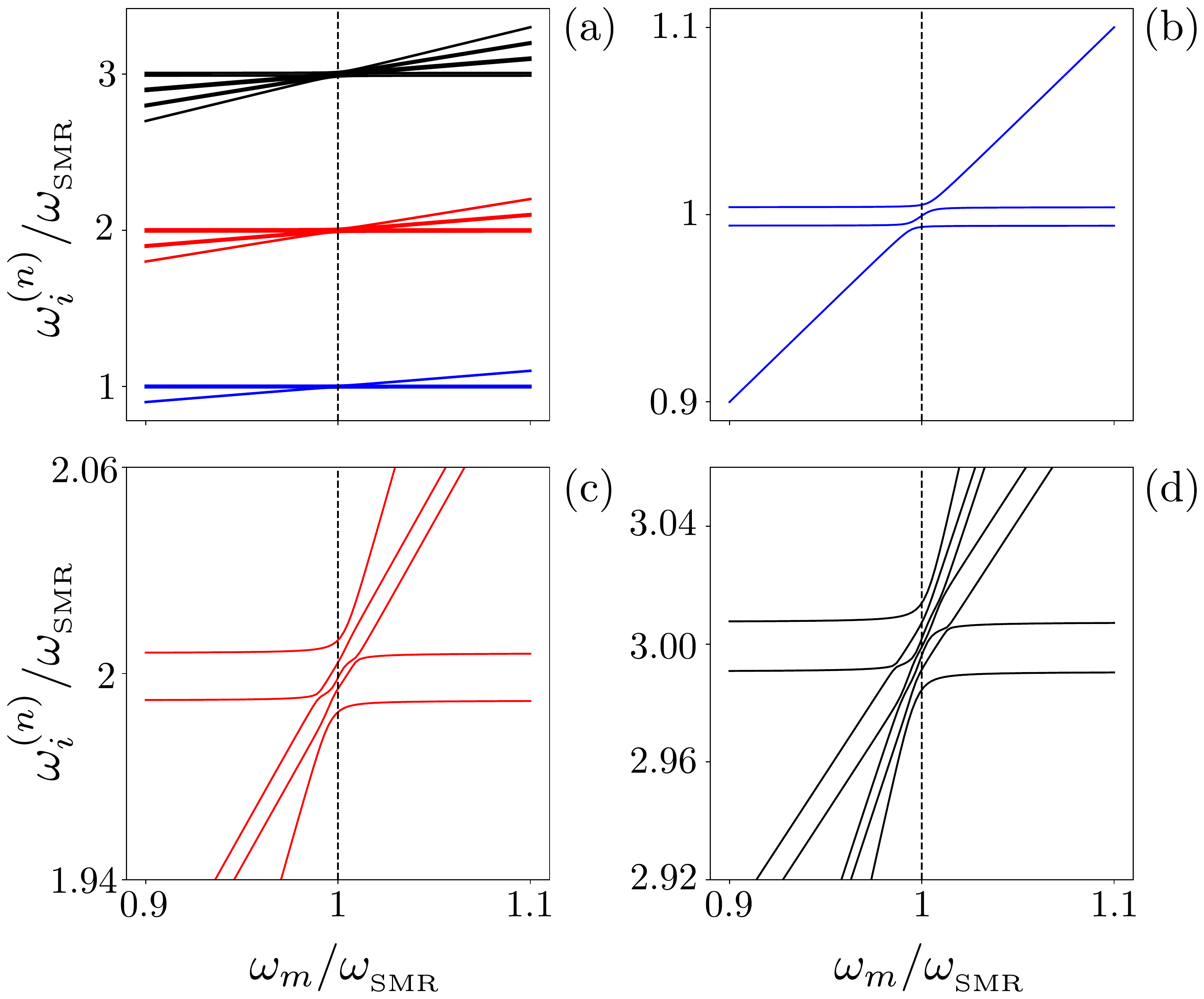}
\caption{Energy levels $\omega_n$ versus the QD frequency
$\omega_m$ in units of the SMR frequency  $\omega_{\SMR}$ for the
Hamiltonian Eq.~(\ref{H_pm}) with the parameters given in
Eq.~(\ref{setA1}) and $g=7.5\gamma$. The three  manifolds of the
lowest energy levels in panel (a) are zoomed in panels (b-d) near
the resonance $\omega_m=\omega_{\SMR}$ to reveal the anti-crossing
of energy levels. Here, $\omega^{(n)}_i$ (with $n=1,2,3$) denotes
the frequencies of the $n$th manifold.} \label{fig02}
\end{figure}

Figure~\ref{fig02} shows the structure of the energy spectrum for
the hybrid system Hamiltonian~\eqref{H_pm}. To study the
sub-Poissonian light generation in hybrid modes, we apply to the
SMR and QD modes a balanced linear coupling transformation, which
is formally equivalent to a balanced (50/50) beam splitter (BS).
This transformation creates the hybrid (or cross) photon-phonon
modes:
\begin{equation}
c=\frac{a+b}{\sqrt{2}},\quad d=\frac{a-b}{\sqrt{2}},
\label{eq:c,d}
\end{equation}
for the system described by $H_+$ and related Hamiltonians. Note
that if this BS transformation is modified as $a\rightarrow -i a$
and $a^\dagger \rightarrow i a^\dagger$ [which compensate the
extra $\pi/2$ phase introduced below Eq.~(\ref{Hprime_pm})] then
all our predictions of number correlations shown in various
figures for the hybrid mode $c$ (in addition to those for the
modes $a$, $b$, and $d$) are the same as those for the model
described by $H_-$.

Thus, the Hamiltonian $H'$ after the BS transformation reads
\begin{eqnarray}
  H'_{\rm BS} &=& \Delta_c c^\dagger c+\Delta_d d^\dagger d + \Delta_q \sigma_{+}\sigma_{-}
  +\delta (c^\dagger d +d^\dagger c)
\nonumber\\
  &&
  +\frac{1}{\sqrt{2}}\big[\eta_{a}(c+c^\dagger)
  +\eta_{a}(d+d^\dagger)
   + g(c^\dagger \sigma_{-}+c\sigma_{+}) +
   g(d^\dagger \sigma_{-} +d \sigma_{+})\big],
  \label{Eq:crossCavities}
 \end{eqnarray}
which describes the qubit interacting with two hybrid modes $c$
and $d$, where $\Delta_{c,d}=(\omega_{\SMR}+
\omega_m)/2-\omega_p\pm f$ and $\delta=(\omega_{\SMR}-
\omega_m)/2$. It is seen that the two modes $c$ and $d$ have no
direct coupling if $\omega_m=\omega_{\SMR}$.

The dynamics of an open system in the presence of losses under the
Markov approximation can be described within the Lindblad approach
for a system reduced density matrix $\rho$ satisfying the standard
master equation,
\begin{equation}
    \frac{\partial \rho}{\partial t}=
    -i[H, \rho]+\kappa_a \mathcal{D}[a]\rho+\kappa_b \mathcal{D}[b]\rho+\gamma \mathcal{D}[\sigma]\rho,
    \label{ME}
\end{equation}
which is given in terms of the Lindblad superoperator
$\mathcal{D}[O]\rho = \frac{1}{2}(2O\rho O^\dagger-\rho O^\dagger
O-O^\dagger O\rho)$, where $\kappa_a$, $\kappa_b,$ and $\gamma$
are the decay rates for the SMR, QD, and qubit, respectively.

All our numerical calculations and their analyses are given for
the system parameters, which satisfy the conditions for the
strong-coupling regime and for small detunings between the SMR,
QD, and qubit. Thus, we can safely apply the standard master
equation given in Eq.~(\ref{ME}). Of course, if one considers
Eq.~(\ref{Hprime_pm}) for the system in the USC or deep-coupling
regimes, then the master equation in Eq.~(\ref{ME}), should be
replaced by a generalised one, e.g., of Refs.~\cite{Ridolfo2012,
Garziano2015, Kockum2019, Mercurio2022}.

We also note that the application even of a single classical drive
to the Jaynes-Cummings model in the strong-coupling regime
effectively creates counter-rotating terms, which can induce a
variety of USC effects, as shown explicitly in
Ref.~\cite{Sanchez2020}. Thus, to confirm the validity of our
results, we have applied the generalised formalism described in
Ref.~\cite{Ridolfo2012}, which is valid for arbitrary light-matter
coupling regimes, including the weak-, strong-, and USC regimes.
In particular, we calculated the correlation functions
$g^{(n)}(0)$ defined in terms of the positive- ($X^+_n$) and
negative- [$X^-_n=(X^+_n)^{\dagger}$] frequency components of the
canonical position operators: $X_a = a + a^{\dagger}$ for photons,
$X_b=b+b^{\dagger}$ for phonons, and $X_c=c+c^{\dagger}$ for
hybrid-mode bosons in the qubit-SMR-QD dressed basis. We
calculated the steady states of the system by solving numerically
the generalised master equation of Ref.~\cite{Ridolfo2012} for the
Hamiltonians $H'$ and $H''$. As expected from general
considerations, our numerical calculations for the parameters set
in Eqs. (\ref{setA1})--(\ref{setA3}) using the standard and
generalised formalisms based on $H'$ (as well as $H''$) give
effectively the same results.

In our simulations, we assume that the system is prepared in the
ground state $\ket{n=0, g}\ket{m=0}$ (i.e., with no photons in the
SMR, no phonons in the QD, and the qubit is in the ground state),
such that a given pump laser can drive the SMR photons in the
microwave frequency range. Note that the choice of initial states
affects the short-time evolution of our system, but has no effect
on the steady-state solutions in the time limit, assuming the
single-photon and single-phonon damping channels, as described in
Eq.~(\ref{ME}). However, as shown in Ref.~\cite{Adam2014}, initial
states of a system can indeed affect  steady states of the system,
thus can also change PB, in case of quantum engineered dissipation
channels allowing for, e.g., two-photon dissipation only.

In the following sections we show that it is possible to observe
both PB and PIT in the hybrid mode in the weak, mediate, and
strong coupling regimes compared to the decay rates of the SMR,
QD, and qubit. In particular, we show that the system can generate
the hybrid photon-phonon modes with strongly sub-Poissonian (or
super-Poissonian) statistics by mixing the SMR and QD modes with
strongly super-Poissonian (or sub-Poissonian) statistics.

\begin{table*}[tbp]
\begin{tabular}{ cccccc}
 \hline\hline
Case & $f_{abc}$ & PNS in mode $a$ & PNS in mode $b$  & PNS in mode $c$ & colour  \\
\hline
1& $(-,-,-)$ &  sub-Poissonian    &  sub-Poissonian   & sub-Poissonian    & aquamarine\\
2& $(-,-,+)$ &  sub-Poissonian    &  sub-Poissonian   & super-Poissonian  & lime \\
3& $(-,+,-)$ &  sub-Poissonian    &  super-Poissonian & sub-Poissonian    & light cyan\\
4& $(+,-,-)$ &  super-Poissonian  &  sub-Poissonian   & sub-Poissonian    & mint cream\\
5& $(-,+,+)$ &  sub-Poissonian    &  super-Poissonian & super-Poissonian  & plum\\
6& $(+,-,+)$ &  super-Poissonian  &  sub-Poissonian   & super-Poissonian  & pink\\
7& $(+,+,-)$ &  super-Poissonian  &  super-Poissonian & sub-Poissonian    & yellow\\
8& $(+,+,+)$ &  super-Poissonian  &  super-Poissonian & super-Poissonian  & cyan \\
 \hline
 \hline
\end{tabular}
\caption{Different predictions of the super- and sub-Poissonian
particle (i.e., photon, phonon or hybrid photon-phonon)-number
statistics (PNS) corresponding, respectively, to PIT and PB, for
the photon mode $a$, phonon mode $b$, and hybrid photon-phonon
mode $c$, where
$f_{abc}=\big(\sgn[g_a^{(2)}(0)-1],\sgn[g_b^{(2)}(0)-1],\sgn[g_c^{(2)}(0)-1]\big)$
and the last column indicates each prediction of the mode $a$,
$b$, and $c$ in the specific colour that is used in our plots. All
these cases can be seen in Fig.~\ref{fig10}.} \label{table8}
\end{table*}
\begin{table*}[ht]
 \begin{tabular}{cccccc}
 \hline\hline
 Case & Effect & Single-time  & Two-time     & Example      & Figure \\
      &        & correlations & correlations & of $g/\kappa_{a}$ & \\
\hline
 I  & stronger form of PB  & sub-Poissonian PNS   & phonon antibunching & 3.0 & \ref{fig07}(a) \\
    & (`true' PB)       & $\cor{b}<1$       & $\corTau{b} >\cor{b}$& & \\
 II & stronger form of PIT & super-Poissonian PNS & phonon bunching     & 2.1 & \ref{fig07}(b)\\
    & (`true' PIT)      & $\cor{b}>1$       & $\corTau{b} <\cor{b}$& & \\
 III& weaker form of PIT or PB & super-Poissonian PNS & phonon antibunching & 3.8 & \ref{fig07}(c)\\
    & & $\cor{b}>1$       & $\corTau{b} >\cor{b}$& &\\
 IV & weaker form of PB or PIT & sub-Poissonian PNS   & phonon bunching     & 2.2 & \ref{fig07}(d)\\
    & & $\cor{b}<1$       & $\corTau{b} <\cor{b}$& \\
 \hline
 \hline
\end{tabular}
\caption{Different single- and two-time phonon-number correlation
effects induced in the QD mode, which can be observed for
different values of the qubit-SMR coupling strength $g$ with
respect to the SMR decay rate $\kappa_{a}$, e.g., by setting the
other parameters to be the same as in Eq.~(\ref{setA3}). Here, PNS
stands specifically for the phonon-number statistics of the mode
$b$. Note that we also found examples of Cases I, II, and IV for
the modes $a$ and $c$ using the same system parameters as for the
mode $b$. } \label{table4}
\end{table*}

\section{Hybrid-mode blockade in the SMR-driven system}\label{Section:SMR-driven}

Here we analyse in detail various blockade and PIT effects in the
SMR-driven dissipative system described by the Hamiltonian $H'$
and the master equation~(\ref{ME}) for the parameters specified in
Eq.~(\ref{setA1}).

Photon/phonon-number statistics of the modes generated by our
hybrid system can be described quantitatively by calculating the
zero-delay-time $k$th-order correlation function ($k$th-order
intensity autocorrelation function),
\begin{equation}
 g_z^{(k)}(0)=\lim_{t\to \infty}\frac{\langle z^{\dagger k}(t) z^k(t) \rangle}
 {\langle z^\dagger(t) z(t)\rangle ^k},
 \label{gk0}
\end{equation}
where $z=a,b,c,d$ and $k=2,3,...$. In the special case of $k=2$,
which is of  particular interest in testing single-PB and
single-PIT, the three different types of the boson-number
statistics can be considered: the Poissonian [if $g^{(2)}(0)=1$],
super-Poissonian [if $g^{(2)}(0)>1$], and sub-Poissonian
(otherwise). Analogously, one can define higher-order Poissonian,
sub-Poissonian, and super-Poissonian statistics for $k>2$. Such
higher-order criteria are not only crucial in analysing multi-PB
and multi-PIT effects~\cite{Hamsen2017,Huang2018,Kowalewska2019},
but they are also important in testing whether a specific PB
effect is a `true' PB, which can be used for generating single
photons or phonons. These higher-order statistics are studied in
Methods.

\begin{figure}[ht]
\centering
\includegraphics[width=0.8\linewidth]{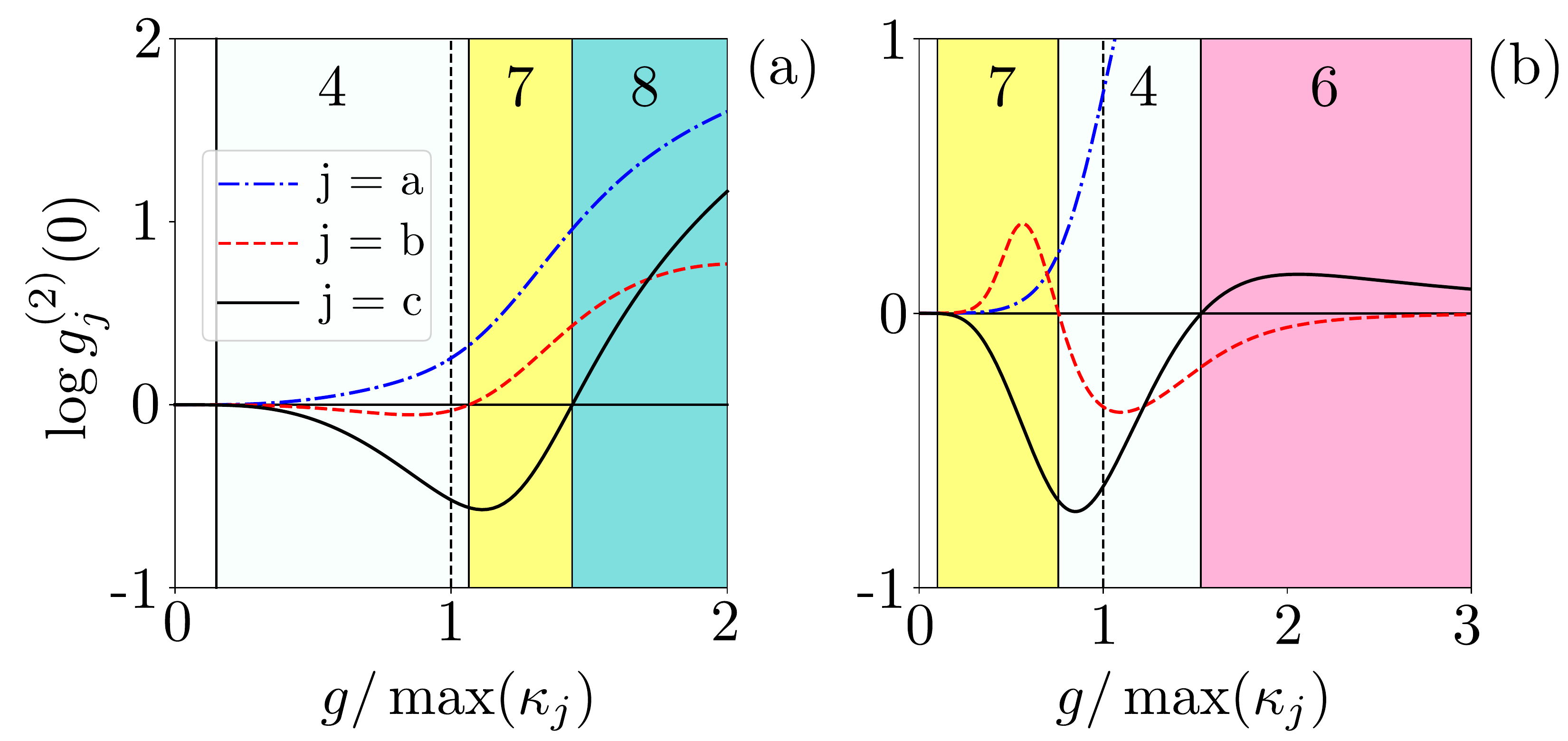}
\caption{Second-order correlation functions $\cor{i}$ (in the
common logarithmic scale) versus the ratio of qubit-SMR coupling
strength and the largest decay rate. Different predictions of the
sub- and super-Poissonian boson number statistics, which can be
interpreted, respectively, as the PB and PIT effects, of the
photonic ($a$), phononic ($b$), and hybrid ($c$) modes assuming:
(a) the SMR-driven system with parameters specified in
Eq.~(\ref{setA1}) and (b) the QD-driven system with
Eq.~(\ref{setA2}). All the shown Cases (i.e., 4, 6, 7, and 8)
correspond to those listed in Table~\ref{table8}. The broken line
at $g=\max\kappa_j$ is the border line between the strong- and
weak-coupling regimes.} \label{fig03}
\end{figure}

Figure~\ref{fig03}(a) shows $g^{(2)}(0)$ as a function of the
qubit-SMR coupling for the SMR-driven system with the parameters
specified in Eq.~(\ref{setA1}). The regions, when the
sub-Poissonian statistics in the hybrid mode $c$ is accompanied by
the super-Poissonian statistics in the modes $a$ and $b,$ are
indicated by the yellow background in this and other figures. This
area in yellow colour is referred to as Case~7 in
Table~\ref{table8}, in which we observe strongly super-Poissonian
photons (phonons) in the SMR (QD); whereas  a single excitation is
observed in the hybrid mode. The system parameters, which lead to
Case 7, are found by numerical simulations and are discussed
below.

Note that Fig.~\ref{fig03}(a) shows these effects in the strong
coupling regime~\cite{Kockum2019}, i.e., when the qubit-SMR
coupling constant $g$ is larger than the system damping rates:
$g/\kappa_{\rm \max}>1$, where $\kappa_{\rm \max}=\max\{\kappa_a,
\kappa_b, \gamma\}$. On the other hand, Fig.~\ref{fig03}(b) shows
the same yellow region in the weak-coupling regime, i.e., when
$g/\kappa_{\rm \max}<1$, but this figure was calculated for the
QD-driven system, which is discussed in the next section.

By considering the values of Eq.~\eqref{setA1}, the SMR decay rate
is $\kappa_{a}=1.5\gamma$, given that the  mode $a$ is always in
the strong qubit-SMR coupling regime in the region of our
interest. This results in Rabi-type oscillations of $g^{(2)}(0)$
that occur in the SMR mode $a$ and the hybrid mode $c$. In
Fig.~\ref{fig03}(a) both weak and strong coupling regimes are
shown corresponding to $g$ smaller or larger than the maximum
decay rate of the whole system.

Given the set of parameters in Eq.~\eqref{setA1}, we are in the
good-cavity regime~\cite{KuhnBook}, because
$\kappa_{a}<\{\kappa_b,g,f\}$. In the  range $g/2\pi \in
(4.5,42)$~MHz, the hybrid mode $c$ has the sub-Poissonian
statistics, while the SMR mode has the super-Poissonian statistics
in all the shown cases and a very weak sub-Poissonian statistics
occur for phonons in the QD mode $b$, but still corresponding to
Case~4 in Table~\ref{table8}. This behaviour changes to the
super-Poissonian statistics in the mode $b$, which corresponds to
Case~7, as shown in Fig.~\ref{fig03}(a). There is a transition for
the mode $c$ from the sub-Poissonian to super-Poissonian
statistics, which corresponds to switching from Case~7 to Case~8
in the strong-coupling regime, where the other two modes are both
super-Poissonian. Observing $g^{(2)}(0)>1$ witnesses PIT and the
quantum nature of this effect is explored further below.

In order to better probe and understand the dynamics of the system
in specific parameter regimes, we analyse also the delay-time
second-order photon correlation function, defined as
\begin{eqnarray}
g_z^{(2)}(\tau) &=& \lim_{t\to \infty}\frac{\langle{\cal T} :
n_z(t+\tau)n_z(t) :\rangle}{\langle n_z(t)\rangle^2}
= \lim_{t\to \infty}\frac{\langle z^\dagger(t) z^\dagger
(t+\tau)z(t+\tau)z(t) \rangle}{\langle z^\dagger(t) z(t)\rangle^2}
, \label{g2tau}
\end{eqnarray}
where $n_z(t)=z^\dagger(t) z(t)$ is the boson number in the modes
$z=a,b,c,d$, and the operator products are written in normal order
(::) and in time order ${\cal T}$. With $g_z^{(2)}(\tau)$ another
quantum optical number-correlation phenomenon can be investigated.
Specifically, in case of photons, it is referred to as photon
antibunching if $g^{(2)}(0)<g^{(2)}(\tau)$, photon unbunching if
$g^{(2)}(0)\approx g^{(2)}(\tau)$, and photon bunching if
$g^{(2)}(0)>g^{(2)}(\tau)$, which is usually defined for short or
very short delay times $\tau$~\cite{MandelBook}. It is worth
noting that photon antibunching was first experimentally observed
in the 1970s by Kimble, Dagenais, and Mandel~\cite{Kimble1977}.
This was historically the first experimental demonstration of the
quantum nature of an electromagnetic field, which cannot be
explained classically, unlike photoelectric bunching.

Analogously, one can also investigate the antibunching and
bunching of phonons and/or hybrid-mode bosons. Note that the term
photon antibunching is often interchangeably used with the
sub-Poissonian photon-number statistics~\cite{Teich1988}. However,
to avoid confusion, one can refer to single-time (or
zero-delay-time) photon antibunching if defined by $g^{(2)}(0)$
and two-time (or delay-time) photon antibunching if defined via
$g^{(2)}(\tau)$.

\begin{figure}[ht]
\centering
\includegraphics[width=\linewidth]{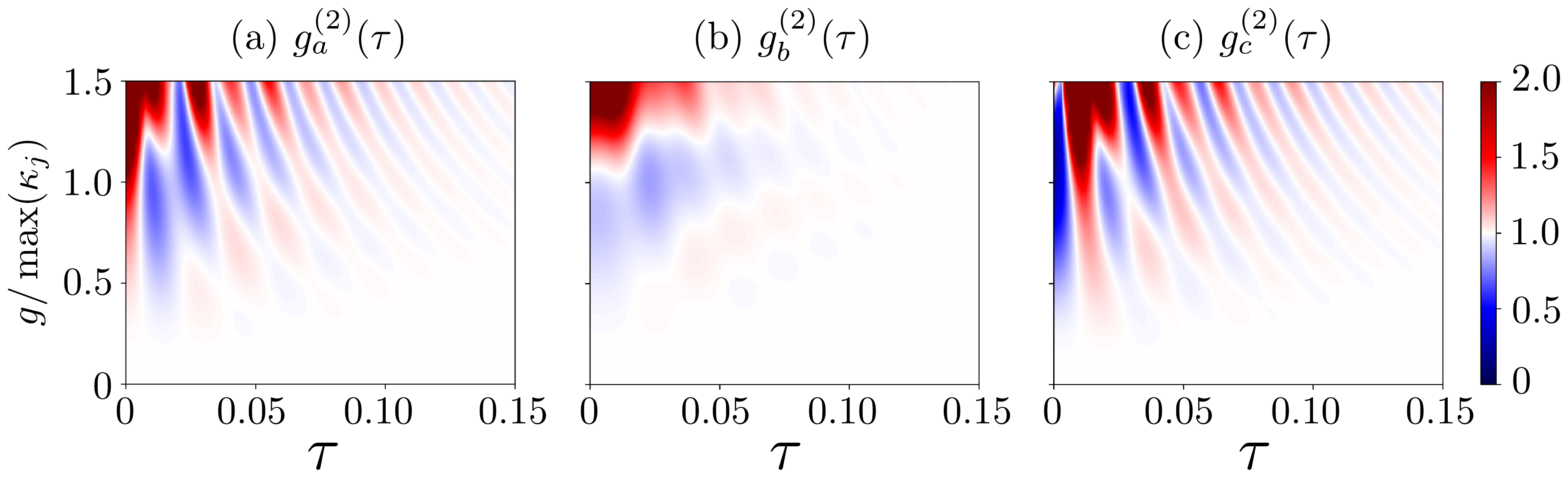}
\caption{Delay-time second-order correlation functions: (a)
$\corTau{a}$ for the photonic mode, (b) $\corTau{b}$ for the
phononic mode, and (c) $\corTau{c}$ for the hybrid mode versus the
coupling strength $g$ (in units of $\kappa_{a}$) and the delay
time $\tau$. We consider here the SMR-driven system with
parameters specified in Eq.~(\ref{setA1}), which enable us to
observe the single-photon resonances in the mode $c$. For clarity,
all the values of the correlation functions $\ge 2$ are truncated
at 2.} \label{fig04}
\end{figure}

\begin{figure}[ht]
\centering
\includegraphics[width=0.9\linewidth]{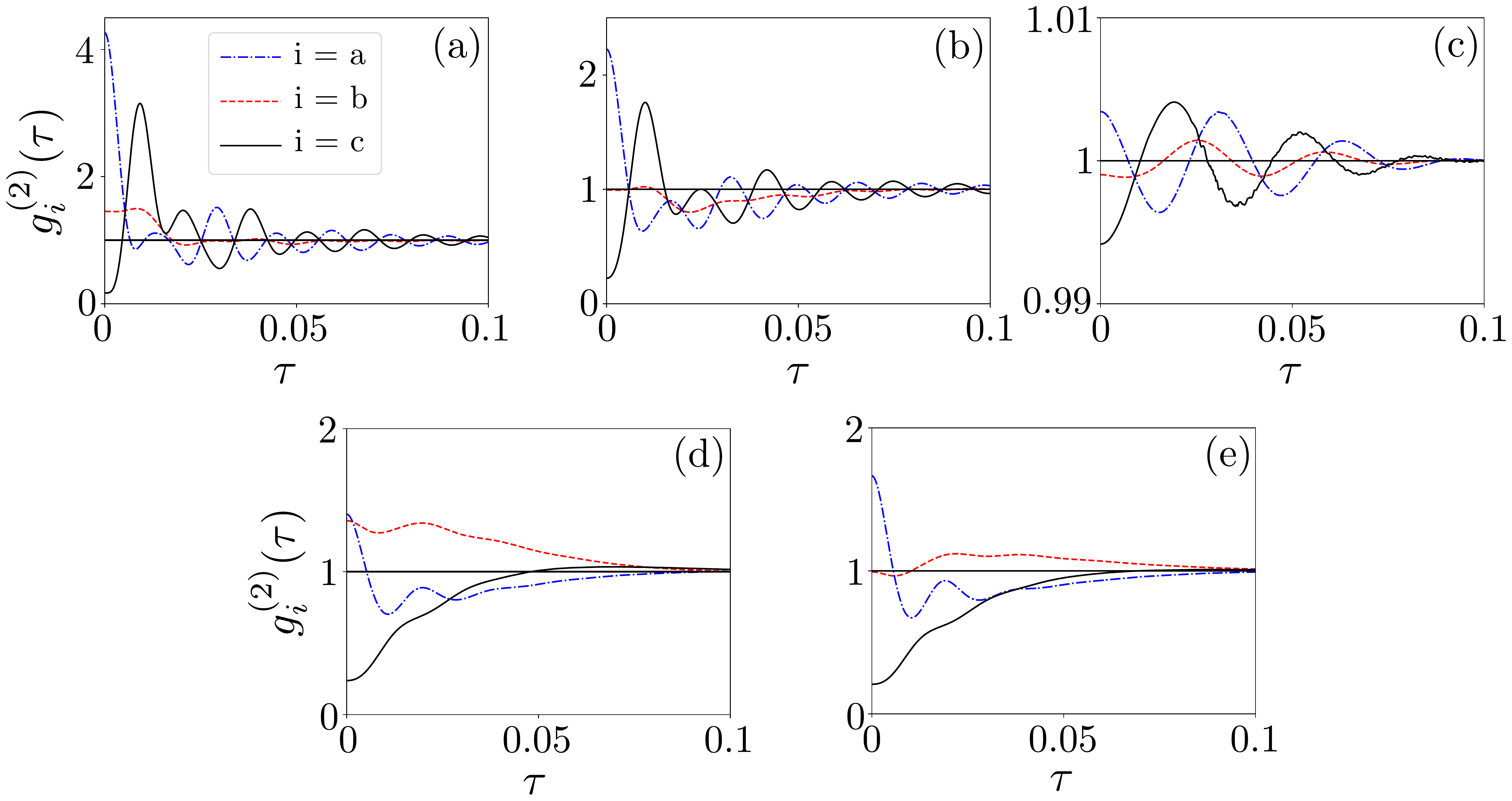}
\caption{Delay-time second-order correlation functions
$\corTau{i}$ for the SMR mode $a$, the QD mode $b$, and the hybrid
mode $c$ modes assuming: (a,b,c) the SMR-driven system specified
in Eq.~(\ref{setA1}) with $f=5.5\gamma$ and
$\kappa_{\max}=\kappa_b=6\gamma$, and (d,e) the QD-driven system
in Eq.~(\ref{setA2}) with $\kappa_{\max}=7.5\gamma$, where we
additionally set:
 (a) $g=1.3\kappa_{\max}=7.8\gamma$,
 (b) $g=1.1\kappa_{\max}=6.6\gamma$,
 (c) $g=0.2\kappa_{\max}=1.2\gamma$,
 (d) $g=0.7\kappa_{\max}=5.25\gamma$, and
 (e) $g=0.758\kappa_{\max}=5.685\gamma$.}
\label{fig05}
\end{figure}

In Fig.~\ref{fig04}, we plotted $g^{(2)}(\tau)$ for the range
$[0,0.15]$ of $g/\kappa_{\max}$. This range is also shown in
Fig.~\ref{fig03}(a), where the examples of Cases~4 and 7 can be
identified. As expected, one can see oscillations in the SMR and
hybrid modes in Figs.~\ref{fig04}(a) and~\ref{fig04}(c),
respectively. These oscillations are induced by the competition
between the qubit-SMR coupling $g$ and the SMR-QD hopping $f$ in
our system. Apparently, by analysing $g^{(2)}(\tau)$ in the
weak-coupling regime, the frequency of the oscillations is smaller
than that in the strong-coupling regime, in which the oscillations
are caused by both couplings $g$ and $f$. Moreover in a very weak
coupling regime, where $g\ll 1$ oscillations occur due to the
hopping strength $f$, with the period
$2\pi/f$~\cite{Verhagen2012}. This means that, in the
weak-coupling regime, also the coupling between the SMR and QD can
generate oscillations in our system, where in this case the period
of oscillations, which are induced by $f=5.5\gamma$, is
approximately equal to $\tau\approx 0.036$, which coincides with
the period deduced from the graph, as seen in Fig.~\ref{fig05}(c).
These detrimental oscillations should be suppressed on a time
scale longer than the SMR lifetime $\tau=1/\kappa_a$ to enable
boson antibunching to survive in the area of our interest.

Various combinations of correlations effects are shown in
Fig.~\ref{fig05}. All panels in Fig.~\ref{fig05} show that the
photon mode $a$ is super-Poissonian and bunched, while the hybrid
mode $c$ is sub-Poissonian and antibunched. However, the
properties of the phonon mode $b$ are different in every panel.
Specifically, the mode $b$ is in panel:
 (a) super-Poissonian and unbunched [defined as
$g_b^{(2)}(0)\approx g_b^{(2)}(\tau)$ for non-zero but short delay
times $\tau$],
 (b) Poissonian and unbunched,
 (c) sub-Poissonian and unbunched,
 (d) super-Poissonian and bunched, and
 (e) Poissonian and bunched,
as usually considered for very short delay times $\tau$. Note that
panels (a,b,c) are for the SMR-driven system, while the remaining
panels (d,e) are for the QD-driven system, which are discussed in
detail in the next section.

In particular, it is seen that by decreasing the coupling at
$g/\kappa_b=1.1$ in Fig.~\ref{fig05}(b), the QD mode $b$ is
unbunched with the Poissonian statistics, while the hybrid mode
$c$ exhibits antibunching $g^{(2)}(0) < g^{(2)}(\tau)$ and the
sub-Poissonian statistics $g^{(2)}(0)<1$, in both cases. The role
of the auxiliary mode $b$ is, in a sense, to convert the
super-Poissonian into sub-Poissonian statistics in the mode $c$.

\begin{figure}[ht]
\centering
\includegraphics[width=\linewidth]{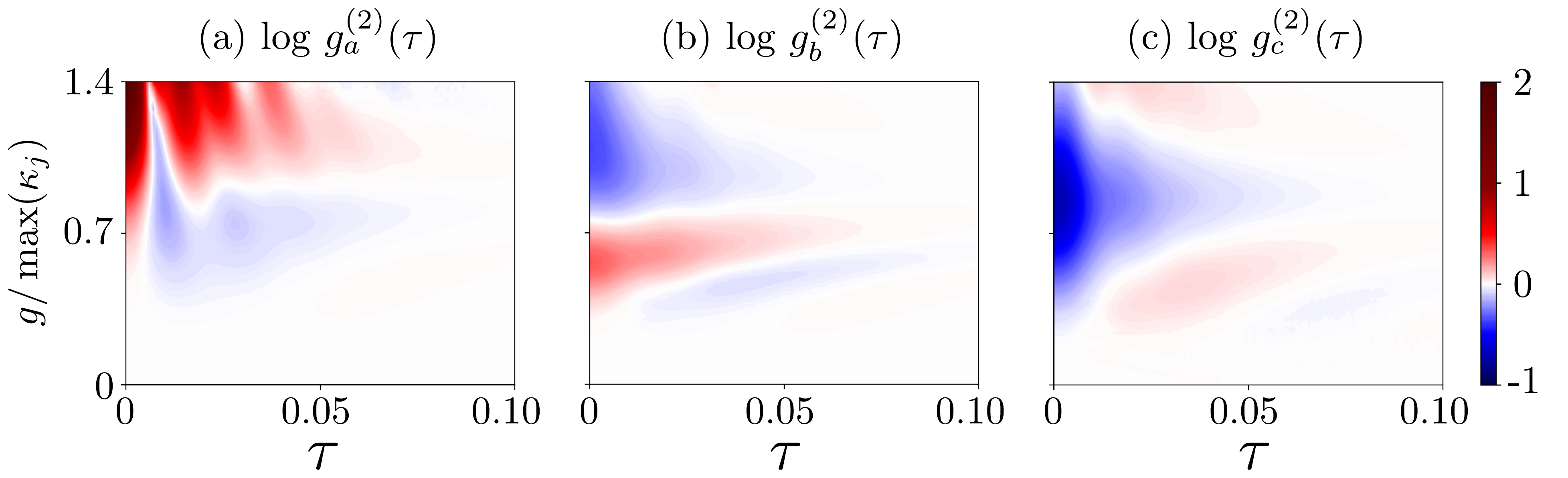}
\caption{Same as in Fig.~\ref{fig04}, but for the QD-driven system
with parameters given in Eq.~(\ref{setA2}). We observe here
single-PRs and the corresponding single-PB effects.} \label{fig06}
\end{figure}

\begin{figure}[ht]
\centering
\includegraphics[width=\linewidth]{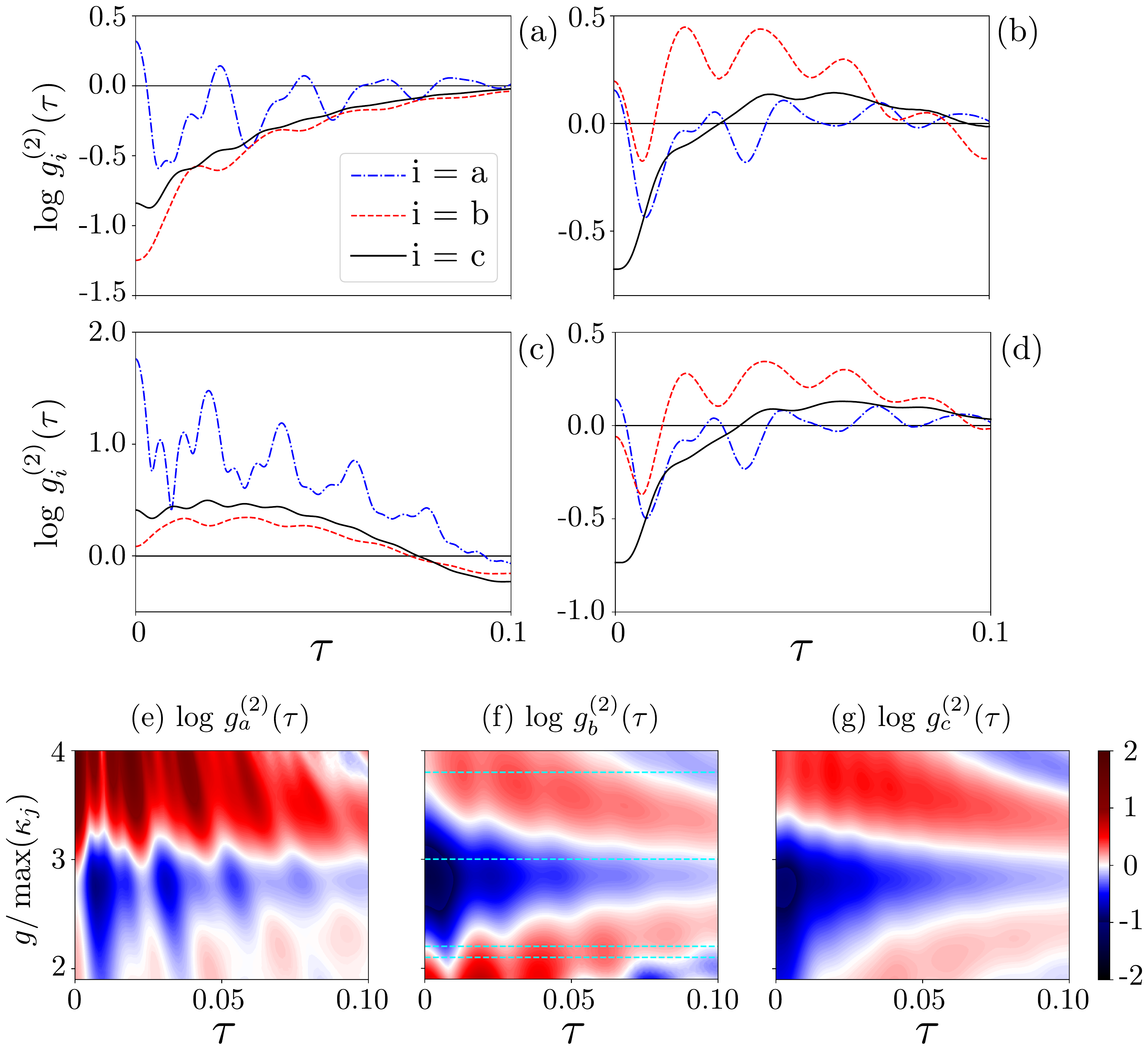}
\caption{(a--d) Delay-time second-order correlation functions
$\corTau{i}$ (in the logarithmic scale) for the SMR mode $a$, the
QD mode $b$, and the hybrid mode $c$ modes in the QD-driven system
assuming that $g/\kappa_{\max}$ is equal to: (a) 3, (b) 2.1, (c)
3.8, and (d) 2.2. The four different predictions of correlations
for the QD mode $b$ correspond to all the cases listed in
Table~\ref{table4}. (e--f) Same as in Fig.~\ref{fig04}, but for
the parameters given in Eq.~(\ref{setA3}). Note that panels (a--d)
show the cross-sections of the 3D plot in (f) at the values of
$g/\kappa_{\max}$ marked by broken lines.} \label{fig07}
\end{figure}

The destructive interference of both modes $a$ and $b$, at the
balanced linear coupler, can result in the sub-Poissonian
statistics of the hybrid modes. We  observe this effect  even in
the weak-nonlinearity (or weak-coupling) regime, which witnesses
unconventional PB, as discussed in detail in Methods. It is worth
noting that in this study we are aiming at observing
$g^{(2)}(\tau)<1$ not only at $\tau=0$, but also for non-zero
delay times (e.g., $\tau\in[0,0.1]$), as in standard experimental
demonstrations of the sub-Poissonian statistics reported in, e.g.,
Refs.~\cite{Lang2011,Walls1994}. Thus, the cases shown in
Figs.~\ref{fig04}(a) and \ref{fig04}(c) can hardly be considered
as convincing demonstrations of boson antibunching, because of the
oscillations, which occur in $g_{a,c}^{(2)}(\tau)$ with increasing
$\tau$. More convincing demonstrations of these effects without
such oscillations (or by considerably suppressing them) are
presented in Figs.~\ref{fig06} and \ref{fig07}, as analysed in
detail in the next section.


To explain the super-Poissonian photon-number statistics and
photon bunching in the mode $a$ for the system pumped in the SMR
mode, let us analyse Fig.~\ref{fig05}(a) with $g\approx\kappa_{m}$
concerning the anharmonicity of the energy levels in these cases.

The $g$ term in Eq.~(\ref{H_pm}) corresponds to the standard
Jaynes-Cummings model with the familiar eigenvalues~\cite{Gu2017}:
\begin{equation}
  E_n^{\pm} \equiv E(\ket{n,\pm}) = n\omega_{\SMR}
  \pm \frac{1}{2} \sqrt{\Delta_1^2 + \Omega_{n}^2}
  \label{eq:JC_eigenvalues}
\end{equation}
with the corresponding eigenstates:
\begin{eqnarray}
\ket{n,+} &\equiv& \cos \left( \tfrac{\theta_n}{2} \right)
\ket{n}\ket{e} + \sin \left( \tfrac{\theta_n}{2} \right)
\ket{n+1}\ket{g},
\nonumber \\
\ket{n,-} &\equiv& - \sin \left( \tfrac{\theta_n}{2} \right)
\ket{n}\ket{e}+ \cos \left( \tfrac{\theta_n}{2} \right)
\ket{n+1}\ket{g}, \label{eq:JC_eigenstates}
\end{eqnarray}
which are often referred to as dressed states or dressed-state
dublets, where $\theta_n = \Omega_{n} / \Delta_1$ is the mixing
angle, $\Delta_1 = \omega_{q}-\omega_{\SMR}$ is the detuning
between the SMR and qubit. Moreover, $\Omega_{n}=2g\sqrt{n+1}$ can
be interpreted as the $n$-photon Rabi frequency on resonance, so,
in particular, $\Omega_{0}=2g$ is the vacuum Rabi frequency. Thus,
the energy spectrum is clearly anharmonic, which is a necessary
condition to observe PB. Note that the Jaynes-Cummings interaction
can be effectively described in the dispersive limit (i.e., far
off resonance) as a Kerr nonlinearity (for a detailed derivation
see, e.g.,~\cite{Adam2016}), which is the standard nonlinearity
assumed in many predictions of PB effects.

To demonstrate the anharmonic energy levels of the complete
Hamiltonian $H_{+}$ on resonance (see Fig.~\ref{fig02}), we assume
a weak drive coupling strength $\eta_{a}$. Given that, the system
Hilbert space can be truncated. We assume that the polariton
number is at most equal to two in this weak-drive regime. The
ground state is $\ket{\psi_0}=\ket{0,0,g}$ with the corresponding
eigenvalue $E_0=0$.  The three eigenvalues of the first manifold
(with eigenstates containing a single polariton), as shown in
Fig.~\ref{fig02}(b), are:
\begin{equation}
 E^{(1)}_{1,3}=\Delta\mp\sqrt{g^2+f^2},\quad E^{(1)}_2 =\Delta,
 \label{Eq:10}
\end{equation}
while the five eigenvalues of the second manifold (with
eigenstates containing two polaritons), which are shown in
Fig.~\ref{fig02}(c), read:
\begin{eqnarray}
 E^{(2)}_{1,2}&=&\frac{1}{2}\big[4\Delta-\sqrt{2(3g^2+5f^2\pm f_1)}\big],
 \nonumber\\
 E^{(2)}_3&=&2\Delta,\nonumber\\
 E^{(2)}_{4,5}&=&\frac{1}{2}\big[4\Delta+\sqrt{2(3g^2+5f^2\mp f_1)}\big],
 \label{Eq:12}
\end{eqnarray}
where $f_1=\sqrt{3f^2(10g^2+3f^2)+g^4}$. In particular, by
assuming $f=5$ and $g=7.5$, the eigenenergies of the first and
second manifolds are, respectively: (i) $\Delta$, $\Delta\pm
9.01388\approx \Delta\pm 9$, and (ii) $2\Delta$, $2\Delta\pm
5.82965\approx 2\Delta\pm 6$, and $2\Delta\pm 16.11725\approx
2\Delta\pm 16$.

\begin{figure}[ht]
\centering
\includegraphics[width=0.8\linewidth]{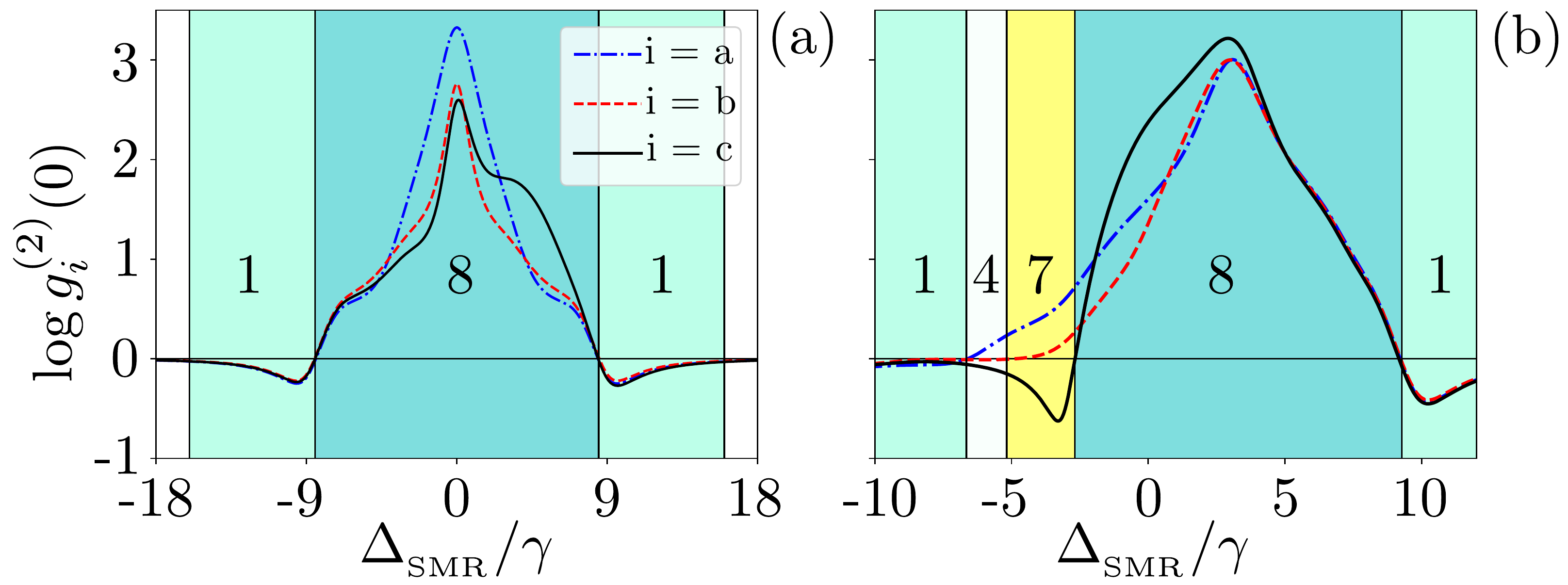}
\caption{Correlation functions $\log\cor{i}$ versus the frequency
detuning $\Delta_{\SMR}$ (in units of the qubit decay rate
$\gamma$) between the drive and SMR for: (a) the resonance case
$\omega_{\SMR}= \omega_{m}=\omega_q$ (so also $\Delta_{\SMR}
=\Delta_{m}=\Delta_q$) and (b) the nonresonance case
$\omega_{\SMR}\neq \omega_{m}\neq\omega_q$, where
$\omega_b/\gamma=1560$ MHz. Note that by changing the pump
frequency, different detunings appear with respect to the modes
$a$ and $b$, and qubit. We set $g=7.5\gamma$ and other parameters
are given in Eq.~(\ref{setA1}). The numbering of the coloured
regions correspond to the cases listed in Table~\ref{table8}.}
\label{fig08}
\end{figure}

A simple way to probe the pumped mode is to record the
second-order correlation $g^{(2)}(0)$ as a function of
$\Delta_{\SMR}$, where the pump frequency $\omega_p$ is changing
(see Fig.~\ref{fig08}). To do so, we first consider the resonance
case as $\omega_{\SMR}=\omega_{m}=\omega_q=\omega$ in
Eq.~\eqref{Hamiltonian1} and $\omega-\omega_p=\Delta$. As depicted
in Fig.~\ref{fig08}(a), one can see local minima with negative
values in $\log g^{(2)}(0)$  for the three modes, which indicate
Case~1 in Table~\ref{table8}, at $\Delta_{\SMR}/\gamma=\pm 9$,
which correspond to $\Delta=\pm \sqrt{g^2+f^2}\approx\pm 9$, given
Eq.~(\ref{Eq:10}). This means that the pump frequency is located
at the two dressed state dublets: $\ket{\psi^-_1}$ with energies
$E^{(1)}_1$ and $E^{(1)}_3$. And we are off-resonance from the
second energy manifold, which implies the possibility of observing
PB at these frequencies.

Furthermore, our simulations predict a maximum of $\log
g^{(2)}(0)\approx 3$  showing a strong super-Poissonian statistics
in the three modes (corresponding to Case~8 in Table~\ref{table8})
as $\Delta_{\SMR} \rightarrow 0$. In particular, at
$\Delta_{\SMR}/\gamma\approx\pm 6$, the pump frequency is near
$E^{(2)}_1\approx 6$ and $E^{(2)}_4\approx-6$, respectively, of
the second manifold, in which the probability of the two-photon
resonance is maximised, as a signature of PIT. It signifies that
the pump is in resonance with one of the levels in the second
manifold of the hybrid system energy levels, here specifically
$E^{(2)}_1$ and $E^{(2)}_4$. One can see in Fig.~\ref{fig08},
peaks (global maxima in the analysed range) of $\log
g_n^{(2)}(0)>0$ for $n=a,b,c$ at $\Delta_{\SMR}=0$. In particular,
the probability of absorbing a single photon decreases here.
However, if a photon is absorbed, it enhances the probability of
capturing subsequent photons, this effect produces the
super-Poissonian statistics, which is due to the fact that the
probability of observing a single photon is also very small
($P_{10g}\ll 1$) and smaller than the probability of observing two
photons~\cite{Kubanek2008,Faraon2008}.

It is seen that, by tuning the drive frequency to the transition
$E_2-E_0$ in the energy spectrum of the total nonlinear system,
the probability of admitting two photons increases. This results
in the super-Poissonian statistics, which is opposite to the case,
when the drive frequency is tuned to the transition $E_1-E_0$,
when the probability of admitting subsequent photons decreases
resulting in PB.

By assuming the off-resonance condition, $\omega_{\SMR}\neq
\omega_{m}\neq \omega_q$, we show in Fig.~\ref{fig08}(b) the
correlation functions for the three modes ($a,b,c$) as a function
of $\Delta_{\SMR}$ in the case, when the drive is tuned in-between
the dressed state eigenenergies of the hybrid system.

\begin{figure}[ht]
\centering
\includegraphics[width=0.8\linewidth]{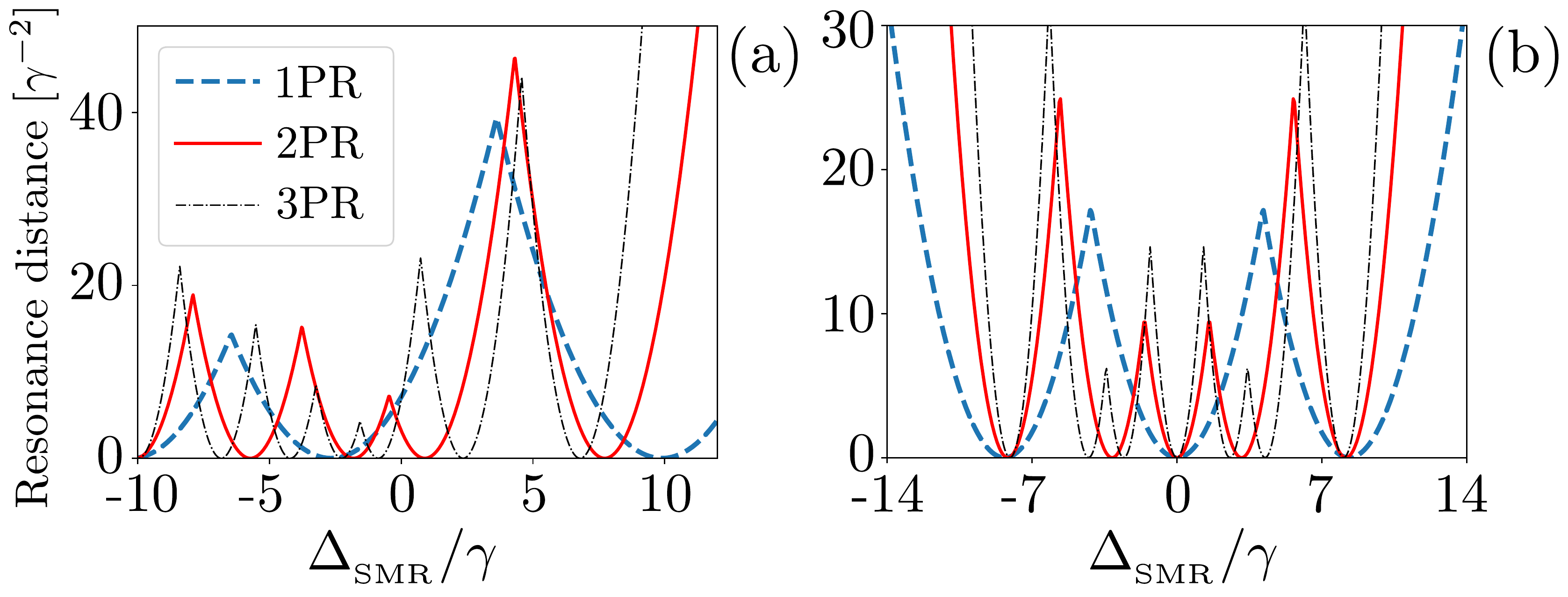}
\caption{Resonance distances, as defined in Eq.~\eqref{PR}, versus
the frequency detuning $\Delta_{\SMR}$ (in units of the qubit
decay rate $\gamma$) between the drive and SMR for: (a) the
SMR-driven system with parameters specified in Eq.~(\ref{setA1})
with $g=7.58\gamma$ and (b) the QD-driven system with
Eq.~(\ref{setA2}) with $g=4.5\gamma$.} \label{fig09}
\end{figure}

The PB and PIT effects observed in Fig.~\ref{fig08} can be
explained by considering some measures of the distances from
resonances, as shown in Fig.~\ref{fig09}(a). The distances of the
single-, two-, and three-photon resonances (PRs) are defined here,
respectively, as:
\begin{eqnarray}
  D_{\rm 1PR} = \min_i|\omega_p-\omega^{(1)}_i|^2,
 \quad D_{\rm 2PR} = \min_i|2\omega_p-\omega^{(2)}_i|^2,
 \quad  D_{\rm 3PR} = \min_i|3\omega_p-\omega^{(3)}_i|^2,
\label{PR}
\end{eqnarray}
where $\omega_p$ is the frequency of the pump that is tuned with
respect to the energy of the hybrid system. Here $\omega^{(n)}_i$
are the frequencies (labelled with subscript $i$) in the $n$th
manifold, so the minimalization is performed over $\omega^{(n)}_i$
for a given manifold $n$. Figure~\ref{fig09} shows the resonance
distances versus $\Delta_{\SMR}$, where $\omega_p$ is tuned with
respect to the energy of the whole system. The dip in $g^{(2)}(0)$
at $\Delta_{\SMR}/\gamma=10$ [see Fig.~\ref{fig08}(b)], which is
characteristic for PB, corresponds to the resonance for a single
excitation, as seen from $D_{\rm 1PR}$, and is off-resonance for
higher excitations at that frequency [see Fig.~\ref{fig09}(a)].
The second-order correlation function $g^{(2)}_c(0)$ for the
hybrid mode has a dip as a signature of PB around
$\Delta_{\SMR}/\gamma=-3.4$, while the modes $a$ and $b$ exhibit
the super-Poissonian statistics (witnessing PIT), as shown in
Fig.~\ref{fig08}(b). This effect is witnessed as a dip in $D_{\rm
1PR}$ and it is off-resonance for $D_{\rm 2PR}$ and $D_{\rm 3PR}$,
as illustrated in Fig.~\ref{fig09}(a), while the modes $a$ and $b$
exhibit PIT. This type of unconventional PB is discussed further
in sections below.

By decreasing $\Delta_{\SMR}/\gamma$ from 0 to -2, the correlation
function  $g^{(2)}_a(0)$ for the SMR mode in Fig.~\ref{fig08}(a)
resembles a shoulder in shape. We observe PIT at this point or
region, as expected from our findings in the resonance-distant
diagram in Fig.~\ref{fig09}(a). Indeed, there is a dip in $D_{\rm
2PR}$ for higher resonances at this point, which explains the
occurrence of PIT.

Let us consider now $\Delta_{\SMR}/\gamma \rightarrow 3$ in
Fig.~\ref{fig08}(b) for the pump frequency in resonance with the
qubit, $\Delta_q=0$, which is close to the resonance frequency of
the hybrid mode. In this case multi-photon transitions are
induced, which result in PIT at $\Delta_{\SMR}/\gamma=3$, and we
observe a peak in $\log g^{(2)}(0)>0$ at this frequency in
Fig.~\ref{fig08}(b). Clearly, we are here in resonance with
higher-energy levels, while the drive strength is very small,
$\eta_{a}/\gamma=0.7$. The probability of observing a single
photon is also small as the peak for $\Delta_c= 0$, but if a
single photon is absorbed, then the probability of capturing
subsequent photons increases, as for PIT.

The analysed system parameters are found by optimising our system
to observe the super-Poissonian statistics in the SMR and QD
modes. At the sub-Poissonian statistics area of $g^{(2)}(0)$, it
is possible to observe in Fig.~\ref{fig14} (in Methods) that
$g^{(3)}(0)>1$ and/or $g^{(4)}(0)>1$, which are signatures of
higher-order photon/phonon resonances and multi-PIT (see Methods).
Actually, by calculating the second-order correlation function to
witness the PB and PIT phenomena, higher-order correlation
functions can be used to test whether a given effect is indeed:
(1)~single-PB or single-PIT, (2)~multi-PB or multi-PIT, or
(3)~nonstandard versions of these effects, as discussed in Methods
and, e.g., in Refs.~\cite{Huang2018,Kowalewska2019}. As mentioned
above, these parameters allow us to achieve the sub-Poissonian
statistics for a relatively long delay times.

\section{Hybrid-mode blockade in the QD-driven system}\label{Section:QD-driven}

In this section, we analyse steady-state boson-correlation
effects, including the hybrid-mode blockade and PIT, in the
QD-driven dissipative system, as described by the Hamiltonian
$H''$ and the master equation~(\ref{ME}) for the parameters
specified mostly in Eqs.~(\ref{setA2}) and (\ref{setA3}).

To eliminate or at least to suppress the undesired oscillations in
$g^{(2)}(\tau)$, we assume in this section that our system is
driven classically at the QD. Moreover, we assume that the SMR is
in the bad-cavity regime, as $\kappa_{\SMR}\gg g^2/\kappa_{\SMR}
\gg \gamma$~\cite{KuhnBook}. So, we apply the effective system
Hamiltonian in the rotating frame, as given by
Eq.~(\ref{Hamiltonian2}). Even if the lifetime
$\tau_{\SMR}=1/\kappa_{\SMR}$ of the SMR is much shorter than that
assumed in the SMR-driven system, which was discussed in the
former section, the hybrid mode, as we show below, reveals no
oscillations for quite long delay times, which is due to driving
the QD.

To study boson-number statistics of our system, we compute the
second-order correlation function $g^{(2)}(0)$ for the optimised
parameters, which enables us to demonstrate Cases~4, 6, and 7 of
Table~\ref{table8} in Fig.~\ref{fig03}(b). In Case~7, which is of
our special interest, the modes $a$ and $b$ are super-Poissonian,
as $\log g^{(2)}(0)>0$, while the hybrid mode $c$ is
sub-Poissonian, as $\log g_c^{(2)}(0)<0$. By increasing the
coupling $g$ between the SMR and qubit, the mode $b$ becomes
sub-Poissonian, as being affected by the nonlinearity of the mode
$a$.

To check the second criterion for PB, the second-order correlation
function $g^{(2)}(\tau)$ is considered below. Figure~\ref{fig06}
shows $g^{(2)}(\tau)$ corresponding to $g^{(2)}(0)$ plotted in
Fig.~\ref{fig03}(b) showing Cases~4, 6, and 7. As expected, boson
antibunching is observed for the hybrid mode, as shown in
Fig.~\ref{fig06}(c), while the SMR mode reveals bunching, as
illustrated in Fig.~\ref{fig06}(a). Moreover both phonon
antibunching and bunching, in addition to unbunching [i.e.,
$g_b^{(2)}(0)\approx g_b^{(2)}(\tau)$ for $\tau \gtrapprox 0$],
have been observed in the studied region of the QD mode, as shown
in Fig.~\ref{fig06}(b). It is clear from Fig.~\ref{fig06} that the
antibunching of bosons in the three modes survives in some
specific coupling regime (around $g=0.7\kappa_m$) for a relatively
long delay time $\tau
> 1/\kappa$ and oscillations in  $g_c^{(2)}(\tau)$ are absent in
the hybrid mode $c$. Moreover, boson bunching is observed, when
$g_a^{(2)}(\tau)$ drops rapidly for delay times greater than the
cavity photon lifetime, as considered in Figs.~\ref{fig05}(d) and
\ref{fig05}(e).

\begin{figure}[ht]
\centering
\includegraphics[width=\linewidth]{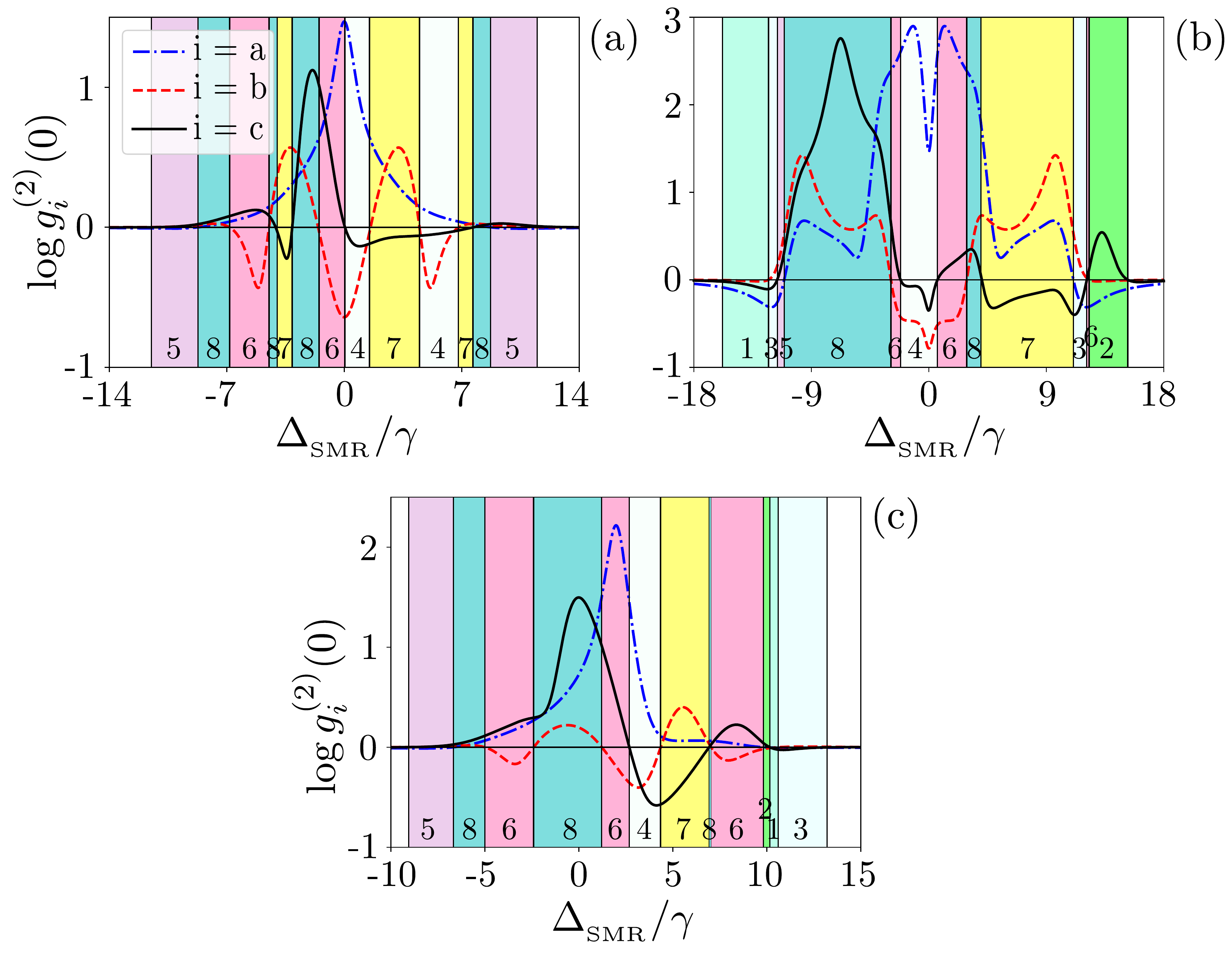}
\caption{Correlation functions $\log\cor{i}$ versus the frequency
detuning $\Delta_{\SMR}$ (in units of the qubit decay rate
$\gamma$) between the drive and SMR for the QD-driven system for:
(a,b) the resonant case with $\omega_{\SMR}= \omega_{m}=\omega_q$
(so also $\Delta_{\SMR}=\Delta_{m}=\Delta_q$), and (c) the
nonresonant case with $\omega_{\SMR}\neq \omega_{m}\neq\omega_q$.
Parameters are set in: Eq.~(\ref{setA2}) with $g=4.5\gamma$ for
(a,c), and Eq.~(\ref{setA3}) with $g=9.5\gamma$ for (b). Eight
different predictions, which correspond to all the cases listed in
Table~\ref{table8}, are marked for the sub- and super-Poissonian
number statistics in the photonic ($a$), phononic ($b$), and
hybrid photon-phonon ($c$) modes.} \label{fig10}
\end{figure}
\begin{figure}[ht]
\centering
\includegraphics[width=0.4\linewidth]{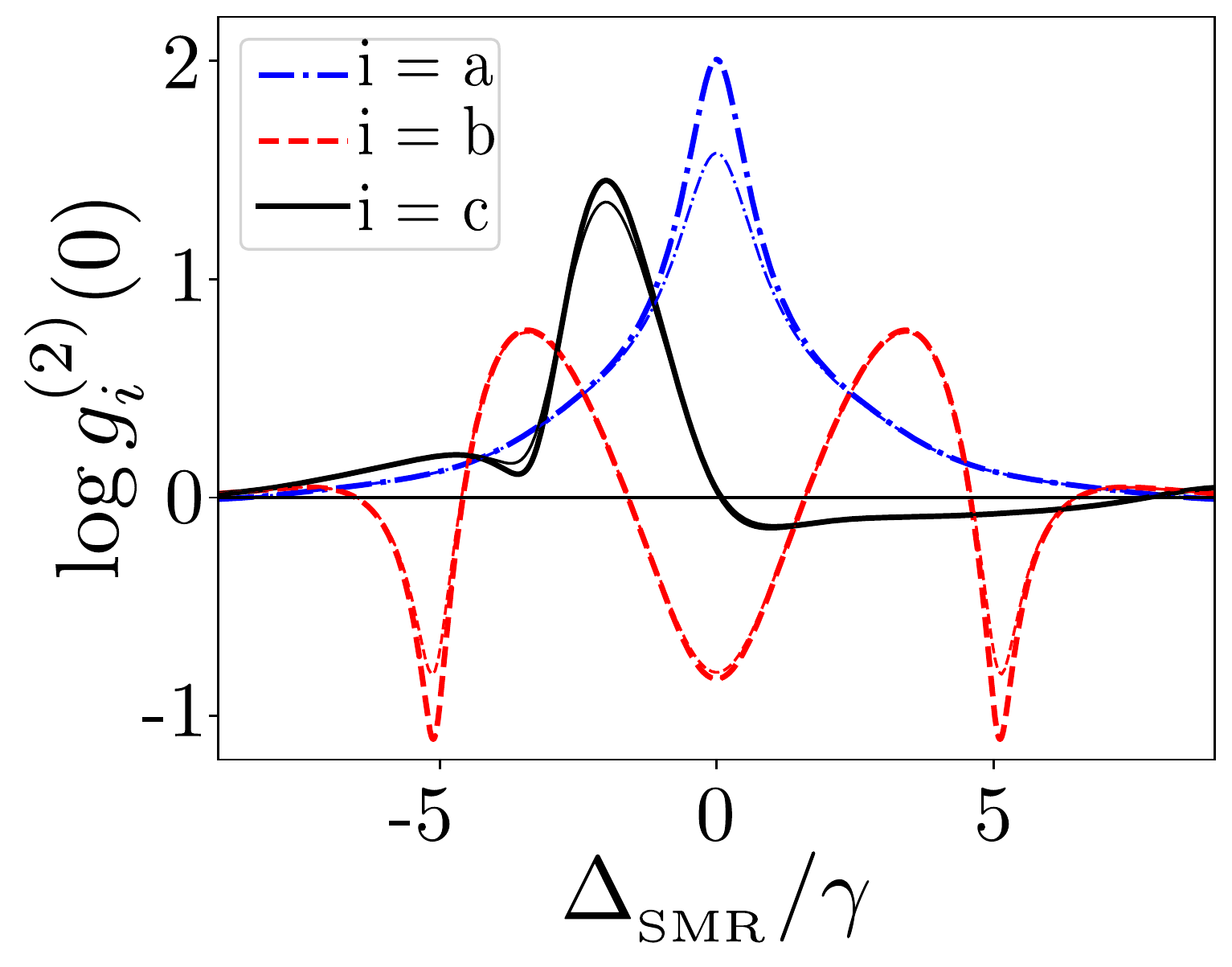}
\caption{Correlation functions $\log \cor{i}$ versus the frequency
detuning $\Delta_{\SMR}$ in units of $\gamma$ for the QD-driven
system for the resonant case with
$\omega_{\SMR}=\omega_{m}=\omega_q=\gamma\times1560$~MHz (so also
$\Delta_a=\Delta_b=\Delta_q$). The thin curves in each mode are
obtained using the master equation in Eq.~(\ref{ME}) and the thick
curves are obtained from the non-Hermitian Hamiltonian method
using Eqs.~(\ref{corr1}) and (\ref{corr2}). Parameters are set in
Eq.~(\ref{setA2}) except $g=4.5\gamma$ and
$\kappa_a=\kappa_b=6\gamma$.} \label{fig11}
\end{figure}

To understand the delay-time dependence of the hybrid mode $c$, we
consider Eq.~\eqref{Hamiltonian2}, when the SMR, QD, and qubit
have the same resonance frequency,
$\omega_{\SMR}=\omega_{m}=\omega_q=\omega$ and $g=4.5\gamma$. As
illustrated in Fig.~\ref{fig10}(a), there are three dips (local
minima) in $g^{(2)}_b(0)<0$ for the mode $b$ of the QD, where we
assumed $g<\min\{\kappa_a, \kappa_b\}$ and $f>g$. For these
parameters, only a weak nonlinearity is induced in the mode $b$.
Thus, the anharmonicity of energy levels cannot explain the PB
effect observed as a dip at these three dips [see
Fig.~\ref{fig09}(b)]. Actually, these dips in $\log g^{(2)}_b(0)$
are due to single-photon resonant transitions, which correspond to
unconventional PB, as explained by the non-Hermitian effective
Hamiltonian method in the next section and in Methods.

Figure~\ref{fig10}(c) shows $\log g_i^{(2)}(0)$ for the three
modes as a function of $\Delta_{\SMR}$. In this case, we assume
that the resonance frequencies of the SMR, QD, and qubit are not
the same, and the detuning of each mode with respect to $\omega_p$
is different. It is shown that, when
$\Delta_{\SMR}/\gamma\rightarrow 2$, multiphoton transitions (and
so PIT or multi-PB) can be induced in the mode $a$, where the pump
frequency is in the resonance with the qubit, $\omega_p=\omega_q$.
This effect is seen in Fig.~\ref{fig14} (in Methods) corresponding
to a local maximum in higher-order moments $g_{i}^{(3)}(0)$ and
$g_{i}^{(4)}(0)$. Likewise the resonance case, unconventional PB
in the modes $b$ and $c$ can be explained by the method applied in
the next section.

\begin{figure}[ht]
\centering
\includegraphics[width=0.7\linewidth]{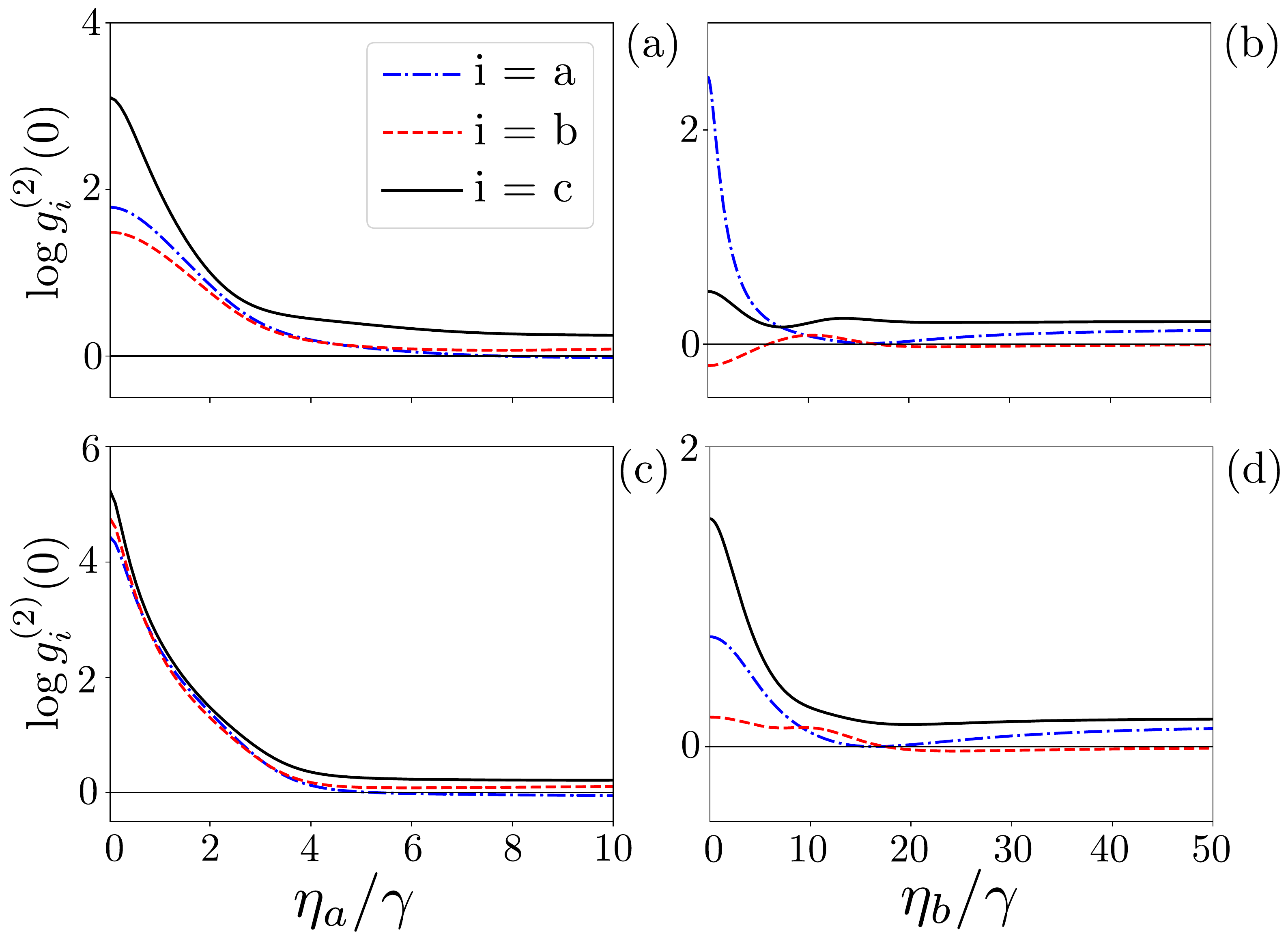}
\caption{Second-order correlation functions $\log \cor{i}$ versus
the drive strengths: (a,c) $\eta_{a}$ for the SMR-driven system
and (b,d) $\eta_{b}$ for the QD-driven system. Parameters are
given in: (a) Eq.~(\ref{setA1}) with $g=7.5\gamma$ and
$\omega_p=1554\gamma$, which implies $\Delta_q=-3\gamma$,
$\Delta_b=6\gamma$, $\Delta_a=0$; (b) Eq.~(\ref{setA2}) with
$\omega_p=1568\gamma$, which implies $\Delta_q=0$,
$\Delta_b=-8\gamma$, and $\Delta_a=2\gamma$; (c) Eq.~(\ref{setA1})
with $g=7.5\gamma$ and $\omega_p=1551\gamma$, which implies
$\Delta_q=0$, $\Delta_b=9\gamma$, and $\Delta_a=3\gamma$; and (d)
Eq.~(\ref{setA2}) with $\omega_p=1570\gamma$, which implies
$\Delta_q=-2\gamma$, $\Delta_b=-10\gamma$, and $\Delta_a=0$.}
\label{fig12}
\end{figure}

In Fig.~\ref{fig12}, we study how the second-order correlation
functions reveal the PIT regime, which corresponds to Case~8 in
Table~\ref{table8}, as a function of the SMR-pump strength
$\eta_a$ [in panels (a) and (c)] and the QD-pump strength $\eta_b$
[in panels (b) and (d)]. The hybrid mode $c$ is super-Poissonian
for all the shown cases and pump strengths. The modes $a$ and $b$
are super-Poissonian [except the mode $a$ in panel (b)] for small
pump strengths $\eta_{a,b}$. By increasing the driving power at
least to some values, which can be identified in the figures for
specific modes, we observe that the correlation functions
$g^{(2)}(0)$ also decrease for all the modes (except the mentioned
case). This property confirms the nonclassicality of the predicted
PIT in the hybrid system according to an additional criterion of
`true' PIT of Ref.~\cite{Majumdar2012pra}.

\section{Unconventional blockade explanation via non-Hermitian
Hamiltonian approach}\label{Section:UPB}

In this Section, we apply the analytical mathematical formalism of
Ref.~\cite{Bamba2011}, based on an non-Hermitian Hamiltonian, to
identify the quantum interference effect that is responsible for
inducing unconventional PB, i.e., strongly sub-Poissonian
statistics in the weak-coupling regime or the weak-nonlinearity
regime. We stress that this is an approximate approach, where the
effect of quantum jumps is ignored~\cite{Minganti2019,
Minganti2020}.

By considering the system studied in the former section under the
weak-pump condition, we can truncate the Hilbert spaces for the
modes $a$ and $b$ and the qubit at their two excitations in total.
This allows us to consider the total-system Hilbert space of
dimension $3\times3\times2=18$. Moreover, the weak-pump condition
implies that $C_{00g}\gg C_{10g}, C_{01g}, C_{00e}\gg C_{11g},
C_{10e}, C_{01e}, C_{20g}, C_{02g}$. Thus, the steady-state of the
coupled system can be expressed as
\begin{eqnarray}
\ket{\Psi_{abq}(t)}&=&C_{00g}\ket{00g}+e^{-i\omega_d t}
\Big(C_{00e}\ket{00e}+C_{10g}\ket{10g} +C_{01g}\ket{01g}\Big)
\nonumber\\
&&+e^{-2 i \omega_d t} \Big(C_{10e}\ket{10e}+C_{01e}\ket{01e}
+C_{11g}\ket{11g}+C_{20g}\ket{20g}+C_{02g}\ket{02g}\Big),
\label{Eq:13}
\end{eqnarray}
where $\ket{n_a,n_b,g/e}$ is the Fock state with $n_a$ photons in
the SMR, $n_b$ phonons in the QD, and the lower ($\ket{g}$) or
upper ($\ket{e}$) state of the qubit. The effective non-Hermitian
Hamiltonian of the system can be written as
\begin{eqnarray}
H_{\rm eff}&=& H''-i\frac{\kappa_a}{2}a^\dagger
a-i\frac{\kappa_b}{2}b^\dagger b -i\frac{\gamma}{2}\sigma_{+}
\sigma_{-}, \label{Hamiltonian_BS}
\end{eqnarray}
where $H''$ is given by Eq.~\eqref{Hamiltonian2}. Analogously, one
can consider the  non-Hermitian Hamiltonian with $H'$, given by
Eq.~\eqref{Hamiltonian1}.

In the weak-pump regime, the mean number of photons and phonons in
the SMR and QD can be approximated as $\langle n_a\rangle \approx
| C_{10g}|^2$ and $\langle n_b\rangle \approx | C_{01g}|^2$,
respectively. As derived in detail in Methods, the second-order
correlation functions for generated photons and phonons, under the
same weak-pump conditions, can be given by:
\begin{eqnarray}
g^{(2)}_a(0)&=&\frac{\langle a^\dagger a^\dagger a a
\rangle}{\langle a^\dagger a\rangle^2}\approx\frac{2|
C_{20g}|^2}{| C_{10g}|^4},\nonumber\\
g^{(2)}_b(0)&=&\frac{\langle b^\dagger b^\dagger b
b\rangle}{\langle b^\dagger b\rangle^2}\approx\frac{2|
C_{02g}|^2}{| C_{01g}|^4}, \label{corr1}
\end{eqnarray}
where the superposition coefficients $C_{n,m,g}$ are given in
Eqs.~\eqref{Eq:19} and \eqref{Eq:20}.

The hybrid photon-phonon modes, which are defined in
Eq.~\eqref{eq:c,d}, are the output modes of the balanced linear
coupler with the SMR and QD modes at its inputs. As shown in
Methods, we find, analogously to Eq.~(\ref{corr1}), the
second-order correlation function for the hybrid mode~$c$ reads:
\begin{eqnarray}
g^{(2)}_c(0)&=&\frac{\langle c^\dagger c^\dagger c c
\rangle}{\langle c^\dagger c\rangle^2}\approx\frac{2|
C^\prime_{20g}|^2}{| C^\prime_{10g}|^4}, \label{corr2}
\end{eqnarray}
where the superposition coefficients $C'_{n,m,e/g}$ are given in
Eqs.~\eqref{Eq:19_2} and \eqref{Eq:20}, and the sixth formula in
Eq.~\eqref{Eq:15}.

This approach enables us to explain unconventional PB generated in
the hybrid system, which is the result of a destructive quantum
interference effect that assures, together with other conditions,
that the probability amplitude of having two photons in the SMR
and QD is negligible. This method can also be used to find some
optimal parameters to observe PB in the system.

Figure~\ref{fig11} presents a comparison of our predictions based
on the precise numerical solutions of the master equation in
Eq.~(\ref{ME}), as shown by thin curves, with those calculated
from Eqs.~(\ref{corr1}) and (\ref{corr2}) using the non-Hermitian
Hamiltonian approach, as shown by thick curves. The locations of
the maxima and minima of the correlation functions are found
similar according to both formalisms. However, these extremal
values can differ more distinctly, especially for the two global
minima in the sub-Poissonian statistics of the mode $b$ and the
super-Poissonian maximum of the mode $a$. The differences result
from the effect of quantum jumps, which are properly included in
the master-equation approach and totally ignored in the
non-Hermitian Hamiltonian approach.

\section{Different types of blockade and tunnelling effects}\label{Section:Tables}

The sub-Poissonian statistics of a bosonic field, as described by
$g^{(2)}(0)\ll 1$, is not a sufficient criterion for observing a
`true' PB, which can be a good single-photon or single-phonon
source. In fact, other criteria, such boson antibunching,
$g^{(2)}(0)< g^{(2)}(\tau)$, and the sub-Poissonian statistics of
higher-order correlation functions, $g^{(n)}(0)\ll 1$, should also
be satisfied (see Methods). Anyway, most of the studies of PB, and
especially those on unconventional PB, are limited to testing the
second-order sub-Poissonian statistics described by $g^{(2)}(0)<
1$.

As explicitly discussed in Refs.~\cite{Zou90, MandelBook,
Teich1988, Adam2010}, photon antibunching and sub-Poissonian
statistics are different photon-number correlation effects. So,
the four cases listed in Table~\ref{table4}, can be considered as
different types of PB and PIT. We show that all these effects  can
be observed in the studied system. For brevity, Table~\ref{table4}
is limited to phononic effects. PB, as defined in Case~I and often
referred to as a `true' PB, can be a good single-photon sources;
but, as mentioned above, other higher-order criteria should also
be satisfied.

To show these four different effects, we use the parameters set in
Eq.~(\ref{setA3}), where $\kappa_{b} \ll\kappa_{a}$ at the
$\kappa_b=0.002\, \gamma$, which indicates that the quality factor
is $Q\approx 200,$ and so $\eta_{b}/\kappa_{b}\approx100$ in the
case of a strong pump driving the QD mode with $\eta_{b}=0.22\,
\gamma$. Apart from the previously mentioned phenomena, such as
observing the super-Poissonian statistics and bunching in the SMR
and QD modes, while a hybrid mode exhibiting the sub-Poissonian
statistics and boson antibunching, we find the four types of
PB/PIT in the mode $b$ in different coupling regimes, as shown in
Table~\ref{table4}, which includes the examples of specific
experimentally feasible values of $g/\kappa_{a}$.

Case~I corresponds to a stronger form of PB, which we refer to as
a `true' PB, when the nonclassical nature of bosons is revealed by
both their antibunching and sub-Poissonian statistics. Case~II
corresponds to a stronger form of PIT, which can be called a
`true' PIT, when bosons exhibit both classical effects: the
super-Poissonian statistics and bunching. In Case~III, one can
talk about a weaker form of PIT or, equivalently, another weaker
type of  PB, as such bosons are characterised by the classical
super-Poissonian statistics  and their nonclassical nature is
revealed by antibunching. Case~IV represents another weaker form
of PB or, equivalently, of PIT, which is characterised by the
nonclassical sub-Poissonian statistics of classically bunched
bosons. These results imply that one cannot say in general that
the antibunching of bosons leads to their sub-Poissonian
statistics and vice versa~\cite{Zou90,Teich1988}.

Therefore, $g^{(2)}(\tau)>g^{(2)}(0)$ does not necessarily imply
$g^{(2)}(0)<1,$ as in Case~III, which can be seen in
Figs.~\ref{fig07}(c) and \ref{fig07}(f). In addition, as another
example related to Case~IV, let us consider a Fock state $\ket{n}$
with $n\geq 2$, for which $g^{(2)}(0)=1-1/n$, such that if $n=2$
then $g^{(2)}(0)=0.5,$ so $g^{(2)}(0)<1$ and it is not accompanied
by boson antibunching, but bunching in this case.

Our focus in this paper is on the generation of PB in the hybrid
mode, while the other two modes exhibit PIT. Note that this a very
special case of Table~\ref{table8}, which shows that eight
combinations of boson number correlation phenomena in the modes
$a$, $b$, and $c$ can be generated in our system, as specified by
the numbered coloured regions in various figures corresponding to
the cases in Table~\ref{table8}. Thus, we found all the eight
possible combinations of the PIT and PB effects in the hybrid
system for the parameters specified in Eqs.~(\ref{setA1}),
(\ref{setA2}), and~(\ref{setA3}).

\section{Detection of the hybrid-mode correlation functions}
\label{Section:Detection}

\begin{figure}[ht]
\centering
\includegraphics[width=0.6\linewidth]{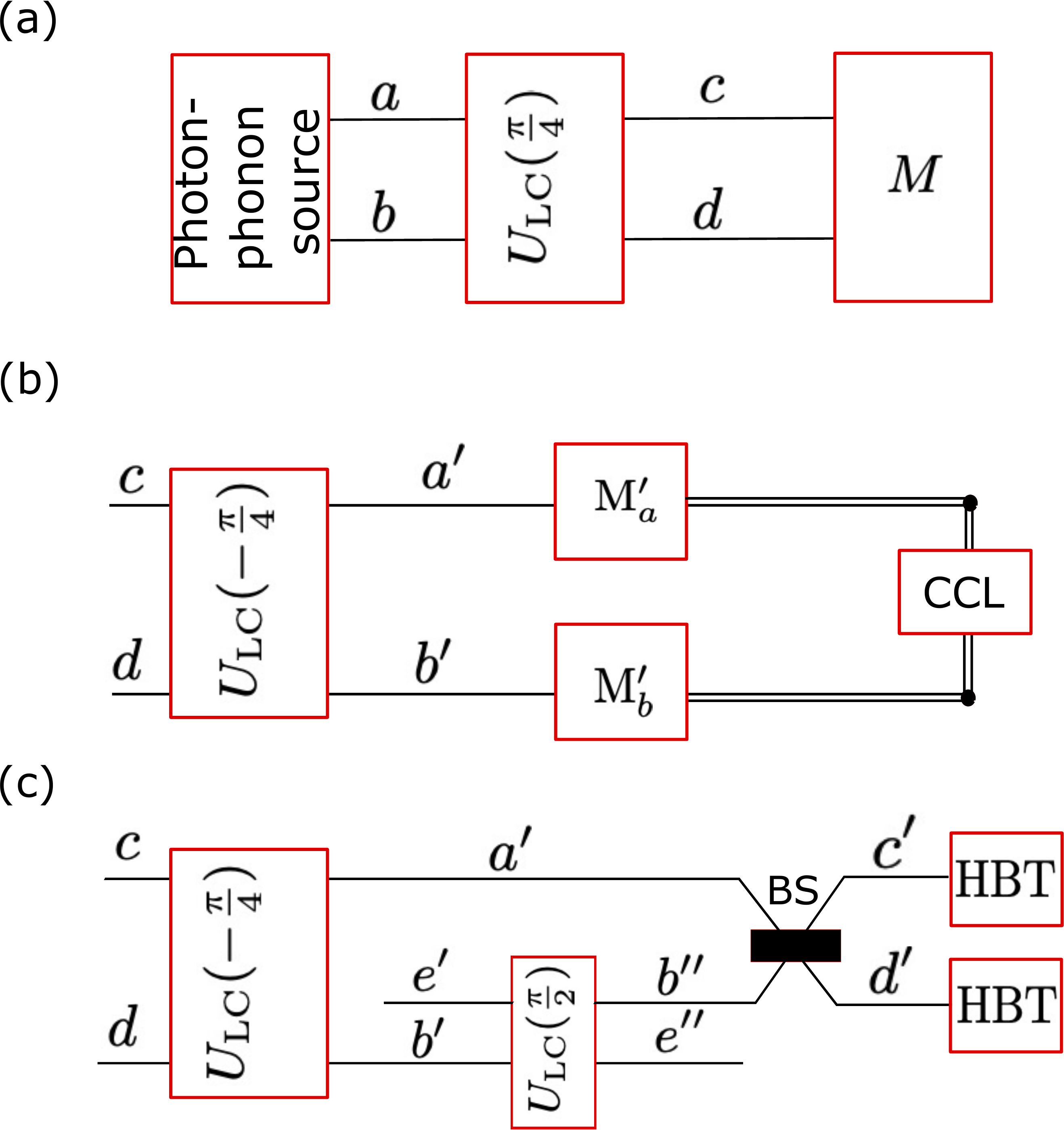}
\caption{Schematics of the proposed detection schemes: (a) General
scheme for the generation of the photonic mode $a$, phononic mode
$b$, and hybrid modes $c,d$, and their detection in the
measurement unit $M$, which is shown in specific implementations
using: (b) Detection Method~1 and (c) Detection Method~2. Key:
$U_{\rm LC}(\theta)$ stands for the linear-coupler transformation,
which in special cases corresponds to multi-level SWAP (for
$\theta=\pi/2$) and Hadamard-like (for $\theta=\pm\pi/4$) gates;
BS is the balanced beam splitter, which corresponds to $U_{\rm
LC}(\pi/4)$, $M'_a$ ($M'_b$) is a measurement unit for detecting
photons (phonons), CCL is a coincidence and count logic unit, HBT
stands for the standard Hanbury-Brown and Twiss optical
interferometer. Mode $e'$ ($e''$) is in the photonic (phononic)
vacuum state.} \label{fig13}
\end{figure}

Here, we describe two detection schemes for measuring the
intensity autocorrelation functions for the hybrid photon-phonon
modes $c$ and $d$, as shown in Fig.~\ref{fig13}.

The measurements of $g^{2}(\tau)$ for the photonic mode $a$ and
the phononic mode $b$ are quite standard and are usually based on
the Hanbury-Brown and Twiss (HBT) optical interferometry and its
generalised version for phonons~\cite{Hong2017}, respectively.
However, the measurement $M$ [as schematically shown in
Fig.~\ref{fig13}(a)] of $g^{(2)}(\tau)$, or even $g^{(2)}(0)$, for
the hybrid photonic-phononic modes $c$ and $d$ is quite
challenging if applied directly. Here we propose two detection
methods, as shown in Figs.~\ref{fig13}(b) and \ref{fig13}(c), for
indirect measuring of $g_{c,d}^{(2)}(0)$.

The first operation of the measurement unit $M$ in both schemes is
a linear-coupler transformation of the hybrid modes $(c,d)$ into
$(a',b')$, which, assuming that the process is perfect, should be
equal to the original purely photonic ($a$) and phononic ($b$)
modes.

We consider a linear coupler (formally equivalent to a beam
splitter) described by a unitary operation $U_{\rm LC}(\theta)$,
which transforms the input operators $a$ and $b$ into:
\begin{eqnarray}
  c(\theta) &=& U^\dagger_{\rm LC}(\theta) a U_{\rm LC}(\theta)
  = a \sin\theta + b \cos\theta,
  \nonumber \\
  d(\theta) &=& U^\dagger_{\rm LC}(\theta) b U_{\rm LC}(\theta)
  = a \cos\theta - b \sin\theta,
\label{LC2}
\end{eqnarray}
for a real parameter $\theta$, where $T=\cos^2\theta$ and
$R=1-T=\sin^2\theta$ are the transmission and reflection
coefficients of the linear coupler, respectively. The studied
hybrid modes are the special cases of Eq.~(\ref{LC2}) for $c\equiv
c(\theta=\pi/4)$ and $d\equiv d(\theta=\pi/4)$. Clearly, the first
transformation $U_{\rm LC}(-\pi/4)$ in Figs.~\ref{fig13}(b) and
\ref{fig13}(c), is the transformation inverse to that in
Fig.~\ref{fig13}(a).

\subsection{Detection method 1 based on measuring photons and phonons}

The correlation functions $g_{c,d}^{(2)}(0)$ in the hybrid
photon-phonon modes can be measured indirectly, as indicated in
Fig.~\ref{fig13}(b), by measuring the observables:
\begin{equation}
  f_{kl}=(a^\dagger)^k a^l,\quad g_{mn}=(b^\dagger)^m b^n,
  \label{fkl}
\end{equation}
where $k,l,m,n=0,1,2$, by using the relations:
\begin{eqnarray}
  \bracket{c^{\dagger} c} &=& \frac12 \Big(
  \bracket{f_{11}}
  + \bracket{g_{11}}
  + \bracket{f_{01}g_{10}}
  + \bracket{f_{10}g_{01}}
  \Big),
\label{fg1}
\end{eqnarray}
\begin{eqnarray}
  \bracket{c^{\dagger 2} c^2} = \frac14 \Big(
  \bracket{f_{22}}
  + 4\bracket{f_{11}g_{11}}
  + \bracket{g_{22}}
  + 2\bracket{f_{01}g_{21}}
  + 2\bracket{f_{10}g_{12}}
  + \bracket{f_{20}g_{02}}
  +\bracket{f_{02}g_{20}}
  + 2\bracket{f_{21}g_{01}}
  + 2\bracket{f_{12}g_{10}}
  \Big),
\label{fg2}
\end{eqnarray}
and analogous relations for the hybrid mode $d$. The measurement
units $M'_a$ and $M'_b$ in this method, as shown in
Fig.~\ref{fig13}(b), describe the measurements of photons and
phonons, respectively. It is seen that, in this approach, to
determine $g_{c,d}^{(2)}(0)$, one has to measure the following
observables: $f_{01}$, $f_{10}$, $f_{11}$, $f_{02}$, $f_{20}$,
$f_{12}$, $f_{21}$, and $f_{22}$. Almost each observable $f_{kl}$
should be measured simultaneously with a specific observable
$g_{mn}$, which can be realised by a coincidence and count logic
(CCL) unit in Fig.~\ref{fig13}(b).

The measurements of all the required photonic observables $f_{kl}$
can be performed by using, e.g., the Shchukin-Vogel method, which
is based on balanced homodyne correlation
measurements~\cite{Shchukin2005}. According to that method, a
photonic signal is superimposed on a balanced beam splitter with a
local oscillator, which is in a coherent state
$\ket{\alpha=|\alpha|\exp(\phi)}$ with  a tunable phase $\phi$. A
desired mean value of the observable $f_{kl}$ can be obtained by
linear combinations of the coincidence counts registered by
specific detectors for different local-oscillator phases $\phi$.
This part of the method corresponds to a Fourier transform. The
simplest nontrivial configuration, which enables the measurement
of the observables $f_{10}$, $f_{10}$, $f_{20}$, and $f_{02}$,
requires four detectors and three balanced BSs, where additional
input ports are left empty, i.e., allowing only for the quantum
vacuum noise. By replacing the four detectors with four balanced
BSs with altogether eight detectors at their outputs, one can
measure any observable $f_{kl}$ for $k+l\le 4$. These include the
desired observables $f_{21}$, $f_{12}$, and $f_{22}$. Of course,
the observable $f_{22}$ can be measured in a simpler way via the
HBT interferometry. The measurement of phononic observable
$g_{mn}$ can be performed analogously just by replacing the
balanced BSs by balanced phonon-mode linear couplers and using
phonon detectors as, e.g., in Ref.~\cite{Hong2017}. The
measurement of two-mode moments $\bracket{f_{kl}g_{mn}}$ is, at
least conceptually, a simple generalisation of the single-mode
methods relying on proper coincidences in photonic and phononic
detectors. Note that a multimode optical version of the original
single-mode method was described in Ref.~\cite{Shchukin2006}.

\subsection{Detection method~2 based on measuring only photons}

Figure~\ref{fig13}(c) shows another realisation of the measurement
unit $M$, to determine $g_{c,d}^{(2)}(0)$, and even
$g_{c,d}^{(2)}(\tau)$. This method is, arguably, simpler and more
effective than detection method~1, because it is based on
measuring only photons and using standard HBT interferometry. Our
approach was inspired by Ref.~\cite{Didier2011}, where the
measurement of single-mode phonon blockade was described via an
optical method instead of a magnetomotive technique, which was
described in Ref.~\cite{Liu2010}, where phonon blockade was first
predicted.

Our measurement setup realises the following three
transformations: (i) converting the phononic mode $b'$ into a
photonic mode $b''$, (ii) mixing the optical modes $a'$ and $b''$
on a balanced BS to generate the modes $c'$ and $d'$, which, in an
ideal case, have the same boson-number statistics as the original
hybrid photon-phonon modes $c$ and $d$; and finally, (iii)
applying the conventional optical HBT interferometry for these two
optical modes. In unit (i), this conversion corresponds to a
multi-level SWAP gate, which can be implemented by a
photonic-phononic linear coupler for $\theta=\pi/2,$ assuming that
the auxiliary input mode $e'$ is in the photonic vacuum state,
while the output mode $e''$ is in the phononic vacuum state. In
unit (iii), the balanced BS action on the optical modes $a'$ and
$b''$ in Fig.~\ref{fig13}(c) corresponds to the transformation of
the balanced linear coupler on the photonic ($a$) and phononic
($b$) modes, as shown in Fig.~\ref{fig13}(a).

Clearly, the linear-coupler transformation $U_{\rm LC}(\theta)$ is
applied not only to the modes $(a,b)$, but also to other modes.
Thus, Eq.~(\ref{LC2}) should be adequately modified by replacing
$(a,b)$ by $(c,d)$, $(e',b')$, and $(a',b'')$. For brevity, we
omit their explicit obvious definitions here. Note that $U_{\rm
LC}(\pi/2)$ and $U_{\rm LC}(\pi/4)$ correspond to a multi-level
SWAP and Hadamard-like gates, respectively; while the balanced BS
in Fig.~\ref{fig13}(c) corresponds to $U_{\rm LC}(\pi/4)$.

\section{Discussion}\label{Section:Conclusions}

We proposed a novel type of boson blockade, as referred to as
hybrid photon-phonon blockade, which is a generalisation of the
standard photon and phonon blockade effects. We predicted the new
effect in a hybrid mode obtained by linear coupling of photonic
and phononic modes. We described how hybrid photon-phonon blockade
can be generated and detected in a driven nonlinear optomechanical
superconducting system. Specifically, we considered the system
composed of linearly coupled microwave and mechanical resonators
with a superconducting qubit inserted in one of them.

We studied boson-number correlations in the photon, phonon, and
hybrid modes in the system. By analysing steady-state second-order
correlation functions, we found such parameter regimes of the
system for which four different types of boson blockade and/or
boson-induced tunnelling can be observed. Thus, we showed that
bosons generated in the studied system can exhibit the
sub-Poissonian (or super-Poissonian) boson-number statistics
accompanied by boson antibunching in some cases or bunching in
others. These results can be interpreted as four different types
of blockade or tunnelling effects, as summarised in
Table~\ref{table4}.

By tuning the pump frequency with respect to the energy levels of
the hybrid system, which is driven via the SMR, we showed that it
is possible to observe PB and PIT that can be explained by a large
energy-level anharmonicity in the strong-coupling (or
large-nonlinearity) regime. However, the time evolution of the
second-order correlation function $g^{(2)}(\tau)$ oscillates due
to the coupling $g$ between the SMR and qubit as well as the
hopping $f$ between the SMR and QD. We showed that it is possible
to induce PB in the hybrid mode $c$ that survives for much longer
delay times by driving the QD instead of the SMR.

We also predicted unconventional PB in the three modes in the
weak-coupling (or weak-nonlinearity) regime using a non-Hermitian
Hamiltonian approach based on neglecting quantum jumps. Our
analytical approximate predictions are in a relatively good
agreement with our precise master-equation solutions (including
quantum jumps).

Moreover, as summarised in Table~\ref{table8}, we showed the
possibility to observe eight different combinations of either PB
or PIT in the three modes ($a$, $b$, and $c$) in different
coupling regimes of this system. Thus, in particular, we found
that the tunnelling effects in the photonic and phononic modes can
lead, by their simple linear mixing, to the hybrid photon-phonon
blockade effect.

Finally, we discussed two methods of detecting hybrid-mode
correlations. One of them is based on measuring various moments of
photons and phonons via balanced homodyne correlation
measurements. While the other method is based on converting
phonons of the hybrid mode into photons, by using a linear coupler
acting as a multi-level SWAP gate, and then applying the standard
optical HBT interferometry.

We believe that our study of the interplay between photons and
phonons can lead to developing new experimental methods for
controlling and testing the quantum states of mechanical systems
with atom-cavity-mechanics polaritons. We hope that our work can
also stimulate research on quantum engineering with hybrid
photon-phonon modes.

\section*{Acknowledgements}

This work was supported by the Polish National Science Centre
(NCN) under the Maestro Grant No. DEC-2019/34/A/ST2/00081. J.P.
acknowledges the support from M\v{S}MT \v{C}R projects No.
CZ.02.2.69/0.0/0.0/18\_053/0016919.

\section*{Methods}

\subsection{Parameters used in our simulations}\label{App:Parameters}

Our figures, as indicated in their captions, are plotted for the
SMR-driven dissipative system described by the Hamiltonian $H'$,
given in Eq.~(\ref{Hamiltonian1}), assuming:
\begin{eqnarray}
  A_1 &=& \{\Delta_a=-3\gamma, \Delta_b=3\gamma, \Delta_q =
-6\gamma, f=5\gamma,
 \eta_a= 0.7\gamma, \eta_b= 0, \kappa_a=1.5\gamma,
\kappa_b=6\gamma\}, \label{setA1}
\end{eqnarray}
and for the QD-driven dissipative system for the Hamiltonian
$H''$, given in Eq.~(\ref{Hamiltonian2}), assuming either
\begin{eqnarray}
  A_2 &=& \{ \Delta_a=5\gamma, \Delta_b=-5\gamma, \Delta_q =
3\gamma, f=7\gamma,
\eta_a= 0, \eta_b= 0.5\gamma,  \kappa_a=7.5\gamma,
\kappa_b=6\gamma \}, \label{setA2}
\end{eqnarray}
or
\begin{eqnarray}
  A_3 &=& \{\Delta_a=4\gamma, \Delta_b=-4\gamma, \Delta_q =
7\gamma, f=6.4\gamma,
\eta_a= 0, \eta_b= 0.22\gamma, \kappa_a=3.5\gamma,
\kappa_b=0.002\gamma\},\quad
 \label{setA3}
\end{eqnarray}
where $\gamma=10\pi$~MHz. Minor modifications of these parameters
are specified in figure captions.

\begin{figure}[ht]
\centering
\includegraphics[width=\linewidth]{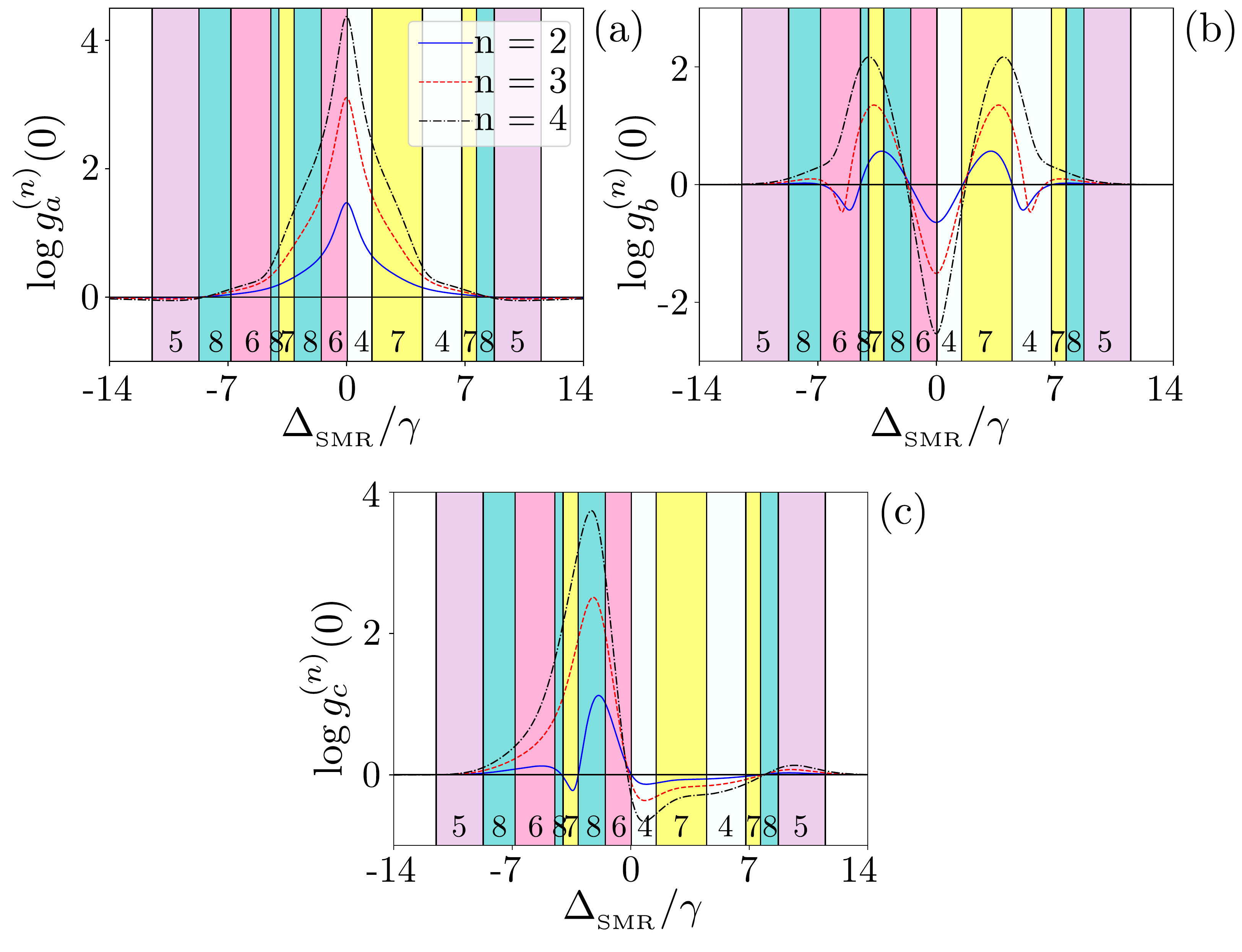}
\caption{Correlation functions $\log g_{i}^{(n)}(0)$ of various
orders [second (solid curves), third (dashed), and forth
(dot-dashed)] versus the detuning between the drive and SMR (in
units of the qubit decay rate $\gamma$) for the QD-driven system
for: (a) the photonic mode $a$, (b) the phononic mode $b$, and (c)
the hybrid mode $c$. All parameters and colourful regions are the
same as in Figs.~\ref{fig10}(a).} \label{fig14}
\end{figure}
\subsection{Higher-order correlation effects}

Here we briefly study the $k$th-order boson-number correlation
functions $g_z^{(k)}(0)$, as defined in Eq.~(\ref{gk0}) for
$k=3,4,$ in comparison to the standard second-order function
$g_z^{(2)}(0)$ for the photon ($z=a$), phonon ($b$), and hybrid
photon-phonon ($c$) modes.

Figure~\ref{fig14} shows our results for $g_z^{(3)}(0)$ (dashed
curves) and $g_z^{(4)}(0)$ (dot-dashed curves) in comparison to
$g_z^{(2)}(0)$ (solid curves) for $z=a,b,c$ in corresponding
panels. Note that the same curves for $g_z^{(2)}(0)$ are also
shown in Figs.~\ref{fig10}(a), but we repeat them for a better
comparison with $g_z^{(3,4)}(0)$. It is seen that the eight cases
of Table~\ref{table8} can be divided into a number of subcases
depending on $g_z^{(3)}(0)$ and $g_z^{(4)}(0)$. Such a
classification is quite complex as includes, in principle,
$8^3=512$ cases. So, instead of that, we present another
much-simplified classification of eight cases only, as shown in
Table~\ref{table8app} using the auxiliary function $g_{234}$
defined as:
\begin{equation}
  g_{234} = \Big[\sgn\log g_z^{(2)}(0),\sgn\log g_z^{(3)}(0),\sgn\log g_z^{(4)}(0)\Big].
\label{g234}
\end{equation}
In particular $[-,-,-]$ means that the second-, third- and
fourth-order sub-Poissonian photon number-statistics are observed
in a given mode, which are the necessary conditions for observing
a `true' single-PB. This case can be easily identified in both
panels of Fig.~\ref{fig14}. One can also find the case when
$[+,+,+]$, which corresponds to the super-Poissonian statistics of
orders $k=2$, 3, and 4, which might be interpreted, as the induced
tunnelling by one, two, and three photons. However, we can also
find intermediate four out of six cases, which can be interpreted
as non-standard types single-PB and/or single-PIT, and in some
cases can be identified as multi-PB~\cite{Adam2013, Hamsen2017,
Huang2018, Kowalewska2019, Li2019}. However, a detailed
classification of such multi-PB and their interpretation is not at
the focus of this paper. The presented results show only the
possibility of generating in our system a plethora of various
photon-phonon correlation effects, which can be revealed by
higher-order correlation functions for the experimentally feasible
parameters.

\begin{table}[tbp]
\begin{tabular}{ ccccc}
 \hline\hline
Case & $g_{234}$ & Mode $a$ & Mode $b$  & Mode $c$  \\
\hline
1& $(-,-,-)$ & $\surd$  & $\surd$  & $\surd$ \\
2& $(-,-,+)$ & $\times$  & $\surd$  & $\surd$$^*$  \\
3& $(-,+,-)$ & $\times$  & $\times$  & $\times$  \\
4& $(+,-,-)$ & $\surd$  & $\surd$  & $\surd$  \\
5& $(-,+,+)$ &  $\times$ &$\surd$   &$\surd$   \\
6& $(+,-,+)$ &  $\times$ &  $\times$ & $\times$  \\
7& $(+,+,-)$ & $\surd$  &$\surd$   & $\surd$  \\
8& $(+,+,+)$ & $\surd$  & $\surd$  & $\surd$ \\
 \hline
 \hline
\end{tabular}
\caption{Different predictions of the $n$th-order super- and
sub-Poissonian statistics with $n=2,3,4$ for the photon ($z=a$),
phonon ($b$), and hybrid photon-phonon ($c$) modes, where
$g_{234}$ is defined in Eq.~(\ref{g234}). The cases marked with
$\surd$ can be identified under both (i) nonresonance conditions,
as shown in Fig.~\ref{fig14}(b), and (ii) resonance conditions, as
shown in Fig.~\ref{fig14}(a), except the case marked with $^*$.}
\label{table8app}
\end{table}

\subsection{Analytical approach via non-Hermitian Hamiltonian
in Eq.~\eqref{Hamiltonian_BS}}\label{App:UPB}

Here, we follow the method of Ref.~\cite{Bamba2011} to derive the
coefficients $C_{n,m,k}$ and $C^\prime_{n,m,k}$ for $n,m
\in{0,1,2}$ and $k=e,g$, which appear in Eqs.~\eqref{corr1} and
\eqref{corr2}.

First we recall that the balanced linear coupler (or a balanced
beam splitter) transformation, which leads to Eq.~\eqref{LC2}, if
applied to the input Fock states $\ket{n_a, n_b}$ for $n_a+n_b\le
2$ yields:
\begin{eqnarray}
\ket{10}\rightarrow \frac{1}{\sqrt{2}}(\ket{10}-\ket{01}), \quad
\ket{01}\rightarrow \frac{1}{\sqrt{2}}(\ket{10}+\ket{01}), \quad
\ket{11}\rightarrow \frac{1}{\sqrt{2}}(\ket{20}-\ket{02}),\nonumber\\
\ket{02}\rightarrow
\frac{1}{2}(\ket{20}+\sqrt{2}\ket{11}+\ket{02}), \quad
\ket{20}\rightarrow \frac{1}{2}(\ket{20}
-\sqrt{2}\ket{11}+\ket{02}).
\end{eqnarray}{}
So, for the input state $\ket{\Psi_{abq}(t)}$, given in
Eq.~(\ref{Eq:13}), the output state of the balanced linear coupler
can be represented as follows:
\begin{eqnarray}
\ket{\Psi_{cdq}(t)}&=&C_{00g}\ket{00g}+e^{-i\omega_d
t}\Big(C_{00e}\ket{00e}+C^\prime_{10g}\ket{10g}
+C^\prime_{01g}\ket{01g}\Big) \nonumber\\&& +e^{-2 i \omega_d
t}\Big(C^\prime_{10e}\ket{10e}
+C^\prime_{01e}\ket{01e} 
+C^\prime_{11g}\ket{11g}+C^\prime_{20g}\ket{20g}
+C^\prime_{02g}\ket{02g}\Big), \nonumber\\
 \label{Eq:14}
\end{eqnarray}
where the superposition coefficients are:
\begin{eqnarray}
C^\prime_{10g}&=&\frac{1}{\sqrt{2}}(C_{10g}+C_{01g}),\nonumber\\
C^\prime_{01g}&=&\frac{1}{\sqrt{2}}(C_{10g}-C_{01g}),\nonumber\\
C^\prime_{10e}&=&\frac{1}{\sqrt{2}}(C_{10e}+C_{01e}),\nonumber\\
C^\prime_{01e}&=&\frac{1}{\sqrt{2}}(C_{10e}-C_{01e}),\nonumber\\
C^\prime_{11g}&=&\frac{1}{\sqrt{2}}(C_{20g}-C_{02g}),\nonumber\\
C^\prime_{20g}&=&\frac{1}{2}(C_{20g}+\sqrt{2}C_{11g}+C_{02g}),\nonumber\\
C^\prime_{02g}&=&\frac{1}{2}(C_{20g}-\sqrt{2}C_{11g}+C_{02g}).
\label{Eq:15}
\end{eqnarray}
We can calculate the coefficients $C_{n_a,n_b,g/e}$
iteratively~\cite{Bamba2011}. For a single excitation and assuming
the resonance case $\Delta_{\SMR}=\Delta_{m}=\Delta_{q}=\Delta$
and $\kappa_{a}=\kappa_b=\kappa$, the steady-state superposition
coefficients can be calculated from:
\begin{eqnarray}
0&=&\left(\Delta-\frac{i\kappa}{2}\right)C_{01g}+f C_{10g}+\eta C_{00g},\nonumber\\
0&=&\left(\Delta-\frac{i\kappa}{2}\right)C_{10g}+f C_{01g}+g C_{00e}, \nonumber\\
0&=&\left(\Delta-\frac{i\gamma}{2}\right)C_{00e}+g C_{10g},
\label{Eq:16}
\end{eqnarray}
where $\eta=\eta_b$, $\Delta=\omega_i-\omega_p$ and
$\omega_{\SMR}=\omega_m=\omega_q=\omega$. Moreover, we assume the
weak-driving regime. So, in the first iteration,  the
contributions from the states with more than a single excitation,
such as $C_{01e}$, $C_{11g}$, ..., are negligible. From
Eq.~\eqref{Eq:16}, by comparing the coefficients with a single
excitation, we can see that $C_{10g}$ and $C_{00e}$ are much
larger than $C_{01g}$, because of a weak-pump amplitude $\eta$,
and they can be written as
\begin{eqnarray}
C_{10g}=\frac{f(24\Delta-2i\kappa)C_{01g}}{(24g^2-24\Delta^2+14i\kappa\Delta+\kappa^2)}\, ,\nonumber\\
C_{00e}=-\frac{24f g
C_{01g}}{(24g^2-24\Delta^2+14i\kappa\Delta+\kappa^2)}.
\end{eqnarray}
In the second iteration, to include states with two excitations in
total, the steady-state coefficients can be calculated from:
\begin{eqnarray}
0 &=& 2\Delta_{\kappa} C_{11g}+\sqrt{2}f C_{20g}+\sqrt{2}f C_{02g}+g C_{01e}+\eta C_{10g}, \nonumber\\
0 &=&\Delta_{\kappa} C_{10e}+\Delta_{\gamma} C_{10e}+f C_{01e}+\sqrt{2}g C_{20g}, \nonumber\\
0 &=&\Delta_{\kappa} C_{01e}+\Delta_{\gamma} C_{01e}+f C_{10e}+g C_{11g}+\eta C_{00e}, \nonumber\\
0 &=&2\Delta_{\kappa} C_{20g}+\sqrt{2}f C_{11g}+\sqrt{2}g C_{10e}, \nonumber\\
0 &=&2\Delta_{\kappa} C_{02g}+\sqrt{2}f C_{11g}+\sqrt{2}\eta C_{01g}, \nonumber\\
\label{Eq:18}
\end{eqnarray}
where $\Delta_{\kappa}=\Delta-i\kappa/2$ and
$\Delta_{\gamma}=\Delta-i\gamma/2$. As can be seen from
Eq.~(\ref{Eq:18}), we have
\begin{eqnarray}
C_{02g}= -(\sqrt{2}f C_{11g}+\sqrt{2}\eta
C_{01g})/(2\Delta_{\kappa}).
\end{eqnarray}
So, to minimise $C_{02g}$, the minimalization of $C_{11g}$ and
$C_{01g}$ is also required. Destructive interference between the
direct and indirect excitation paths in the energy ladders of the
total system can enable us minimising $C_{02g}$. This explains the
occurrence of the dip in $g^{(2)}(0)$ in the mode $b$, as a
signature of PB. As clearly seen in Fig.~\ref{fig11}(a), the
optimal PB in this mode occurs at $\Delta_{\SMR}/g=\pm 1.2$. The
above equations lead us to analytical optimal conditions for the
system parameters to maximise the sub-Poissonian character of the
QD mode and, thus, to optimise the parameters for observing PB in
the mode $b$. Given Eq.~\eqref{Eq:19} for a single excitation and
Eq.~\eqref{Eq:20} for two excitations, which are calculated from
Eq.~\eqref{Eq:18}, we show that the second-order correlation
function calculated by this method and the master equation method
both give very similar predictions, as shown in Fig.~\ref{fig11},
where the thick curves are calculated based on the non-Hermitian
Hamiltonian approach and the thin curves correspond to the
master-equation approach for the modes $a$, $b$, and $c$.

Thus we find
\begin{eqnarray}
C_{01g}&=&(\Delta_{\kappa} \Delta_{\gamma}-g^2)\eta X^{-1}_{5},\nonumber\\
C_{10g}&=&-\Delta_{\gamma}f\eta X_{5}^{-1},\nonumber\\
\label{Eq:19}
\end{eqnarray}
which yields
\begin{eqnarray}
C^\prime_{10g}&=&\frac{(\Delta_{\kappa}
\Delta_{\gamma}-\Delta_{\gamma}f-g^2)\eta}{\sqrt{2}X_5
},\nonumber\\
\label{Eq:19_2}
\end{eqnarray}
Analogously, we find
\begin{eqnarray}
C_{02g}&=&\frac{\eta^2[-2\Delta_{\kappa}^3\Delta_{\gamma}X_1+\Delta_{\kappa}^2 X_2g^2-X_6 g^4+g^6]}{\sqrt{2}X_5 (X_3-X_4)},\nonumber\\
C_{20g}&=&-\frac{\eta^2 f^2[2\Delta_{\kappa} \Delta_{\gamma}X_1+(2\Delta_{\kappa} - \Delta_{\gamma})\Delta_{\kappa \gamma}g^2-g^4]}{\sqrt{2}X_5 (X_3-X_4)},\nonumber\\
C_{11g}&=&\frac{\eta^2 f(2\Delta_{\kappa}^2 \Delta_{\gamma}X_1+X_7 g^2+\Delta_{\gamma}g^4)}{X_5 (X_3-X_4)},\nonumber\\
\label{Eq:20}
\end{eqnarray}
where $\Delta_{\kappa \gamma}=\Delta_{\kappa} + \Delta_{\gamma}$
and the  auxiliary functions $X_n$ read:
 $X_1=\Delta_{\kappa \gamma}^2-f^2$,
 $X_2=\Delta_{\kappa \gamma}(2\Delta_{\kappa} +5\Delta_{\gamma})-4f^2$,
 $ X_3=2\Delta_{\kappa}(\Delta_{\kappa}^2- f^2)X_1$,
 $X_4=\big[3\Delta_{\kappa}^2\Delta_{\kappa \gamma}+(\Delta_{\kappa}-\Delta_{\gamma})f^2\big]g^2-\Delta_{\kappa} g^4$,
 $X_5=\Delta_{\kappa}^2\Delta_{\gamma}- \Delta_{\gamma}f^2-\Delta_{\kappa} g^2$,
 $X_6=3\Delta_{\kappa}^2+4\Delta_{\kappa} \Delta_{\gamma}+f^2$,
 and
 $X_7=\Delta_{\kappa}(2f^2-3\Delta_{\gamma}\Delta_{\kappa \gamma})$.
These formulas, together with $C'_{20g}$ in Eq.~(\ref{Eq:15}),
enable us to calculate analytically the correlation functions in
Eqs.~(\ref{corr1}) and~(\ref{corr2}).



\begin{thebibliography}{83}%
\makeatletter
\providecommand \@ifxundefined [1]{%
 \@ifx{#1\undefined}
}%
\providecommand \@ifnum [1]{%
 \ifnum #1\expandafter \@firstoftwo
 \else \expandafter \@secondoftwo
 \fi
}%
\providecommand \@ifx [1]{%
 \ifx #1\expandafter \@firstoftwo
 \else \expandafter \@secondoftwo
 \fi
}%
\providecommand \natexlab [1]{#1}%
\providecommand \enquote  [1]{``#1''}%
\providecommand \bibnamefont  [1]{#1}%
\providecommand \bibfnamefont [1]{#1}%
\providecommand \citenamefont [1]{#1}%
\providecommand \href@noop [0]{\@secondoftwo}%
\providecommand \href [0]{\begingroup \@sanitize@url \@href}%
\providecommand \@href[1]{\@@startlink{#1}\@@href}%
\providecommand \@@href[1]{\endgroup#1\@@endlink}%
\providecommand \@sanitize@url [0]{\catcode `\\12\catcode
`\$12\catcode
  `\&12\catcode `\#12\catcode `\^12\catcode `\_12\catcode `\%12\relax}%
\providecommand \@@startlink[1]{}%
\providecommand \@@endlink[0]{}%
\providecommand \url  [0]{\begingroup\@sanitize@url \@url }%
\providecommand \@url [1]{\endgroup\@href {#1}{\urlprefix }}%
\providecommand \urlprefix  [0]{URL }%
\providecommand \Eprint [0]{\href }%
\providecommand \doibase [0]{http://dx.doi.org/}%
\providecommand \selectlanguage [0]{\@gobble}%
\providecommand \bibinfo  [0]{\@secondoftwo}%
\providecommand \bibfield  [0]{\@secondoftwo}%
\providecommand \translation [1]{[#1]}%
\providecommand \BibitemOpen [0]{}%
\providecommand \bibitemStop [0]{}%
\providecommand \bibitemNoStop [0]{.\EOS\space}%
\providecommand \EOS [0]{\spacefactor3000\relax}%
\providecommand \BibitemShut  [1]{\csname bibitem#1\endcsname}%
\let\auto@bib@innerbib\@empty
\bibitem [{\citenamefont {Imamo{\u g}lu}\ \emph {et~al.}(1997)\citenamefont
  {Imamo{\u g}lu}, \citenamefont {Schmidt}, \citenamefont {Woods},\ and\
  \citenamefont {Deutsch}}]{Imamoglu1997}%
  \BibitemOpen
  \bibinfo {author} {A.~Imamo{\u g}lu}, \bibinfo {author} {H.~Schmidt},
  \bibinfo {author} {G.~Woods},\ and\ \bibinfo {author} {M.~Deutsch},\ \emph
  {\bibinfo {title} {Strongly Interacting Photons in a Nonlinear Cavity}},\
  \href {https://doi.org/10.1103/physrevlett.79.1467} {\bibfield  {journal}
  {\bibinfo  {journal} {Phys. Rev. Lett.}\ }\textbf {\bibinfo {volume} {79}},\
  \bibinfo {pages} {1467} (\bibinfo {year} {1997})}\BibitemShut {NoStop}%
\bibitem [{\citenamefont {Miranowicz}\ \emph {et~al.}(2001)\citenamefont
  {Miranowicz}, \citenamefont {Leo{\'{n}}ski},\ and\ \citenamefont
  {Imoto}}]{Adam2001}%
  \BibitemOpen
  \bibinfo {author} {A.~Miranowicz}, \bibinfo {author} {W.~Leo{\'{n}}ski},\
  and\ \bibinfo {author} {N.~Imoto},\ \emph {\bibinfo {title} {Quantum-optical
  states in finite-dimensional {H}ilbert space. {I}. {G}eneral formalism}},\
  \href {http://dx.doi.org/10.1002/0471231479.ch3} {\bibfield  {journal}
  {\bibinfo  {journal} {Adv. Chem. Phys.}\ }\textbf {\bibinfo {volume}
  {119(I)}},\ \bibinfo {pages} {155} (\bibinfo {year} {2001})}\BibitemShut
  {NoStop}%
\bibitem [{\citenamefont {Leo{\'{n}}ski}\ and\ \citenamefont
  {Miranowicz}(2001)}]{Leonski2001}%
  \BibitemOpen
  \bibinfo {author} {W.~Leo{\'{n}}ski}\ and\ \bibinfo {author}
  {A.~Miranowicz},\ \emph {\bibinfo {title} {Quantum-optical states in
  finite-dimensional {H}ilbert space. {II}. State generation}},\ \href
  {http://dx.doi.org/10.1002/0471231479.ch4} {\bibfield  {journal} {\bibinfo
  {journal} {Adv. Chem. Phys.}\ }\textbf {\bibinfo {volume} {119(I)}},\
  \bibinfo {pages} {195} (\bibinfo {year} {2001})}\BibitemShut {NoStop}%
\bibitem [{\citenamefont {Leo{\'{n}}ski}\ and\ \citenamefont
  {Kowalewska-Kud{\l}aszyk}(2011)}]{Leonski2011}%
  \BibitemOpen
  \bibinfo {author} {W.~Leo{\'{n}}ski}\ and\ \bibinfo {author}
  {A.~Kowalewska-Kud{\l}aszyk},\ \emph {\bibinfo {title} {Quantum Scissors:
  Finite-Dimensional States Engineering}},\ \href
  {https://doi.org/10.1016/b978-0-444-53886-4.00003-4} {\bibfield  {journal}
  {\bibinfo  {journal} {Prog. Opt.}\ }\textbf {\bibinfo {volume} {56}},\
  \bibinfo {pages} {131} (\bibinfo {year} {2011})}\BibitemShut {NoStop}%
\bibitem [{\citenamefont {Birnbaum}\ \emph {et~al.}(2005)\citenamefont
  {Birnbaum}, \citenamefont {Boca}, \citenamefont {Miller}, \citenamefont
  {Boozer}, \citenamefont {Northup},\ and\ \citenamefont
  {Kimble}}]{Birnbaum2005}%
  \BibitemOpen
  \bibinfo {author} {K.~M. Birnbaum}, \bibinfo {author} {A.~Boca}, \bibinfo
  {author} {R.~Miller}, \bibinfo {author} {A.~D. Boozer}, \bibinfo {author}
  {T.~E. Northup},\ and\ \bibinfo {author} {H.~J. Kimble},\ \emph {\bibinfo
  {title} {Photon blockade in an optical cavity with one trapped atom}},\ \href
  {https://doi.org/10.1038/nature03804} {\bibfield  {journal} {\bibinfo
  {journal} {Nature (London)}\ }\textbf {\bibinfo {volume} {436}},\ \bibinfo
  {pages} {87} (\bibinfo {year} {2005})}\BibitemShut {NoStop}%
\bibitem [{\citenamefont {Faraon}\ \emph {et~al.}(2008)\citenamefont {Faraon},
  \citenamefont {Fushman}, \citenamefont {Englund}, \citenamefont {Stoltz},
  \citenamefont {Petroff},\ and\ \citenamefont
  {Vu{\v{c}}kovi{\'{c}}}}]{Faraon2008}%
  \BibitemOpen
  \bibinfo {author} {A.~Faraon}, \bibinfo {author} {I.~Fushman}, \bibinfo
  {author} {D.~Englund}, \bibinfo {author} {N.~Stoltz}, \bibinfo {author}
  {P.~Petroff},\ and\ \bibinfo {author} {J.~Vu{\v{c}}kovi{\'{c}}},\ \emph
  {\bibinfo {title} {Coherent generation of non-classical light on a chip via
  photon-induced tunnelling and blockade}},\ \href
  {https://doi.org/10.1038/nphys1078} {\bibfield  {journal} {\bibinfo
  {journal} {Nat. Phys.}\ }\textbf {\bibinfo {volume} {4}},\ \bibinfo {pages}
  {859} (\bibinfo {year} {2008})}\BibitemShut {NoStop}%
\bibitem [{\citenamefont {Lang}\ \emph {et~al.}(2011)\citenamefont {Lang} \emph
  {et~al.}}]{Lang2011}%
  \BibitemOpen
  \bibinfo {author} {C.~Lang} et~al.,\ \emph {\bibinfo {title} {Observation of
  Resonant Photon Blockade at Microwave Frequencies Using Correlation Function
  Measurements}},\ \href {https://doi.org/10.1103/physrevlett.106.243601}
  {\bibfield  {journal} {\bibinfo  {journal} {Phys. Rev. Lett.}\ }\textbf
  {\bibinfo {volume} {106}},\ \bibinfo {pages} {243601} (\bibinfo {year}
  {2011})}\BibitemShut {NoStop}%
\bibitem [{\citenamefont {Hoffman}\ \emph {et~al.}(2011)\citenamefont {Hoffman}
  \emph {et~al.}}]{Hoffman2011}%
  \BibitemOpen
  \bibinfo {author} {A.~J. Hoffman} et~al.,\ \emph {\bibinfo {title}
  {Dispersive Photon Blockade in a Superconducting Circuit}},\ \href
  {https://doi.org/10.1103/physrevlett.107.053602} {\bibfield  {journal}
  {\bibinfo  {journal} {Phys. Rev. Lett.}\ }\textbf {\bibinfo {volume} {107}},\
  \bibinfo {pages} {053602} (\bibinfo {year} {2011})}\BibitemShut {NoStop}%
\bibitem [{\citenamefont {Reinhard}\ \emph {et~al.}(2011)\citenamefont
  {Reinhard} \emph {et~al.}}]{Reinhard2011}%
  \BibitemOpen
  \bibinfo {author} {A.~Reinhard} et~al.,\ \emph {\bibinfo {title} {Strongly
  correlated photons on a chip}},\ \href
  {https://doi.org/10.1038/nphoton.2011.321} {\bibfield  {journal} {\bibinfo
  {journal} {Nature Photonics}\ }\textbf {\bibinfo {volume} {6}},\ \bibinfo
  {pages} {93} (\bibinfo {year} {2011})}\BibitemShut {NoStop}%
\bibitem [{\citenamefont {M\"{u}ller}\ \emph {et~al.}(2015)\citenamefont
  {M\"{u}ller} \emph {et~al.}}]{Muller2015}%
  \BibitemOpen
  \bibinfo {author} {K.~M\"{u}ller} et~al.,\ \emph {\bibinfo {title} {Coherent
  Generation of Nonclassical Light on Chip via Detuned Photon Blockade}},\
  \href {https://doi.org/10.1103/physrevlett.114.233601} {\bibfield  {journal}
  {\bibinfo  {journal} {Phys. Rev. Lett.}\ }\textbf {\bibinfo {volume} {114}},\
  \bibinfo {pages} {233601} (\bibinfo {year} {2015})}\BibitemShut {NoStop}%
\bibitem [{\citenamefont {Hamsen}\ \emph {et~al.}(2017)\citenamefont {Hamsen},
  \citenamefont {Tolazzi}, \citenamefont {Wilk},\ and\ \citenamefont
  {Rempe}}]{Hamsen2017}%
  \BibitemOpen
  \bibinfo {author} {C.~Hamsen}, \bibinfo {author} {K.~N. Tolazzi}, \bibinfo
  {author} {T.~Wilk},\ and\ \bibinfo {author} {G.~Rempe},\ \emph {\bibinfo
  {title} {Two-Photon Blockade in an Atom-Driven Cavity {QED} System}},\ \href
  {https://doi.org/10.1103/physrevlett.118.133604} {\bibfield  {journal}
  {\bibinfo  {journal} {Phys. Rev. Lett.}\ }\textbf {\bibinfo {volume} {118}},\
  \bibinfo {pages} {133604} (\bibinfo {year} {2017})}\BibitemShut {NoStop}%
\bibitem [{\citenamefont {Snijders}\ \emph {et~al.}(2018)\citenamefont
  {Snijders} \emph {et~al.}}]{Snijders2018}%
  \BibitemOpen
  \bibinfo {author} {H.~Snijders} et~al.,\ \emph {\bibinfo {title} {Observation
  of the Unconventional Photon Blockade}},\ \href
  {https://doi.org/10.1103/physrevlett.121.043601} {\bibfield  {journal}
  {\bibinfo  {journal} {Phys. Rev. Lett.}\ }\textbf {\bibinfo {volume} {121}},\
  \bibinfo {pages} {043601} (\bibinfo {year} {2018})}\BibitemShut {NoStop}%
\bibitem [{\citenamefont {Vaneph}\ \emph {et~al.}(2018)\citenamefont {Vaneph}
  \emph {et~al.}}]{Vaneph2018}%
  \BibitemOpen
  \bibinfo {author} {C.~Vaneph} et~al.,\ \emph {\bibinfo {title} {Observation
  of the Unconventional Photon Blockade in the Microwave Domain}},\ \href
  {https://doi.org/10.1103/physrevlett.121.043602} {\bibfield  {journal}
  {\bibinfo  {journal} {Phys. Rev. Lett.}\ }\textbf {\bibinfo {volume} {121}},\
  \bibinfo {pages} {043602} (\bibinfo {year} {2018})}\BibitemShut {NoStop}%
\bibitem [{\citenamefont {Majumdar}\ \emph {et~al.}(2012)\citenamefont
  {Majumdar}, \citenamefont {Bajcsy},\ and\ \citenamefont
  {Vu{\v{c}}kovi{\'{c}}}}]{Majumdar2012pra}%
  \BibitemOpen
  \bibinfo {author} {A.~Majumdar}, \bibinfo {author} {M.~Bajcsy},\ and\
  \bibinfo {author} {J.~Vu{\v{c}}kovi{\'{c}}},\ \emph {\bibinfo {title}
  {Probing the ladder of dressed states and nonclassical light generation in
  quantum-dot{\textendash}cavity {QED}}},\ \href
  {https://doi.org/10.1103/physreva.85.041801} {\bibfield  {journal} {\bibinfo
  {journal} {Phys. Rev. A}\ }\textbf {\bibinfo {volume} {85}},\ \bibinfo
  {pages} {041801} (\bibinfo {year} {2012})}\BibitemShut {NoStop}%
\bibitem [{\citenamefont {Peyronel}\ \emph {et~al.}(2012)\citenamefont
  {Peyronel}, \citenamefont {Firstenberg}, \citenamefont {Liang}, \citenamefont
  {Hofferberth}, \citenamefont {Gorshkov}, \citenamefont {Pohl}, \citenamefont
  {Lukin},\ and\ \citenamefont {Vuleti{\'c}}}]{Peyronel2012}%
  \BibitemOpen
  \bibinfo {author} {T.~Peyronel}, \bibinfo {author} {O.~Firstenberg}, \bibinfo
  {author} {Q.-Y. Liang}, \bibinfo {author} {S.~Hofferberth}, \bibinfo {author}
  {A.~V. Gorshkov}, \bibinfo {author} {T.~Pohl}, \bibinfo {author} {M.~D.
  Lukin},\ and\ \bibinfo {author} {V.~Vuleti{\'c}},\ \emph {\bibinfo {title}
  {{Quantum nonlinear optics with single photons enabled by strongly
  interacting atoms}}},\ \href {https://doi.org/10.1038/nature11361} {\bibfield
   {journal} {\bibinfo  {journal} {Nature (London)}\ }\textbf {\bibinfo
  {volume} {488}},\ \bibinfo {pages} {57} (\bibinfo {year} {2012})}\BibitemShut
  {NoStop}%
\bibitem [{\citenamefont {Dayan}\ \emph {et~al.}(2008)\citenamefont {Dayan},
  \citenamefont {Parkins}, \citenamefont {Aoki}, \citenamefont {Ostby},
  \citenamefont {Vahala},\ and\ \citenamefont {Kimble}}]{Dayan2008}%
  \BibitemOpen
  \bibinfo {author} {B.~Dayan}, \bibinfo {author} {A.~S. Parkins}, \bibinfo
  {author} {T.~Aoki}, \bibinfo {author} {E.~P. Ostby}, \bibinfo {author} {K.~J.
  Vahala},\ and\ \bibinfo {author} {H.~J. Kimble},\ \emph {\bibinfo {title} {{A
  photon turnstile dynamically regulated by one atom}}},\ \href
  {https://science.sciencemag.org/content/319/5866/1062} {\bibfield  {journal}
  {\bibinfo  {journal} {Science}\ }\textbf {\bibinfo {volume} {319}},\ \bibinfo
  {pages} {1062} (\bibinfo {year} {2008})}\BibitemShut {NoStop}%
\bibitem [{\citenamefont {Tian}\ and\ \citenamefont
  {Carmichael}(1992)}]{Tian1992}%
  \BibitemOpen
  \bibinfo {author} {L.~Tian}\ and\ \bibinfo {author} {H.~J. Carmichael},\
  \emph {\bibinfo {title} {Quantum trajectory simulations of two-state behavior
  in an optical cavity containing one atom}},\ \href
  {https://doi.org/10.1103/physreva.46.r6801} {\bibfield  {journal} {\bibinfo
  {journal} {Phys. Rev. A}\ }\textbf {\bibinfo {volume} {46}},\ \bibinfo
  {pages} {R6801} (\bibinfo {year} {1992})}\BibitemShut {NoStop}%
\bibitem [{\citenamefont {Leo{\'{n}}ski}\ and\ \citenamefont
  {Tana{\'{s}}}(1994)}]{Leonski1994}%
  \BibitemOpen
  \bibinfo {author} {W.~Leo{\'{n}}ski}\ and\ \bibinfo {author}
  {R.~Tana{\'{s}}},\ \emph {\bibinfo {title} {Possibility of producing the
  one-photon state in a kicked cavity with a nonlinear Kerr medium}},\ \href
  {https://doi.org/10.1103/physreva.49.r20} {\bibfield  {journal} {\bibinfo
  {journal} {Phys. Rev. A}\ }\textbf {\bibinfo {volume} {49}},\ \bibinfo
  {pages} {R20} (\bibinfo {year} {1994})}\BibitemShut {NoStop}%
\bibitem [{\citenamefont {Miranowicz}\ \emph {et~al.}(1996)\citenamefont
  {Miranowicz}, \citenamefont {Leo\'nski}, \citenamefont {Dyrting},\ and\
  \citenamefont {Tana\'s}}]{Adam1996}%
  \BibitemOpen
  \bibinfo {author} {A.~Miranowicz}, \bibinfo {author} {W.~Leo\'nski}, \bibinfo
  {author} {S.~Dyrting},\ and\ \bibinfo {author} {R.~Tana\'s},\ \emph {\bibinfo
  {title} {Quantum state engineering in finite-dimensional Hilbert space}},\
  \href {http://www.physics.sk/aps/pubs/1996/aps_1996_46_3_451.pdf} {\bibfield
  {journal} {\bibinfo  {journal} {Acta Phys. Slov.}\ }\textbf {\bibinfo
  {volume} {46}},\ \bibinfo {pages} {451} (\bibinfo {year} {1996})}\BibitemShut
  {NoStop}%
\bibitem [{\citenamefont {Paul}(1982)}]{Paul1982}%
  \BibitemOpen
  \bibinfo {author} {H.~Paul},\ \emph {\bibinfo {title} {Photon
  antibunching}},\ \href {https://link.aps.org/doi/10.1103/RevModPhys.54.1061}
  {\bibfield  {journal} {\bibinfo  {journal} {Rev. Mod. Phys.}\ }\textbf
  {\bibinfo {volume} {54}},\ \bibinfo {pages} {1061} (\bibinfo {year}
  {1982})}\BibitemShut {NoStop}%
\bibitem [{\citenamefont {Teich}\ and\ \citenamefont
  {Saleh}(1988)}]{Teich1988}%
  \BibitemOpen
  \bibinfo {author} {M.~C. Teich}\ and\ \bibinfo {author} {B.~E.~A. Saleh},\
  \emph {\bibinfo {title} {Photon Bunching and Antibunching}},\ \href
  {https://doi.org/10.1016/S0079-6638(08)70174-4} {\bibfield  {journal}
  {\bibinfo  {journal} {Prog. Opt.}\ }\textbf {\bibinfo {volume} {26}},\
  \bibinfo {pages} {1} (\bibinfo {year} {1988})}\BibitemShut {NoStop}%
\bibitem [{\citenamefont {Kozierowski}(1980)}]{Kozierowski1980}%
  \BibitemOpen
  \bibinfo {author} {M.~Kozierowski},\ \emph {\bibinfo {title} {Photon
  antibunching in nonlinear optical phenomena}},\ \href@noop {} {\bibfield
  {journal} {\bibinfo  {journal} {Kvantovaya Elektronika}\ }\textbf {\bibinfo
  {volume} {6}},\ \bibinfo {pages} {695} (\bibinfo {year} {1980})}\BibitemShut
  {NoStop}%
\bibitem [{\citenamefont {Michler}(2000)}]{Michler2000}%
  \BibitemOpen
  \bibinfo {author} {P.~Michler},\ \emph {\bibinfo {title} {A Quantum Dot
  Single-Photon Turnstile Device}},\ \href
  {https://doi.org/10.1126/science.290.5500.2282} {\bibfield  {journal}
  {\bibinfo  {journal} {Science}\ }\textbf {\bibinfo {volume} {290}},\ \bibinfo
  {pages} {2282} (\bibinfo {year} {2000})}\BibitemShut {NoStop}%
\bibitem [{\citenamefont {Wang}\ \emph
  {et~al.}(2016{\natexlab{a}})\citenamefont {Wang}, \citenamefont {Miranowicz},
  \citenamefont {Li},\ and\ \citenamefont {Nori}}]{Wang2016photon}%
  \BibitemOpen
  \bibinfo {author} {X.~Wang}, \bibinfo {author} {A.~Miranowicz}, \bibinfo
  {author} {H.-R. Li},\ and\ \bibinfo {author} {F.~Nori},\ \emph {\bibinfo
  {title} {Multiple-output microwave single-photon source using superconducting
  circuits with longitudinal and transverse couplings}},\ \href
  {https://doi.org/10.1103/physreva.94.053858} {\bibfield  {journal} {\bibinfo
  {journal} {Phys. Rev. A}\ }\textbf {\bibinfo {volume} {94}},\ \bibinfo
  {pages} {053858} (\bibinfo {year} {2016}{\natexlab{a}})}\BibitemShut
  {NoStop}%
\bibitem [{\citenamefont {Shamailov}\ \emph {et~al.}(2010)\citenamefont
  {Shamailov}, \citenamefont {Parkins}, \citenamefont {Collett},\ and\
  \citenamefont {Carmichael}}]{Shamailov2010}%
  \BibitemOpen
  \bibinfo {author} {S.~Shamailov}, \bibinfo {author} {A.~Parkins}, \bibinfo
  {author} {M.~Collett},\ and\ \bibinfo {author} {H.~Carmichael},\ \emph
  {\bibinfo {title} {Multi-photon blockade and dressing of the dressed
  states}},\ \href {https://doi.org/10.1016/j.optcom.2009.10.062} {\bibfield
  {journal} {\bibinfo  {journal} {Opt. Commun.}\ }\textbf {\bibinfo {volume}
  {283}},\ \bibinfo {pages} {766} (\bibinfo {year} {2010})}\BibitemShut
  {NoStop}%
\bibitem [{\citenamefont {Miranowicz}\ \emph {et~al.}(2013)\citenamefont
  {Miranowicz}, \citenamefont {Paprzycka}, \citenamefont {Liu}, \citenamefont
  {Bajer},\ and\ \citenamefont {Nori}}]{Adam2013}%
  \BibitemOpen
  \bibinfo {author} {A.~Miranowicz}, \bibinfo {author} {M.~Paprzycka}, \bibinfo
  {author} {Y.-X. Liu}, \bibinfo {author} {J.~Bajer},\ and\ \bibinfo {author}
  {F.~Nori},\ \emph {\bibinfo {title} {Two-photon and three-photon blockades in
  driven nonlinear systems}},\ \href
  {https://doi.org/10.1103/physreva.87.023809} {\bibfield  {journal} {\bibinfo
  {journal} {Phys. Rev. A}\ }\textbf {\bibinfo {volume} {87}},\ \bibinfo
  {pages} {023809} (\bibinfo {year} {2013})}\BibitemShut {NoStop}%
\bibitem [{\citenamefont {Chakram}\ \emph {et~al.}(2020)\citenamefont {Chakram}
  \emph {et~al.}}]{Chakram2020}%
  \BibitemOpen
  \bibinfo {author} {S.~Chakram} et~al.,\ \emph {\bibinfo {title} {{Multimode
  photon blockade}}},\ \href@noop {} {\bibfield  {journal} {\bibinfo  {journal}
  {arXiv preprint}\ } (\bibinfo {year} {2020})},\ \Eprint
  {http://arxiv.org/abs/arXiv:2010.15292} {arXiv:2010.15292} \BibitemShut
  {NoStop}%
\bibitem [{\citenamefont {Liew}\ and\ \citenamefont {Savona}(2010)}]{Liew2010}%
  \BibitemOpen
  \bibinfo {author} {T.~C.~H. Liew}\ and\ \bibinfo {author} {V.~Savona},\ \emph
  {\bibinfo {title} {Single Photons from Coupled Quantum Modes}},\ \href
  {https://doi.org/10.1103/physrevlett.104.183601} {\bibfield  {journal}
  {\bibinfo  {journal} {Phys. Rev. Lett.}\ }\textbf {\bibinfo {volume} {104}},\
  \bibinfo {pages} {183601} (\bibinfo {year} {2010})}\BibitemShut {NoStop}%
\bibitem [{\citenamefont {Huang}\ \emph {et~al.}(2018)\citenamefont {Huang},
  \citenamefont {Miranowicz}, \citenamefont {Liao}, \citenamefont {Nori},\ and\
  \citenamefont {Jing}}]{Huang2018}%
  \BibitemOpen
  \bibinfo {author} {R.~Huang}, \bibinfo {author} {A.~Miranowicz}, \bibinfo
  {author} {J.-Q. Liao}, \bibinfo {author} {F.~Nori},\ and\ \bibinfo {author}
  {H.~Jing},\ \emph {\bibinfo {title} {Nonreciprocal Photon Blockade}},\ \href
  {https://doi.org/10.1103/physrevlett.121.153601} {\bibfield  {journal}
  {\bibinfo  {journal} {Phys. Rev. Lett.}\ }\textbf {\bibinfo {volume} {121}},\
  \bibinfo {pages} {153601} (\bibinfo {year} {2018})}\BibitemShut {NoStop}%
\bibitem [{\citenamefont {Li}\ \emph {et~al.}(2019)\citenamefont {Li},
  \citenamefont {Huang}, \citenamefont {Xu}, \citenamefont {Miranowicz},\ and\
  \citenamefont {Jing}}]{Li2019}%
  \BibitemOpen
  \bibinfo {author} {B.~Li}, \bibinfo {author} {R.~Huang}, \bibinfo {author}
  {X.~Xu}, \bibinfo {author} {A.~Miranowicz},\ and\ \bibinfo {author}
  {H.~Jing},\ \emph {\bibinfo {title} {Nonreciprocal unconventional photon
  blockade in a spinning optomechanical system}},\ \href
  {https://doi.org/10.1364/prj.7.000630} {\bibfield  {journal} {\bibinfo
  {journal} {Photon. Res.}\ }\textbf {\bibinfo {volume} {7}},\ \bibinfo {pages}
  {630} (\bibinfo {year} {2019})}\BibitemShut {NoStop}%
\bibitem [{\citenamefont {Yang}\ \emph {et~al.}(2019)\citenamefont {Yang},
  \citenamefont {Xia}, \citenamefont {He}, \citenamefont {Li}, \citenamefont
  {Han}, \citenamefont {Zhang}, \citenamefont {Li}, \citenamefont {Zhang},
  \citenamefont {Xu}, \citenamefont {Yang},\ and\ \citenamefont
  {Zhang}}]{Yang2019}%
  \BibitemOpen
  \bibinfo {author} {P.~Yang}, \bibinfo {author} {X.~Xia}, \bibinfo {author}
  {H.~He}, \bibinfo {author} {S.~Li}, \bibinfo {author} {X.~Han}, \bibinfo
  {author} {P.~Zhang}, \bibinfo {author} {G.~Li}, \bibinfo {author} {P.~Zhang},
  \bibinfo {author} {J.~Xu}, \bibinfo {author} {Y.~Yang},\ and\ \bibinfo
  {author} {T.~Zhang},\ \emph {\bibinfo {title} {{Realization of Nonlinear
  Optical Nonreciprocity on a Few-Photon Level Based on Atoms Strongly Coupled
  to an Asymmetric Cavity}}},\ \href
  {https://link.aps.org/doi/10.1103/PhysRevLett.123.233604} {\bibfield
  {journal} {\bibinfo  {journal} {Phys. Rev. Lett.}\ }\textbf {\bibinfo
  {volume} {123}},\ \bibinfo {pages} {233604} (\bibinfo {year}
  {2019})}\BibitemShut {NoStop}%
\bibitem [{\citenamefont {Miranowicz}\ \emph
  {et~al.}(2014{\natexlab{a}})\citenamefont {Miranowicz}, \citenamefont
  {Bajer}, \citenamefont {Paprzycka}, \citenamefont {Liu}, \citenamefont
  {Zagoskin},\ and\ \citenamefont {Nori}}]{Adam2014}%
  \BibitemOpen
  \bibinfo {author} {A.~Miranowicz}, \bibinfo {author} {J.~Bajer}, \bibinfo
  {author} {M.~Paprzycka}, \bibinfo {author} {Y.-X. Liu}, \bibinfo {author}
  {A.~M. Zagoskin},\ and\ \bibinfo {author} {F.~Nori},\ \emph {\bibinfo {title}
  {State-dependent photon blockade via quantum-reservoir engineering}},\ \href
  {https://doi.org/10.1103/physreva.90.033831} {\bibfield  {journal} {\bibinfo
  {journal} {Phys. Rev. A}\ }\textbf {\bibinfo {volume} {90}},\ \bibinfo
  {pages} {033831} (\bibinfo {year} {2014}{\natexlab{a}})}\BibitemShut
  {NoStop}%
\bibitem [{\citenamefont {Huang}\ \emph {et~al.}(2022)\citenamefont {Huang},
  \citenamefont {\"Ozdemir}, \citenamefont {Liao}, \citenamefont {Minganti},
  \citenamefont {Kuang}, \citenamefont {Nori},\ and\ \citenamefont
  {Jing}}]{Huang2022}%
  \BibitemOpen
  \bibinfo {author} {R.~Huang}, \bibinfo {author} {S.~K. \"Ozdemir}, \bibinfo
  {author} {J.-Q. Liao}, \bibinfo {author} {F.~Minganti}, \bibinfo {author}
  {L.-M. Kuang}, \bibinfo {author} {F.~Nori},\ and\ \bibinfo {author}
  {H.~Jing},\ \emph {\bibinfo {title} {Exceptional Photon Blockade: Engineering
  Photon Blockade with Chiral Exceptional Points}},\ \href
  {https://doi.org/10.1002/lpor.202100430} {\bibfield  {journal} {\bibinfo
  {journal} {Laser Phot. Rev.}\ }\textbf {\bibinfo {volume} {n/a}},\ \bibinfo
  {pages} {2100430} (\bibinfo {year} {2022})}\BibitemShut {NoStop}%
\bibitem [{\citenamefont {Pegg}\ \emph {et~al.}(1998)\citenamefont {Pegg},
  \citenamefont {Phillips},\ and\ \citenamefont {Barnett}}]{Pegg1998}%
  \BibitemOpen
  \bibinfo {author} {D.~T. Pegg}, \bibinfo {author} {L.~S. Phillips},\ and\
  \bibinfo {author} {S.~M. Barnett},\ \emph {\bibinfo {title} {{Optical State
  Truncation by Projection Synthesis}}},\ \href
  {https://link.aps.org/doi/10.1103/PhysRevLett.81.1604} {\bibfield  {journal}
  {\bibinfo  {journal} {Phys. Rev. Lett.}\ }\textbf {\bibinfo {volume} {81}},\
  \bibinfo {pages} {1604} (\bibinfo {year} {1998})}\BibitemShut {NoStop}%
\bibitem [{\citenamefont {\"Ozdemir}\ \emph {et~al.}(2001)\citenamefont
  {\"Ozdemir}, \citenamefont {Miranowicz}, \citenamefont {Koashi},\ and\
  \citenamefont {Imoto}}]{Ozdemir2001}%
  \BibitemOpen
  \bibinfo {author} {S.~K. \"Ozdemir}, \bibinfo {author} {A.~Miranowicz},
  \bibinfo {author} {M.~Koashi},\ and\ \bibinfo {author} {N.~Imoto},\ \emph
  {\bibinfo {title} {{Quantum-scissors device for optical state truncation: A
  proposal for practical realization}}},\ \href
  {https://link.aps.org/doi/10.1103/PhysRevA.64.063818} {\bibfield  {journal}
  {\bibinfo  {journal} {Phys. Rev. A}\ }\textbf {\bibinfo {volume} {64}},\
  \bibinfo {pages} {063818} (\bibinfo {year} {2001})}\BibitemShut {NoStop}%
\bibitem [{\citenamefont {\"Ozdemir}\ \emph {et~al.}(2002)\citenamefont
  {\"Ozdemir}, \citenamefont {Miranowicz}, \citenamefont {Koashi},\ and\
  \citenamefont {Imoto}}]{Ozdemir2002}%
  \BibitemOpen
  \bibinfo {author} {S.~K. \"Ozdemir}, \bibinfo {author} {A.~Miranowicz},
  \bibinfo {author} {M.~Koashi},\ and\ \bibinfo {author} {N.~Imoto},\ \emph
  {\bibinfo {title} {{Pulse-mode quantum projection synthesis: Effects of mode
  mismatch on optical state truncation and preparation}}},\ \href
  {https://link.aps.org/doi/10.1103/PhysRevA.66.053809} {\bibfield  {journal}
  {\bibinfo  {journal} {Phys. Rev. A}\ }\textbf {\bibinfo {volume} {66}},\
  \bibinfo {pages} {053809} (\bibinfo {year} {2002})}\BibitemShut {NoStop}%
\bibitem [{\citenamefont {Babichev}\ \emph {et~al.}(2003)\citenamefont
  {Babichev}, \citenamefont {Ries},\ and\ \citenamefont
  {Lvovsky}}]{Babichev2003}%
  \BibitemOpen
  \bibinfo {author} {S.~A. Babichev}, \bibinfo {author} {J.~Ries},\ and\
  \bibinfo {author} {A.~I. Lvovsky},\ \emph {\bibinfo {title} {{Quantum
  scissors: teleportation of single-mode optical states by means of a nonlocal
  single photon}}},\ \href {https://doi.org/10.1209/epl/i2003-00504-y}
  {\bibfield  {journal} {\bibinfo  {journal} {EPL (Europhys. Lett.)}\ }\textbf
  {\bibinfo {volume} {64}},\ \bibinfo {pages} {1} (\bibinfo {year}
  {2003})}\BibitemShut {NoStop}%
\bibitem [{\citenamefont {Koniorczyk}\ \emph {et~al.}(2000)\citenamefont
  {Koniorczyk}, \citenamefont {Kurucz}, \citenamefont {G\'abris},\ and\
  \citenamefont {Janszky}}]{Koniorczyk2000}%
  \BibitemOpen
  \bibinfo {author} {M.~Koniorczyk}, \bibinfo {author} {Z.~Kurucz}, \bibinfo
  {author} {A.~G\'abris},\ and\ \bibinfo {author} {J.~Janszky},\ \emph
  {\bibinfo {title} {{General optical state truncation and its
  teleportation}}},\ \href
  {https://link.aps.org/doi/10.1103/PhysRevA.62.013802} {\bibfield  {journal}
  {\bibinfo  {journal} {Phys. Rev. A}\ }\textbf {\bibinfo {volume} {62}},\
  \bibinfo {pages} {013802} (\bibinfo {year} {2000})}\BibitemShut {NoStop}%
\bibitem [{\citenamefont {Miranowicz}(2005)}]{Adam2005}%
  \BibitemOpen
  \bibinfo {author} {A.~Miranowicz},\ \emph {\bibinfo {title} {{Optical-state
  truncation and teleportation of qudits by conditional eight-port
  interferometry}}},\ \href {https://doi.org/10.1088/1464-4266/7/5/004}
  {\bibfield  {journal} {\bibinfo  {journal} {J. Opt. B: Quant. Semicl. Opt.}\
  }\textbf {\bibinfo {volume} {7}},\ \bibinfo {pages} {142} (\bibinfo {year}
  {2005})}\BibitemShut {NoStop}%
\bibitem [{\citenamefont {Miranowicz}\ \emph
  {et~al.}(2014{\natexlab{b}})\citenamefont {Miranowicz}, \citenamefont
  {Paprzycka}, \citenamefont {Pathak},\ and\ \citenamefont
  {Nori}}]{Adam2014cats}%
  \BibitemOpen
  \bibinfo {author} {A.~Miranowicz}, \bibinfo {author} {M.~Paprzycka}, \bibinfo
  {author} {A.~Pathak},\ and\ \bibinfo {author} {F.~Nori},\ \emph {\bibinfo
  {title} {Phase-space interference of states optically truncated by quantum
  scissors}},\ \href {http://dx.doi.org/10.1103/PhysRevA.89.033812} {\bibfield
  {journal} {\bibinfo  {journal} {Phys. Rev. A}\ }\textbf {\bibinfo {volume}
  {89}},\ \bibinfo {pages} {033812} (\bibinfo {year}
  {2014}{\natexlab{b}})}\BibitemShut {NoStop}%
\bibitem [{\citenamefont {Reck}\ \emph {et~al.}(1994)\citenamefont {Reck},
  \citenamefont {Zeilinger}, \citenamefont {Bernstein},\ and\ \citenamefont
  {Bertani}}]{Reck1994}%
  \BibitemOpen
  \bibinfo {author} {M.~Reck}, \bibinfo {author} {A.~Zeilinger}, \bibinfo
  {author} {H.~J. Bernstein},\ and\ \bibinfo {author} {P.~Bertani},\ \emph
  {\bibinfo {title} {{Experimental realization of any discrete unitary
  operator}}},\ \href {https://link.aps.org/doi/10.1103/PhysRevLett.73.58}
  {\bibfield  {journal} {\bibinfo  {journal} {Phys. Rev. Lett.}\ }\textbf
  {\bibinfo {volume} {73}},\ \bibinfo {pages} {58} (\bibinfo {year}
  {1994})}\BibitemShut {NoStop}%
\bibitem [{\citenamefont {Miranowicz}\ \emph {et~al.}(2007)\citenamefont
  {Miranowicz}, \citenamefont {\"Ozdemir}, \citenamefont {Bajer}, \citenamefont
  {Koashi},\ and\ \citenamefont {Imoto}}]{Adam2007}%
  \BibitemOpen
  \bibinfo {author} {A.~Miranowicz}, \bibinfo {author} {S.~K. \"Ozdemir},
  \bibinfo {author} {J.~Bajer}, \bibinfo {author} {M.~Koashi},\ and\ \bibinfo
  {author} {N.~Imoto},\ \emph {\bibinfo {title} {{Selective truncations of an
  optical state using projection synthesis}}},\ \href
  {https://doi.org/10.1364/JOSAB.24.000379} {\bibfield  {journal} {\bibinfo
  {journal} {J. Opt. Soc. Am. B}\ }\textbf {\bibinfo {volume} {24}},\ \bibinfo
  {pages} {379} (\bibinfo {year} {2007})}\BibitemShut {NoStop}%
\bibitem [{\citenamefont {Leo\'nski}\ and\ \citenamefont
  {Miranowicz}(2004)}]{Leonski2004}%
  \BibitemOpen
  \bibinfo {author} {W.~Leo\'nski}\ and\ \bibinfo {author} {A.~Miranowicz},\
  \emph {\bibinfo {title} {Kerr nonlinear coupler and entanglement}},\ \href
  {http://dx.doi.org/10.1088/1464-4266/6/3/007} {\bibfield  {journal} {\bibinfo
   {journal} {J. Opt. B}\ }\textbf {\bibinfo {volume} {6}},\ \bibinfo {pages}
  {S37} (\bibinfo {year} {2004})}\BibitemShut {NoStop}%
\bibitem [{\citenamefont {Miranowicz}\ and\ \citenamefont
  {Leo{\'{n}}ski}(2006)}]{Adam2006}%
  \BibitemOpen
  \bibinfo {author} {A.~Miranowicz}\ and\ \bibinfo {author}
  {W.~Leo{\'{n}}ski},\ \emph {\bibinfo {title} {Two-mode optical state
  truncation and generation of maximally entangled states in pumped nonlinear
  couplers}},\ \href {http://dx.doi.org/10.1088/0953-4075/39/7/011} {\bibfield
  {journal} {\bibinfo  {journal} {J. Phys. B}\ }\textbf {\bibinfo {volume}
  {39}},\ \bibinfo {pages} {1683} (\bibinfo {year} {2006})}\BibitemShut
  {NoStop}%
\bibitem [{\citenamefont {Bamba}\ \emph {et~al.}(2011)\citenamefont {Bamba},
  \citenamefont {Imamo{\u{g}}lu}, \citenamefont {Carusotto},\ and\
  \citenamefont {Ciuti}}]{Bamba2011}%
  \BibitemOpen
  \bibinfo {author} {M.~Bamba}, \bibinfo {author} {A.~Imamo{\u{g}}lu}, \bibinfo
  {author} {I.~Carusotto},\ and\ \bibinfo {author} {C.~Ciuti},\ \emph {\bibinfo
  {title} {Origin of strong photon antibunching in weakly nonlinear photonic
  molecules}},\ \href {https://doi.org/10.1103/physreva.83.021802} {\bibfield
  {journal} {\bibinfo  {journal} {Phys. Rev. A}\ }\textbf {\bibinfo {volume}
  {83}},\ \bibinfo {pages} {021802} (\bibinfo {year} {2011})}\BibitemShut
  {NoStop}%
\bibitem [{\citenamefont {Flayac}\ and\ \citenamefont
  {Savona}(2017)}]{Flayac2018review}%
  \BibitemOpen
  \bibinfo {author} {H.~Flayac}\ and\ \bibinfo {author} {V.~Savona},\ \emph
  {\bibinfo {title} {Unconventional photon blockade}},\ \href
  {https://doi.org/10.1103/physreva.96.053810} {\bibfield  {journal} {\bibinfo
  {journal} {Phys. Rev. A}\ }\textbf {\bibinfo {volume} {96}},\ \bibinfo
  {pages} {053810} (\bibinfo {year} {2017})}\BibitemShut {NoStop}%
\bibitem [{\citenamefont {Liu}\ \emph {et~al.}(2010)\citenamefont {Liu},
  \citenamefont {Miranowicz}, \citenamefont {Gao}, \citenamefont {Bajer},
  \citenamefont {Sun},\ and\ \citenamefont {Nori}}]{Liu2010}%
  \BibitemOpen
  \bibinfo {author} {Y.-X. Liu}, \bibinfo {author} {A.~Miranowicz}, \bibinfo
  {author} {Y.~B. Gao}, \bibinfo {author} {J.~Bajer}, \bibinfo {author} {C.~P.
  Sun},\ and\ \bibinfo {author} {F.~Nori},\ \emph {\bibinfo {title}
  {Qubit-induced phonon blockade as a signature of quantum behavior in
  nanomechanical resonators}},\ \href
  {https://link.aps.org/doi/10.1103/PhysRevA.82.032101} {\bibfield  {journal}
  {\bibinfo  {journal} {Phys. Rev. A}\ }\textbf {\bibinfo {volume} {82}},\
  \bibinfo {pages} {032101} (\bibinfo {year} {2010})}\BibitemShut {NoStop}%
\bibitem [{\citenamefont {Didier}\ \emph {et~al.}(2011)\citenamefont {Didier},
  \citenamefont {Pugnetti}, \citenamefont {Blanter},\ and\ \citenamefont
  {Fazio}}]{Didier2011}%
  \BibitemOpen
  \bibinfo {author} {N.~Didier}, \bibinfo {author} {S.~Pugnetti}, \bibinfo
  {author} {Y.~M. Blanter},\ and\ \bibinfo {author} {R.~Fazio},\ \emph
  {\bibinfo {title} {Detecting phonon blockade with photons}},\ \href
  {https://doi.org/10.1103/physrevb.84.054503} {\bibfield  {journal} {\bibinfo
  {journal} {Phys. Rev. B}\ }\textbf {\bibinfo {volume} {84}},\ \bibinfo
  {pages} {054503} (\bibinfo {year} {2011})}\BibitemShut {NoStop}%
\bibitem [{\citenamefont {Wang}\ \emph
  {et~al.}(2016{\natexlab{b}})\citenamefont {Wang}, \citenamefont {Miranowicz},
  \citenamefont {Li},\ and\ \citenamefont {Nori}}]{XinWang2016phonon}%
  \BibitemOpen
  \bibinfo {author} {X.~Wang}, \bibinfo {author} {A.~Miranowicz}, \bibinfo
  {author} {H.-R. Li},\ and\ \bibinfo {author} {F.~Nori},\ \emph {\bibinfo
  {title} {Method for observing robust and tunable phonon blockade in a
  nanomechanical resonator coupled to a charge qubit}},\ \href
  {https://link.aps.org/doi/10.1103/PhysRevA.93.063861} {\bibfield  {journal}
  {\bibinfo  {journal} {Phys. Rev. A}\ }\textbf {\bibinfo {volume} {93}},\
  \bibinfo {pages} {063861} (\bibinfo {year} {2016}{\natexlab{b}})}\BibitemShut
  {NoStop}%
\bibitem [{\citenamefont {Miranowicz}\ \emph {et~al.}(2016)\citenamefont
  {Miranowicz}, \citenamefont {Bajer}, \citenamefont {Lambert}, \citenamefont
  {Liu},\ and\ \citenamefont {Nori}}]{Adam2016}%
  \BibitemOpen
  \bibinfo {author} {A.~Miranowicz}, \bibinfo {author} {J.~Bajer}, \bibinfo
  {author} {N.~Lambert}, \bibinfo {author} {Y.-X. Liu},\ and\ \bibinfo {author}
  {F.~Nori},\ \emph {\bibinfo {title} {Tunable multiphonon blockade in coupled
  nanomechanical resonators}},\ \href
  {https://link.aps.org/doi/10.1103/PhysRevA.93.013808} {\bibfield  {journal}
  {\bibinfo  {journal} {Phys. Rev. A}\ }\textbf {\bibinfo {volume} {93}},\
  \bibinfo {pages} {013808} (\bibinfo {year} {2016})}\BibitemShut {NoStop}%
\bibitem [{\citenamefont {Shi}\ \emph {et~al.}(2018)\citenamefont {Shi},
  \citenamefont {Zhou}, \citenamefont {Xu},\ and\ \citenamefont
  {Liu}}]{Shi2018}%
  \BibitemOpen
  \bibinfo {author} {H.-Q. Shi}, \bibinfo {author} {X.-T. Zhou}, \bibinfo
  {author} {X.-W. Xu},\ and\ \bibinfo {author} {N.-H. Liu},\ \emph {\bibinfo
  {title} {Tunable phonon blockade in quadratically coupled optomechanical
  systems}},\ \href {https://doi.org/10.1038/s41598-018-20568-x} {\bibfield
  {journal} {\bibinfo  {journal} {Sci. Rep.}\ }\textbf {\bibinfo {volume}
  {8}},\ \bibinfo {pages} {2212} (\bibinfo {year} {2018})}\BibitemShut
  {NoStop}%
\bibitem [{\citenamefont {Liu}\ \emph {et~al.}(2014)\citenamefont {Liu},
  \citenamefont {Xu}, \citenamefont {Miranowicz},\ and\ \citenamefont
  {Nori}}]{Liu2014}%
  \BibitemOpen
  \bibinfo {author} {Y.~X. Liu}, \bibinfo {author} {X.~W. Xu}, \bibinfo
  {author} {A.~Miranowicz},\ and\ \bibinfo {author} {F.~Nori},\ \emph {\bibinfo
  {title} {From blockade to transparency: Controllable photon transmission
  through a circuit-{QED} system}},\ \href
  {http://dx.doi.org/10.1103/PhysRevA.89.043818} {\bibfield  {journal}
  {\bibinfo  {journal} {Phys. Rev. A}\ }\textbf {\bibinfo {volume} {89}},\
  \bibinfo {pages} {043818} (\bibinfo {year} {2014})}\BibitemShut {NoStop}%
\bibitem [{\citenamefont {Kowalewska-Kud{\l}aszyk}\ \emph
  {et~al.}(2019)\citenamefont {Kowalewska-Kud{\l}aszyk}, \citenamefont {Abo},
  \citenamefont {Chimczak}, \citenamefont {Pe{\v{r}}ina}, \citenamefont
  {Nori},\ and\ \citenamefont {Miranowicz}}]{Kowalewska2019}%
  \BibitemOpen
  \bibinfo {author} {A.~Kowalewska-Kud{\l}aszyk}, \bibinfo {author} {S.~I.
  Abo}, \bibinfo {author} {G.~Chimczak}, \bibinfo {author} {J.~Pe{\v{r}}ina},
  \bibinfo {author} {F.~Nori},\ and\ \bibinfo {author} {A.~Miranowicz},\ \emph
  {\bibinfo {title} {Two-photon blockade and photon-induced tunneling generated
  by squeezing}},\ \href {https://doi.org/10.1103/physreva.100.053857}
  {\bibfield  {journal} {\bibinfo  {journal} {Phys. Rev. A}\ }\textbf {\bibinfo
  {volume} {100}},\ \bibinfo {pages} {053857} (\bibinfo {year}
  {2019})}\BibitemShut {NoStop}%
\bibitem [{\citenamefont {Aspelmeyer}\ \emph {et~al.}(2014)\citenamefont
  {Aspelmeyer}, \citenamefont {Kippenberg},\ and\ \citenamefont
  {Marquardt}}]{Aspelmeyer2014}%
  \BibitemOpen
  \bibinfo {author} {M.~Aspelmeyer}, \bibinfo {author} {T.~J. Kippenberg},\
  and\ \bibinfo {author} {F.~Marquardt},\ \emph {\bibinfo {title} {Cavity
  optomechanics}},\ \href {https://doi.org/10.1103/revmodphys.86.1391}
  {\bibfield  {journal} {\bibinfo  {journal} {Rev. Mod. Phys.}\ }\textbf
  {\bibinfo {volume} {86}},\ \bibinfo {pages} {1391} (\bibinfo {year}
  {2014})}\BibitemShut {NoStop}%
\bibitem [{\citenamefont {Xu}\ \emph {et~al.}(2019)\citenamefont {Xu},
  \citenamefont {Shi}, \citenamefont {Liao},\ and\ \citenamefont
  {Chen}}]{Xu2019}%
  \BibitemOpen
  \bibinfo {author} {X.-W. Xu}, \bibinfo {author} {H.-Q. Shi}, \bibinfo
  {author} {J.-Q. Liao},\ and\ \bibinfo {author} {A.-X. Chen},\ \emph {\bibinfo
  {title} {Generation of single entangled photon-phonon pairs via an
  atom-photon-phonon interaction}},\ \href
  {https://doi.org/10.1103/physreva.100.053802} {\bibfield  {journal} {\bibinfo
   {journal} {Phys. Rev. A}\ }\textbf {\bibinfo {volume} {100}},\ \bibinfo
  {pages} {053802} (\bibinfo {year} {2019})}\BibitemShut {NoStop}%
\bibitem [{\citenamefont {Xu}\ \emph {et~al.}(2018)\citenamefont {Xu},
  \citenamefont {Shi}, \citenamefont {Chen},\ and\ \citenamefont
  {x.~Liu}}]{Xu2018}%
  \BibitemOpen
  \bibinfo {author} {X.-W. Xu}, \bibinfo {author} {H.-Q. Shi}, \bibinfo
  {author} {A.-X. Chen},\ and\ \bibinfo {author} {Y.~x.~Liu},\ \emph {\bibinfo
  {title} {Cross-correlation between photons and phonons in quadratically
  coupled optomechanical systems}},\ \href
  {https://doi.org/10.1103/physreva.98.013821} {\bibfield  {journal} {\bibinfo
  {journal} {Phys. Rev. A}\ }\textbf {\bibinfo {volume} {98}},\ \bibinfo
  {pages} {013821} (\bibinfo {year} {2018})}\BibitemShut {NoStop}%
\bibitem [{\citenamefont {Zhai}\ \emph {et~al.}(2019)\citenamefont {Zhai},
  \citenamefont {Huang}, \citenamefont {Jing},\ and\ \citenamefont
  {Kuang}}]{Zhai2019}%
  \BibitemOpen
  \bibinfo {author} {C.~Zhai}, \bibinfo {author} {R.~Huang}, \bibinfo {author}
  {H.~Jing},\ and\ \bibinfo {author} {L.-M. Kuang},\ \emph {\bibinfo {title}
  {Mechanical switch of photon blockade and photon-induced tunneling}},\ \href
  {https://doi.org/10.1364/oe.27.027649} {\bibfield  {journal} {\bibinfo
  {journal} {Opt. Express}\ }\textbf {\bibinfo {volume} {27}},\ \bibinfo
  {pages} {27649} (\bibinfo {year} {2019})}\BibitemShut {NoStop}%
\bibitem [{\citenamefont {Santori}\ \emph {et~al.}(2001)\citenamefont
  {Santori}, \citenamefont {Pelton}, \citenamefont {Solomon}, \citenamefont
  {Dale},\ and\ \citenamefont {Yamamoto}}]{Santori2001}%
  \BibitemOpen
  \bibinfo {author} {C.~Santori}, \bibinfo {author} {M.~Pelton}, \bibinfo
  {author} {G.~Solomon}, \bibinfo {author} {Y.~Dale},\ and\ \bibinfo {author}
  {Y.~Yamamoto},\ \emph {\bibinfo {title} {Triggered Single Photons from a
  Quantum Dot}},\ \href {https://doi.org/10.1103/physrevlett.86.1502}
  {\bibfield  {journal} {\bibinfo  {journal} {Phys. Rev. Lett.}\ }\textbf
  {\bibinfo {volume} {86}},\ \bibinfo {pages} {1502} (\bibinfo {year}
  {2001})}\BibitemShut {NoStop}%
\bibitem [{\citenamefont {Ding}\ \emph {et~al.}(2016)\citenamefont {Ding} \emph
  {et~al.}}]{Ding2016}%
  \BibitemOpen
  \bibinfo {author} {X.~Ding} et~al.,\ \emph {\bibinfo {title} {On-Demand
  Single Photons with High Extraction Efficiency and Near-Unity
  Indistinguishability from a Resonantly Driven Quantum Dot in a
  Micropillar}},\ \href {https://doi.org/10.1103/physrevlett.116.020401}
  {\bibfield  {journal} {\bibinfo  {journal} {Phys. Rev. Lett.}\ }\textbf
  {\bibinfo {volume} {116}},\ \bibinfo {pages} {020401} (\bibinfo {year}
  {2016})}\BibitemShut {NoStop}%
\bibitem [{\citenamefont {Grangier}\ \emph {et~al.}(1998)\citenamefont
  {Grangier}, \citenamefont {Walls},\ and\ \citenamefont
  {Gheri}}]{Grangier1998}%
  \BibitemOpen
  \bibinfo {author} {P.~Grangier}, \bibinfo {author} {D.~F. Walls},\ and\
  \bibinfo {author} {K.~M. Gheri},\ \emph {\bibinfo {title} {Comment on
  {\textquotedblleft}Strongly Interacting Photons in a Nonlinear
  Cavity{\textquotedblright}}},\ \href
  {https://doi.org/10.1103/physrevlett.81.2833} {\bibfield  {journal} {\bibinfo
   {journal} {Phys. Rev. Lett.}\ }\textbf {\bibinfo {volume} {81}},\ \bibinfo
  {pages} {2833} (\bibinfo {year} {1998})}\BibitemShut {NoStop}%
\bibitem [{\citenamefont {Kimble}(1998)}]{Kimble1998}%
  \BibitemOpen
  \bibinfo {author} {H.~J. Kimble},\ \emph {\bibinfo {title} {Strong
  Interactions of Single Atoms and Photons in Cavity {QED}}},\ \href
  {https://doi.org/10.1238/physica.topical.076a00127} {\bibfield  {journal}
  {\bibinfo  {journal} {Phys. Scripta}\ }\textbf {\bibinfo {volume} {T76}},\
  \bibinfo {pages} {127} (\bibinfo {year} {1998})}\BibitemShut {NoStop}%
\bibitem [{\citenamefont {Gu}\ \emph {et~al.}(2017)\citenamefont {Gu},
  \citenamefont {Kockum}, \citenamefont {Miranowicz}, \citenamefont {Liu},\
  and\ \citenamefont {Nori}}]{Gu2017}%
  \BibitemOpen
  \bibinfo {author} {X.~Gu}, \bibinfo {author} {A.~F. Kockum}, \bibinfo
  {author} {A.~Miranowicz}, \bibinfo {author} {Y.-X. Liu},\ and\ \bibinfo
  {author} {F.~Nori},\ \emph {\bibinfo {title} {Microwave photonics with
  superconducting quantum circuits}},\ \href
  {https://doi.org/10.1016/j.physrep.2017.10.002} {\bibfield  {journal}
  {\bibinfo  {journal} {Phys. Rep.}\ }\textbf {\bibinfo {volume} {718-719}},\
  \bibinfo {pages} {1} (\bibinfo {year} {2017})}\BibitemShut {NoStop}%
\bibitem [{\citenamefont {Tian}(2009)}]{Tian2009}%
  \BibitemOpen
  \bibinfo {author} {L.~Tian},\ \emph {\bibinfo {title} {Ground state cooling
  of a nanomechanical resonator via parametric linear coupling}},\ \href
  {https://doi.org/10.1103/physrevb.79.193407} {\bibfield  {journal} {\bibinfo
  {journal} {Phys. Rev. B}\ }\textbf {\bibinfo {volume} {79}},\ \bibinfo
  {pages} {193407} (\bibinfo {year} {2009})}\BibitemShut {NoStop}%
\bibitem [{\citenamefont {Kockum}\ \emph {et~al.}(2019)\citenamefont {Kockum},
  \citenamefont {Miranowicz}, \citenamefont {Liberato}, \citenamefont
  {Savasta},\ and\ \citenamefont {Nori}}]{Kockum2019}%
  \BibitemOpen
  \bibinfo {author} {A.~F. Kockum}, \bibinfo {author} {A.~Miranowicz}, \bibinfo
  {author} {S.~D. Liberato}, \bibinfo {author} {S.~Savasta},\ and\ \bibinfo
  {author} {F.~Nori},\ \emph {\bibinfo {title} {Ultrastrong coupling between
  light and matter}},\ \href {https://doi.org/10.1038/s42254-018-0006-2}
  {\bibfield  {journal} {\bibinfo  {journal} {Nat. Rev. Phys.}\ }\textbf
  {\bibinfo {volume} {1}},\ \bibinfo {pages} {19} (\bibinfo {year}
  {2019})}\BibitemShut {NoStop}%
\bibitem [{\citenamefont {Restrepo}\ \emph {et~al.}(2014)\citenamefont
  {Restrepo}, \citenamefont {Ciuti},\ and\ \citenamefont
  {Favero}}]{Restrepo2014}%
  \BibitemOpen
  \bibinfo {author} {J.~Restrepo}, \bibinfo {author} {C.~Ciuti},\ and\ \bibinfo
  {author} {I.~Favero},\ \emph {\bibinfo {title} {Single-polariton
  optomechanics}},\ \href {https://doi.org/10.1103/PhysRevLett.112.013601}
  {\bibfield  {journal} {\bibinfo  {journal} {Phys. Rev. Lett.}\ }\textbf
  {\bibinfo {volume} {112}},\ \bibinfo {pages} {013601} (\bibinfo {year}
  {2014})}\BibitemShut {NoStop}%
\bibitem [{\citenamefont {Larson}\ and\ \citenamefont
  {Mavrogordatos}(2021)}]{Larson2021}%
  \BibitemOpen
  \bibinfo {author} {J.~Larson}\ and\ \bibinfo {author} {T.~Mavrogordatos},\
  \href {https://dx.doi.org/10.1088/978-0-7503-3447-1} {\emph {\bibinfo {title}
  {The Jaynes-Cummings Model and Its Descendants}}}\ (\bibinfo  {publisher}
  {IOP Publishing},\ \bibinfo {year} {2021})\BibitemShut {NoStop}%
\bibitem [{\citenamefont {Ridolfo}\ \emph {et~al.}(2012)\citenamefont
  {Ridolfo}, \citenamefont {Leib}, \citenamefont {Savasta},\ and\ \citenamefont
  {Hartmann}}]{Ridolfo2012}%
  \BibitemOpen
  \bibinfo {author} {A.~Ridolfo}, \bibinfo {author} {M.~Leib}, \bibinfo
  {author} {S.~Savasta},\ and\ \bibinfo {author} {M.~J. Hartmann},\ \emph
  {\bibinfo {title} {Photon Blockade in the Ultrastrong Coupling Regime}},\
  \href {https://doi.org/10.1103/physrevlett.109.193602} {\bibfield  {journal}
  {\bibinfo  {journal} {Phys. Rev. Lett.}\ }\textbf {\bibinfo {volume} {109}},\
  \bibinfo {pages} {193602} (\bibinfo {year} {2012})}\BibitemShut {NoStop}%
\bibitem [{\citenamefont {Garziano}\ \emph {et~al.}(2015)\citenamefont
  {Garziano}, \citenamefont {Stassi}, \citenamefont {Macr\`{\i}}, \citenamefont
  {Kockum}, \citenamefont {Savasta},\ and\ \citenamefont
  {Nori}}]{Garziano2015}%
  \BibitemOpen
  \bibinfo {author} {L.~Garziano}, \bibinfo {author} {R.~Stassi}, \bibinfo
  {author} {V.~Macr\`{\i}}, \bibinfo {author} {A.~F. Kockum}, \bibinfo {author}
  {S.~Savasta},\ and\ \bibinfo {author} {F.~Nori},\ \emph {\bibinfo {title}
  {Multiphoton quantum Rabi oscillations in ultrastrong cavity QED}},\ \href
  {https://link.aps.org/doi/10.1103/PhysRevA.92.063830} {\bibfield  {journal}
  {\bibinfo  {journal} {Phys. Rev. A}\ }\textbf {\bibinfo {volume} {92}},\
  \bibinfo {pages} {063830} (\bibinfo {year} {2015})}\BibitemShut {NoStop}%
\bibitem [{\citenamefont {Mercurio}\ \emph {et~al.}(2022)\citenamefont
  {Mercurio}, \citenamefont {Abo}, \citenamefont {Mauceri}, \citenamefont
  {Russo}, \citenamefont {Macri}, \citenamefont {Miranowicz}, \citenamefont
  {Savasta},\ and\ \citenamefont {Stefano}}]{Mercurio2022}%
  \BibitemOpen
  \bibinfo {author} {A.~Mercurio}, \bibinfo {author} {S.~Abo}, \bibinfo
  {author} {F.~Mauceri}, \bibinfo {author} {E.~Russo}, \bibinfo {author}
  {V.~Macri}, \bibinfo {author} {A.~Miranowicz}, \bibinfo {author}
  {S.~Savasta},\ and\ \bibinfo {author} {O.~D. Stefano},\ \href@noop {} {\emph
  {\bibinfo {title} {Pure Dephasing of Light-Matter Systems in the Ultrastrong
  and Deep-Strong Coupling Regimes}}} (\bibinfo {year} {2022}),\ \Eprint
  {http://arxiv.org/abs/arXiv:2205.05352} {arXiv:2205.05352} \BibitemShut
  {NoStop}%
\bibitem [{\citenamefont {S\'anchez Mu\~noz}\ \emph {et~al.}(2020)\citenamefont
  {S\'anchez Mu\~noz}, \citenamefont {Frisk~Kockum}, \citenamefont
  {Miranowicz},\ and\ \citenamefont {Nori}}]{Sanchez2020}%
  \BibitemOpen
  \bibinfo {author} {C.~S\'anchez Mu\~noz}, \bibinfo {author}
  {A.~Frisk~Kockum}, \bibinfo {author} {A.~Miranowicz},\ and\ \bibinfo {author}
  {F.~Nori},\ \emph {\bibinfo {title} {Simulating ultrastrong-coupling
  processes breaking parity conservation in {J}aynes-{C}ummings systems}},\
  \href {https://link.aps.org/doi/10.1103/PhysRevA.102.033716} {\bibfield
  {journal} {\bibinfo  {journal} {Phys. Rev. A}\ }\textbf {\bibinfo {volume}
  {102}},\ \bibinfo {pages} {033716} (\bibinfo {year} {2020})}\BibitemShut
  {NoStop}%
\bibitem [{\citenamefont {Kuhn}(2015)}]{KuhnBook}%
  \BibitemOpen
  \bibinfo {author} {A.~Kuhn},\ \href@noop {} {\emph {\bibinfo {title} {Cavity
  Induced Interfacing of Atoms and Light}}}\ (\bibinfo  {publisher} {Springer,
  Berlin},\ \bibinfo {year} {2015})\BibitemShut {NoStop}%
\bibitem [{\citenamefont {Mandel}\ and\ \citenamefont
  {Wolf}(1995)}]{MandelBook}%
  \BibitemOpen
  \bibinfo {author} {L.~Mandel}\ and\ \bibinfo {author} {E.~Wolf},\ \href@noop
  {} {\emph {\bibinfo {title} {Optical Coherence and Quantum Optics}}}\
  (\bibinfo  {publisher} {Cambridge University Press},\ \bibinfo {address}
  {Cambridge},\ \bibinfo {year} {1995})\BibitemShut {NoStop}%
\bibitem [{\citenamefont {Kimble}\ \emph {et~al.}(1977)\citenamefont {Kimble},
  \citenamefont {Dagenais},\ and\ \citenamefont {Mandel}}]{Kimble1977}%
  \BibitemOpen
  \bibinfo {author} {H.~J. Kimble}, \bibinfo {author} {M.~Dagenais},\ and\
  \bibinfo {author} {L.~Mandel},\ \emph {\bibinfo {title} {{Photon Antibunching
  in Resonance Fluorescence}}},\ \href
  {http://link.aps.org/doi/10.1103/PhysRevLett.39.691} {\bibfield  {journal}
  {\bibinfo  {journal} {Phys. Rev. Lett.}\ }\textbf {\bibinfo {volume} {39}},\
  \bibinfo {pages} {691} (\bibinfo {year} {1977})}\BibitemShut {NoStop}%
\bibitem [{\citenamefont {Verhagen}\ \emph {et~al.}(2012)\citenamefont
  {Verhagen}, \citenamefont {Del{\'{e}}glise}, \citenamefont {Weis},
  \citenamefont {Schliesser},\ and\ \citenamefont {Kippenberg}}]{Verhagen2012}%
  \BibitemOpen
  \bibinfo {author} {E.~Verhagen}, \bibinfo {author} {S.~Del{\'{e}}glise},
  \bibinfo {author} {S.~Weis}, \bibinfo {author} {A.~Schliesser},\ and\
  \bibinfo {author} {T.~J. Kippenberg},\ \emph {\bibinfo {title}
  {Quantum-coherent coupling of a mechanical oscillator to an optical cavity
  mode}},\ \href {https://doi.org/10.1038/nature10787} {\bibfield  {journal}
  {\bibinfo  {journal} {Nature (London)}\ }\textbf {\bibinfo {volume} {482}},\
  \bibinfo {pages} {63} (\bibinfo {year} {2012})}\BibitemShut {NoStop}%
\bibitem [{\citenamefont {Walls}\ and\ \citenamefont
  {Milburn}(1994)}]{Walls1994}%
  \BibitemOpen
  \bibinfo {author} {D.~F. Walls}\ and\ \bibinfo {author} {G.~J. Milburn},\
  \href {https://doi.org/10.1007/978-3-642-79504-6} {\emph {\bibinfo {title}
  {Quantum Optics}}}\ (\bibinfo  {publisher} {Springer, Berlin},\ \bibinfo
  {year} {1994})\BibitemShut {NoStop}%
\bibitem [{\citenamefont {Kubanek}\ \emph {et~al.}(2008)\citenamefont {Kubanek}
  \emph {et~al.}}]{Kubanek2008}%
  \BibitemOpen
  \bibinfo {author} {A.~Kubanek} et~al.,\ \emph {\bibinfo {title} {Two-Photon
  Gateway in One-Atom Cavity Quantum Electrodynamics}},\ \href
  {https://doi.org/10.1103/physrevlett.101.203602} {\bibfield  {journal}
  {\bibinfo  {journal} {Phys. Rev. Lett.}\ }\textbf {\bibinfo {volume} {101}},\
  \bibinfo {pages} {203602} (\bibinfo {year} {2008})}\BibitemShut {NoStop}%
\bibitem [{\citenamefont {Minganti}\ \emph {et~al.}(2019)\citenamefont
  {Minganti}, \citenamefont {Miranowicz}, \citenamefont {Chhajlany},\ and\
  \citenamefont {Nori}}]{Minganti2019}%
  \BibitemOpen
  \bibinfo {author} {F.~Minganti}, \bibinfo {author} {A.~Miranowicz}, \bibinfo
  {author} {R.~W. Chhajlany},\ and\ \bibinfo {author} {F.~Nori},\ \emph
  {\bibinfo {title} {Quantum exceptional points of non-{H}ermitian
  {H}amiltonians and {L}iouvillians: {T}he effects of quantum jumps}},\ \href
  {https://link.aps.org/doi/10.1103/PhysRevA.100.062131} {\bibfield  {journal}
  {\bibinfo  {journal} {Phys. Rev. A}\ }\textbf {\bibinfo {volume} {100}},\
  \bibinfo {pages} {062131} (\bibinfo {year} {2019})}\BibitemShut {NoStop}%
\bibitem [{\citenamefont {Minganti}\ \emph {et~al.}(2020)\citenamefont
  {Minganti}, \citenamefont {Miranowicz}, \citenamefont {Chhajlany},
  \citenamefont {Arkhipov},\ and\ \citenamefont {Nori}}]{Minganti2020}%
  \BibitemOpen
  \bibinfo {author} {F.~Minganti}, \bibinfo {author} {A.~Miranowicz}, \bibinfo
  {author} {R.~W. Chhajlany}, \bibinfo {author} {I.~I. Arkhipov},\ and\
  \bibinfo {author} {F.~Nori},\ \emph {\bibinfo {title} {Hybrid-{L}iouvillian
  formalism connecting exceptional points of non-{H}ermitian {H}amiltonians and
  {L}iouvillians via postselection of quantum trajectories}},\ \href
  {https://link.aps.org/doi/10.1103/PhysRevA.101.062112} {\bibfield  {journal}
  {\bibinfo  {journal} {Phys. Rev. A}\ }\textbf {\bibinfo {volume} {101}},\
  \bibinfo {pages} {062112} (\bibinfo {year} {2020})}\BibitemShut {NoStop}%
\bibitem [{\citenamefont {Zou}\ and\ \citenamefont {Mandel}(1990)}]{Zou90}%
  \BibitemOpen
  \bibinfo {author} {X.~T. Zou}\ and\ \bibinfo {author} {L.~Mandel},\ \emph
  {\bibinfo {title} {{Photon-antibunching and sub-{P}oissonian photon
  statistics}}},\ \href {http://link.aps.org/doi/10.1103/PhysRevA.41.475}
  {\bibfield  {journal} {\bibinfo  {journal} {Phys. Rev. A}\ }\textbf {\bibinfo
  {volume} {41}},\ \bibinfo {pages} {475} (\bibinfo {year} {1990})}\BibitemShut
  {NoStop}%
\bibitem [{\citenamefont {Miranowicz}\ \emph {et~al.}(2010)\citenamefont
  {Miranowicz}, \citenamefont {Bartkowiak}, \citenamefont {Wang}, \citenamefont
  {Liu},\ and\ \citenamefont {Nori}}]{Adam2010}%
  \BibitemOpen
  \bibinfo {author} {A.~Miranowicz}, \bibinfo {author} {M.~Bartkowiak},
  \bibinfo {author} {X.~Wang}, \bibinfo {author} {Y.-X. Liu},\ and\ \bibinfo
  {author} {F.~Nori},\ \emph {\bibinfo {title} {Testing nonclassicality in
  multimode fields: {A} unified derivation of classical inequalities}},\ \href
  {http://dx.doi.org/10.1103/PhysRevA.82.013824} {\bibfield  {journal}
  {\bibinfo  {journal} {Phys. Rev. A}\ }\textbf {\bibinfo {volume} {82}},\
  \bibinfo {pages} {013824} (\bibinfo {year} {2010})}\BibitemShut {NoStop}%
\bibitem [{\citenamefont {Hong}\ \emph {et~al.}(2017)\citenamefont {Hong} \emph
  {et~al.}}]{Hong2017}%
  \BibitemOpen
  \bibinfo {author} {S.~Hong} et~al.,\ \emph {\bibinfo {title} {Hanbury Brown
  and Twiss interferometry of single phonons from an optomechanical
  resonator}},\ \href {https://doi.org/10.1126/science.aan7939} {\bibfield
  {journal} {\bibinfo  {journal} {Science}\ }\textbf {\bibinfo {volume}
  {358}},\ \bibinfo {pages} {203} (\bibinfo {year} {2017})}\BibitemShut
  {NoStop}%
\bibitem [{\citenamefont {Shchukin}\ and\ \citenamefont
  {Vogel}(2005)}]{Shchukin2005}%
  \BibitemOpen
  \bibinfo {author} {E.~V. Shchukin}\ and\ \bibinfo {author} {W.~Vogel},\ \emph
  {\bibinfo {title} {Nonclassical moments and their measurement}},\ \href
  {https://doi.org/10.1103/physreva.72.043808} {\bibfield  {journal} {\bibinfo
  {journal} {Phys. Rev. A}\ }\textbf {\bibinfo {volume} {72}},\ \bibinfo
  {pages} {043808} (\bibinfo {year} {2005})}\BibitemShut {NoStop}%
\bibitem [{\citenamefont {Shchukin}\ and\ \citenamefont
  {Vogel}(2006)}]{Shchukin2006}%
  \BibitemOpen
  \bibinfo {author} {E.~Shchukin}\ and\ \bibinfo {author} {W.~Vogel},\ \emph
  {\bibinfo {title} {Universal Measurement of Quantum Correlations of
  Radiation}},\ \href {https://doi.org/10.1103/physrevlett.96.200403}
  {\bibfield  {journal} {\bibinfo  {journal} {Phys. Rev. Lett.}\ }\textbf
  {\bibinfo {volume} {96}},\ \bibinfo {pages} {200403} (\bibinfo {year}
  {2006})}\BibitemShut {NoStop}%
\end{thebibliography}

%


%
%
%

\end{document}